\documentclass[aps,prd,onecolumn,groupedaddress,nofootinbib,amssymb]{revtex4}
\usepackage[T1]{fontenc}
\usepackage[latin1]{inputenc}
\usepackage{graphicx,amssymb}
\usepackage[english]{babel}

\begin{document}

\title{Cosmography and large scale structure by $f(R)$-gravity: new results}

\author{Salvatore Capozziello and Vincenzo Salzano}

\affiliation{Dipartimento di Scienze Fisiche, Univ. di Napoli
"Federico II" and  INFN, Sez. di Napoli,   Compl. Univ. di Monte
S. Angelo, Ed. N, via Cinthia, 80126 - Napoli, Italy}
\date{\today}

\begin{abstract}
The so called $f(R)$-gravity has recently attracted a lot of
interest since it could be, in principle,  able to explain the
accelerated expansion of the Universe without adding unknown forms
of dark energy/dark matter but, more simply, extending the General
Relativity by generic functions of the Ricci scalar. However,
apart several phenomenological models, there is no final
$f(R)$-theory capable of fitting all the observations and
addressing all the issues related to the presence of dark energy
and dark matter. An alternative approach could be to "reconstruct"
the form of $f(R)$ starting from data without imposing particular
classes of model. In this review paper, we will consider two
typical cosmological problems where the role of dark energy and
dark matter is crucial. Firstly, assuming generic $f(R)$, we show
that it is possible to relate the cosmographic parameters (namely
the deceleration $q_0$, the jerk $j_0$, the snap $s_0$ and the
lerk $l_0$ parameters) to the present day values of $f(R)$ and its
derivatives $f^{(n)}(R) = d^nf/dR^n$ (with $n = 1, 2, 3$) thus
offering a new tool to constrain such higher order models.  Our
analysis gives the possibility to relate the model-independent
results coming from cosmography to some theoretically motivated
assumptions of $f(R)$ cosmology. Besides, adopting the same
philosophy, we take into account the possibility that galaxy
cluster masses, estimated at X-ray wavelengths, could be
explained, without dark matter, reconstructing the  weak-field
limit of analytic $f(R)$ models. The corrected gravitational
potential, obtained in this approximation, is used to estimate the
total mass of a sample of $12$  well-shaped clusters of galaxies.
Results show that such a gravitational potential provides a fair
fit to the mass of visible matter (i.e. gas + stars) estimated by
X-ray observations, without the need of additional dark matter
while the size of the clusters, as already observed at different
scale for galaxies, strictly depends on the interaction lengths of
the corrections to the Newtonian potential. These two examples
could be paradigmatic to overcome dark energy and dark matter
problems by the extended gravity approach.

\end{abstract}

\maketitle

\section{Introduction}

As soon as astrophysicists realized that Type Ia Supernovae
(SNeIa) were standard candles, it appeared evident that their high
luminosity should make it possible to build a Hubble diagram, i.e.
a plot of the distance\,-\,redshift relation, over cosmologically
interesting distance ranges. Motivated by this attractive
consideration, two independent teams started SNeIa surveys leading
to the unexpected discovery that the Universe expansion is
speeding up rather than decelerating as assumed by the
Cosmological Standard Model
\cite{Perlm97,Perlm99,Riess98,Schmidt98,Garn98}. This surprising
result has now been strengthened by more recent data coming from
SNeIa surveys
\cite{Knop03,Tonry03,Barris04,Riess04,R06,SNLS,ESSENCE,D07}, large
scale structure \cite{Dode02,Perci02,Szal03,Hawk03,pope04} and
cosmic microwave background (CMBR) anisotropy spectrum
\cite{Boom,Stomp01,Netter02,Rebo04,wmap,WMAP,WMAP3}. This large
data set coherently points toward the picture of a spatially flat
Universe undergoing an accelerated expansion driven by a dominant
negative pressure fluid, typically referred to as {\it dark
energy} \cite{copeland}.

While there is a wide consensus on the above scenario depicted by
such good quality data, there is a similarly wide range of
contrasting proposals to solve the dark energy puzzle.
Surprisingly, the simplest explanation, namely the cosmological
constant $\Lambda$ \cite{CarLam,Sahni}, is also the best one from
a statistical point of view \cite{Teg03,Teg06,Sel04}.
Unfortunately, the well known coincidence and 120 orders of
magnitude problems render $\Lambda$ a rather unattractive solution
from a theoretical point of view. Inspired by the analogy with
inflation, a scalar field $\phi$, dubbed {\it quintessence}
\cite{PB03,Pad03}, has then been proposed to give a dynamical
$\Lambda$ term in order to both fit the data and avoid the above
problems. However, such models are still plagued by difficulties
on their own, such as the almost complete freedom in the choice of
the scalar field potential and the fine tuning of the initial
conditions. Needless to say, a plethora of alternative models are
now on the market all sharing the main property to be in agreement
with observations, but relying on completely different physics.

Notwithstanding their differences, all  dark energy models assume
that the observed apparent acceleration is the outcome of some
unknown  ingredient, at fundamental level, to be added to the
cosmic pie. In terms of the Einstein equations, $G_{\mu \nu} =
\chi T_{\mu \nu}$,  the right hand side should include  something
more than the usual matter and radiation components in the
stress\,-\,energy tensor.

As a radically different approach, one can also try to leave
unchanged the source side (actually "observed" since composed by
radiation and baryonic matter), but rather modifying the left hand
side. In a sense, one is therefore interpreting cosmic speed up as
a first signal of the breakdown of the laws of physics as
described by the standard General Relativity (GR). Since this
theory has been experimentally tested only up to the Solar System
scale, there is no a priori theoretical motivation to extend its
validity to extraordinarily larger scales such as the
extragalactic and cosmological ones (up to the last scattering
surface!). Extending GR, not giving up to its positive results at
local scales, opens the way to a large class of alternative
theories of gravity ranging from extra\,-\,dimensions
\cite{DGP,DGP2,DGP3,Lue,Lue2} to nonminimally coupled scalar
fields \cite{stbook,Care04,Petto05,Demia06}. In particular, we are
interested here in fourth order theories
\cite{alle04,capozzcurv,noirev,capfra,noiijmpd,cct,sante,cdtt,Klein02,no03a,odi03,nodi03,cdtt}
based on replacing the scalar curvature $R$ in the
Hilbert--Einstein action with a generic analytic function $f(R)$
which should be reconstructed starting from data and physically
motivated issues. Also referred to as $f(R)$-gravity, some of
these models have been shown to be able to both fit the
cosmological data and evade the Solar System constraints in
several physically interesting cases
\cite{Hu,Starobinsky,Appleby,Odintsov,Tsuji}.

In this review paper, we will face two of the main problems
directly related to the dark energy and dark matter issues:
cosmography and clusters of galaxies. These are typical  examples
where the standard General Relativity and Newtonian potential
schemes fail to describe dynamics since data present accelerated
expansion and missing matter. Our goal is to address them by
$f(R)$-gravity.

\subsection{Cosmography: why?}

It is worth noting that both dark energy models and modified
gravity theories seem to be in agreement with  data. As a
consequence, unless higher precision probes of the expansion rate
and the growth of structure will be available, these two rival
approaches could not be discriminated. This confusion about the
theoretical background suggests that a more conservative approach
to the problem of cosmic acceleration, relying on as less model
dependent quantities as possible, is welcome.  A possible solution
could be to come back to the cosmography \cite{W72} rather than
finding out solutions of the Friedmann equations and testing them.
Being only related to the derivatives of the scale factor, the
cosmographic parameters make it possible to fit the data on the
distance\,-\,redshift relation without any {\it a priori}
assumption on the underlying cosmological model: in this case, the
only assumption is that the metric is the Robertson\,-\,Walker one
(and hence not relying on the solution of cosmological equations).
Almost eighty years after Hubble discovery of the expansion of the
Universe, we can now  extend, in principle, cosmography well
beyond the search for the value of the only Hubble constant. The
SNeIa Hubble diagram extends up to $z = 1.7$ thus invoking the
need for, at least, a fifth order Taylor expansion of the scale
factor in order to give a reliable approximation of the
distance\,-\,redshift relation. As a consequence, it could be, in
principle, possible to estimate up to five cosmographic
parameters, although the still too small data set available does
not allow to get a precise and realistic determination of all of
them.

Once these quantities have been determined, one could use them to
put constraints on the models. In a sense, we can revert the usual
approach, consisting in deriving the cosmographic parameters as a
sort of byproduct of an assumed theory. Here, we follow the other
way around expressing the model characterizing quantities as a
function of the cosmographic parameters. Such a program is
particularly suited for the study of fourth order theories of
gravity. As it is well known, the mathematical difficulties
entering the solution of fourth order field equations make it
quite problematic to find out analytical expressions for the scale
factor and hence predict the values of the cosmographic
parameters. A key role in $f(R)$-gravity is played by the choice
of the $f(R)$ function. Under quite general hypotheses, we will
derive useful relations among the cosmographic parameters and the
present day value of $f^{(n)}(R) = d^nf/dR^n$, with $n = 0,
\ldots, 3$, whatever $f(R)$ is\footnote{As an important remark, we
stress that our derivation will rely on the metric formulation of
$f(R)$ theories, while we refer the reader to
\cite{poplawski06,poplawski07} for a similar work in the Palatini
approach.}. Once the cosmographic parameters will be determined,
this method will allow us to investigate the cosmography of $f(R)$
theories.

It is worth stressing that the definition of the cosmographic
parameters only relies on the assumption of the
Robertson\,-\,Walker metric. As such, it is however difficult to
state a priori to what extent the fifth order expansion provides
an accurate enough description of the quantities of interest.
Actually, the number of cosmographic parameters to be used depends
on the problem one is interested in. As we will see later, we are
here concerned only with the SNeIa Hubble diagram so that we have
to check that the distance modulus $\mu_{cp}(z)$ obtained using
the fifth order expansion of the scale factor is the same (within
the errors) as the one $\mu_{DE}(z)$ of the underlying physical
model. Being such a model of course unknown, one can adopt a
phenomenological parameterization for the dark
energy\footnote{Note that one can always use a phenomenological
dark energy model to get a reliable estimate of the scale factor
evolution even if the correct model is a fourth order one.}
equation of state (EoS) and look at the percentage deviation
$\Delta \mu/\mu_{DE}$ as function of the EoS parameters. We have
carried out such exercise using the CPL model, introduced below,
and verified that $\Delta \mu/\mu_{DE}$ is an increasing function
of $z$ (as expected), but still remains smaller than $2\%$ up to
$z \sim 2$ over a wide range of the CPL parameter space. On the
other hand, halting the Taylor expansion to a lower order may
introduce significant deviation for $z > 1$ that can potentially
bias the analysis if the measurement errors are as small as those
predicted by future SNeIa surveys. We are therefore confident that
our fifth order expansion is both sufficient to get an accurate
distance modulus over the redshift range probed by SNeIa and
necessary to avoid dangerous biases.

\subsection{Clusters of galaxies: why?}

In the second part of this review we will  apply the
$f(R)$-gravity approach to cluster of galaxies. In fact, changing
the gravity sector has consequences not only at cosmological
scales, but also at  galactic and cluster scales so that it is
mandatory to investigate the low energy limit of such theories. A
strong debate is  open with different  results arguing in favor
\cite{dick,sotiriou,cembranos,navarro,allrugg,ppnantro} or against
\cite{dolgov,chiba,olmo} such models at local scales. It is worth
noting that, as a general result, higher order theories of gravity
cause the gravitational potential to deviate from its Newtonian
$1/r$ scaling \cite{b10,hj,hjrev,cb05,sobouti,Mendoza} even if
such deviations may  be  vanishing.

In  \cite{CapCardTro07},  the Newtonian limit of power law $f(R) =
f_0 R^n$ theories has been investigated, assuming that the metric
in the low energy limit ($\Phi/c^2 << 1$) may be taken as
Schwarzschild\,-\,like. It turns out that a power law term
$(r/r_c)^{\beta}$ has to be added to the Newtonian $1/r$ term in
order to get the correct gravitational potential. While the
parameter $\beta$ may be expressed analytically as a function of
the slope $n$ of the $f(R)$ theory, $r_c$ sets the scale where the
correction term starts being significant. A particular range of
values of $n$  has been investigated so that the corrective term
is an increasing function of the radius $r$ thus causing an
increase of the rotation curve with respect to the Newtonian one
and offering the possibility to fit the galaxy rotation curves
without the need of further dark matter components.

A set of low surface brightness (LSB) galaxies with extended and
well measured rotation curves has been considered
\cite{dbb02,db05}. These systems are supposed to be dark matter
dominated, and successfully fitting data without dark matter is a
strong evidence in favor of the approach (see also \cite{Frigerio}
for an independent analysis using another sample of galaxies).
Combined with the hints coming from the cosmological applications,
one should have, in principle, the possibility to address both the
dark energy and dark matter problems resorting to the same well
motivated fundamental theory \cite{fogdm,prl,Koivisto,lobo}.
Nevertheless, the simple power law $f(R)$ gravity is nothing else
but a toy-model which fail if one tries to achieve a comprehensive
model for all the cosmological dynamics, ranging from the early
Universe, to the large scale structure up to the late accelerated
era \cite{prl,Koivisto}.

A fundamental issue is related to  clusters and  superclusters of
galaxies. Such structures, essentially, rule the large scale
structure, and are the intermediate step between galaxies and
cosmology. As the galaxies, they appear dark-matter dominated but
the distribution  of dark matter component seems clustered and
organized in a very different way with respect to galaxies. It
seems that dark matter is ruled by the scale and also its
fundamental nature could depend on the scale. For a comprehensive
review see \cite{Bah96}.

In the philosophy of $f(R)$-gravity, the issue is  to reconstruct
the mass profile of clusters {\it without} dark matter, i.e. to
find out corrections to the Newton potential producing the same
dynamics as dark matter  but starting from a well motivated
theory.

In conclusion, $f(R)$-gravity, as the simplest approach to any
extended or alternative gravity scheme, could be the paradigm to
interpret dark energy and dark matter as curvature effects acting
at scales larger than those where General Relativity has been
actually investigated and probed.

Let us discuss now how cosmography and then galaxy clusters could
be two main examples to realize this program.

\section{The cosmographic apparatus}

The key rule in cosmography is the Taylor series expansion of the
scale factor with respect to the cosmic time. To this aim, it is
convenient to introduce the following functions:
{\setlength\arraycolsep{0.2pt}
\begin{eqnarray}
H(t) &\equiv& + \frac{1}{a}\frac{da}{dt}\, ,
\\
q(t) &\equiv& - \frac{1}{a}\frac{d^{2}a}{dt^{2}}\frac{1}{H^{2}}\,
,
\\
j(t) &\equiv& + \frac{1}{a}\frac{d^{3}a}{dt^{3}}\frac{1}{H^{3}}\,
,
\\
s(t) &\equiv& + \frac{1}{a}\frac{d^{4}a}{dt^{4}}\frac{1}{H^{4}}\,
,
\\
l(t) &\equiv& + \frac{1}{a}\frac{d^{5}a}{dt^{5}}\frac{1}{H^{5}}\,
,
\end{eqnarray}}
which are usually referred to as the \textit{Hubble},
\textit{deceleration}, \textit{jerk}, \textit{snap} and
\textit{lerk} parameters, respectively. It is then a matter of
algebra to demonstrate the following useful relations\,:
{\setlength\arraycolsep{0.2pt}
\begin{eqnarray}
\dot{H} &=& -H^2 (1 + q) \ , \label{eq: hdot}
\\
\ddot{H} &=& H^3 (j + 3q + 2) \ , \label{eq: h2dot}
\\
d^3H/dt^3 &=& H^4 \left [ s - 4j - 3q (q + 4) - 6 \right ] \ ,
\label{eq: h3dot}
\\
d^4H/dt^4 &=& H^5 \left [ l - 5s + 10 (q + 2) j + 30 (q + 2) q +
24 \right ] \ , \label{eq: h4dot}
\end{eqnarray}}
where a dot denotes derivative with respect to the cosmic time
$t$. Eqs.(\ref{eq: hdot})\,-\,(\ref{eq: h4dot}) make it possible
to relate the derivative of the Hubble parameter to the other
cosmographic parameters. The distance\,-\,redshift relation may
then be obtained starting from the Taylor expansion of $a(t)$
along the lines described in \cite{Visser,WM04,CV07}.

\subsection{The scale-factor series}

With these definitions the series expansion to the 5th order in
time of the scale factor will be: {\setlength\arraycolsep{0.2pt}
\begin{eqnarray}
a(t) &=& a(t_{0}) \left\{ H_{0} (t-t_{0}) - \frac{q_{0}}{2}
H_{0}^{2} (t-t_{0})^{2} +  \frac{j_{0}}{3!} H_{0}^{3}
(t-t_{0})^{3} + \frac{s_{0}}{4!} H_{0}^{4} (t-t_{0})^{4} +
\frac{l_{0}}{5!} H_{0}^{5} (t-t_{0})^{5}
+\emph{O}[(t-t_{0})^{6}]\right\}
\end{eqnarray}}
{\setlength\arraycolsep{0.2pt}
\begin{eqnarray}\label{eq:a_series}
\frac{a(t)}{a(t_{0})} &=& 1 + H_{0} (t-t_{0}) -\frac{q_{0}}{2}
H_{0}^{2} (t-t_{0})^{2} +\frac{j_{0}}{3!} H_{0}^{3} (t-t_{0})^{3}
+ \frac{s_{0}}{4!} H_{0}^{4} (t-t_{0})^{4}+ \frac{l_{0}}{5!}
H_{0}^{5} (t-t_{0})^{5} +\emph{O}[(t-t_{0})^{6}]
\end{eqnarray}}
It's easy to see that Eq.(\ref{eq:a_series}) is the inverse of
redshift $z$, being the redshift defined by:
$$
1 + z = \frac{a(t_{0})}{a(t)}
$$
The physical distance travelled by a photon that is emitted at
time $t_{*}$ and absorbed at the current epoch $t_{0}$ is
$$
D = c \int dt = c (t_{0} - t_{*})
$$
Assuming $t_{*} = t_{0} - \frac{D}{c}$ and inserting in
Eq.(\ref{eq:a_series}) we have:
\begin{equation}
1 + z = \frac{a(t_{0})}{a(t_{0}-\frac{D}{c})} =\frac{1}{1 -
\frac{H_{0}}{c}D -
\frac{q_{0}}{2}\left(\frac{H_{0}}{c}\right)^{2}D^{2}  -
\frac{j_{0}}{6}\left(\frac{H_{0}}{c}\right)^{3}D^{3}  +
\frac{s_{0}}{24}\left(\frac{H_{0}}{c}\right)^{4}D^{4} -
\frac{l_{0}}{120}\left(\frac{H_{0}}{c}\right)^{5}D^{5}
+\emph{O}[(\frac{H_{0} D}{c})^{6}]}
\end{equation}
The inverse of this expression will be:
\begin{eqnarray}
1 + z &=& 1 + \frac{H_{0}}{c}D + \left(1 +
\frac{q_{0}}{2}\right)\left(\frac{H_{0}}{c}\right)^{2}D^{2} +
\left(1 + q_{0} +\frac{
j_{0}}{6}\right)\left(\frac{H_{0}}{c}\right)^{3}D^{3} + \left(1 +
\frac{3}{2}q_{0} + \frac{q_{0}^{2}}{4} + \frac{j_{0}}{3} -
\frac{s_{0}}{24}\right)\left(\frac{H_{0}}{c}\right)^{4}D^{4} +
\nonumber \\
&+& \left( 1 + 2 q_{0} + \frac{3}{4} q_{0}^{2} + \frac{q_{0}
j_{0}}{6} + \frac{j_{0}}{2} - \frac{s}{12} +
l_{0}\right)\left(\frac{H_{0}}{c}\right)^{5}D^{5} +
\emph{O}\left[\left(\frac{H_{0} D}{c}\right)^{6}\right]
\end{eqnarray}
Then we reverse the series $z(D) \rightarrow D(z)$ to have the
physical distance $D$ expressed as function of redshift $z$:
{\setlength\arraycolsep{0.2pt}
\begin{eqnarray}
z(D) &=& \mathcal{Z}_{D}^{1} \left(\frac{H_{0} D}{c}\right) +
\mathcal{Z}_{D}^{2} \left(\frac{H_{0} D}{c}\right)^{2} +
\mathcal{Z}_{D}^{3} \left(\frac{H_{0} D}{c}\right)^{3} +
\mathcal{Z}_{D}^{4} \left(\frac{H_{0} D}{c}\right)^{4} +
\mathcal{Z}_{D}^{5} \left(\frac{H_{0} D}{c}\right)^{5} +
\emph{O}\left[\left(\frac{H_{0} D}{c}\right)^{6}\right]
\end{eqnarray}}
with: {\setlength\arraycolsep{0.2pt}
\begin{eqnarray}
\mathcal{Z}_{D}^{1} &=& 1 \\
\mathcal{Z}_{D}^{2} &=& 1 + \frac{q_{0}}{2} \\
\mathcal{Z}_{D}^{3} &=& 1 + q_{0} +\frac{j_{0}}{6} \\
\mathcal{Z}_{D}^{4} &=& 1 + \frac{3}{2}q_{0} + \frac{q_{0}^{2}}{4} + \frac{j_{0}}{3} - \frac{s_{0}}{24} \\
\mathcal{Z}_{D}^{5} &=& 1 + 2 q_{0} + \frac{3}{4} q_{0}^{2} +
\frac{q_{0} j_{0}}{6} + \frac{j_{0}}{2} - \frac{s}{12} + l_{0}
\end{eqnarray}}
From this we have:
\begin{eqnarray}
D(z) &=& \frac{c z}{H_{0}} \left\{ \mathcal{D}_{z}^{0} +
\mathcal{D}_{z}^{1} \ z + \mathcal{D}_{z}^{2} \ z^{2} +
\mathcal{D}_{z}^{3} \ z^{3} + \mathcal{D}_{z}^{4} \ z^{4} +
\emph{O}(z^{5}) \right\}
\end{eqnarray}
with: {\setlength\arraycolsep{0.2pt}
\begin{eqnarray}
\mathcal{D}_{z}^{0} &=& 1 \\
\mathcal{D}_{z}^{1} &=& - \left(1 +\frac{ q_{0}}{2}\right) \\
\mathcal{D}_{z}^{2} &=& 1 + q_{0} + \frac{q_{0}^{2}}{2} - \frac{j_{0}}{6} \\
\mathcal{D}_{z}^{3} &=& - \left(1 + \frac{3}{2}q_{0}+
\frac{3}{2}q_{0}^{2} + \frac{5}{8} q_{0}^{3} - \frac{1}{2} j_{0} -
\frac{5}{12} q_{0} j_{0} - \frac{s_{0}}{24}\right) \\
\mathcal{D}_{z}^{4} &=& 1 + 2 q_{0} + 3 q_{0}^{2} + \frac{5}{2}
q_{0}^{3} + \frac{7}{2} q_{0}^{4} - \frac{5}{3} q_{0}
j_{0} - \frac{7}{8} q_{0}^{2} j_{0} - \frac{1}{8} q_{0} s_{0} - j_{0} +\frac{j_{0}^{2}}{12} - \frac{s_{0}}{6} - \frac{l_{0}}{120} \\
\end{eqnarray}}
In typical applications, one is not interested in the physical
distance $D(z)$, but other definitions:
\begin{itemize}
    \item the luminosity distance:
    \begin{equation}
    d_{L} = \frac{a(t_{0})}{a(t_{0}-\frac{D}{c})} \: (a(t_{0}) r_{0})
    \end{equation}
    \item the angular-diameter distance:
    \begin{equation}
    d_{A} = \frac{a(t_{0}-\frac{D}{c})}{a(t_{0})} \: (a(t_{0}) r_{0})
    \end{equation}
\end{itemize}
where $r_{0}(D)$ is:
\begin{equation}\label{eq:r_sin}
r_{0}(D) = \left\{
\begin{array}{lr}
  \sin ( \int_{t_{0}- \frac{D}{c}}^{t_{0}} \frac{c \ \mathrm{d}t}{a(t)} ) &  k = +1; \\
  &  \\
  \int_{t_{0}- \frac{D}{c}}^{t_{0}} \frac{c \ \mathrm{d}t}{a(t)} &  k = 0; \\
  &  \\
  \sinh ( \int_{t_{0}- \frac{D}{c}}^{t_{0}} \frac{c \ \mathrm{d}t}{a(t)} ) &  k = -1.
\end{array} \right.
\end{equation}
If we make the expansion for short distances, namely if we insert
the series expansion of $a(t)$ in $r_{0}(D)$, we have:
{\setlength\arraycolsep{0.2pt}
\begin{eqnarray}
r_{0}(D) &=& \int_{t_{0} - \frac{D}{c}}^{t_{0}} \frac{c \
\mathrm{d}t}{a(t)} = \int_{t_{0} - \frac{D}{c}}^{t_{0}} \frac{c \
\mathrm{d}t}{a_{0}} \left\{ 1 + H_{0} (t_{0} - t) + \left(1 +
\frac{q_{0}}{2}\right) H_{0}^{2}(t_{0} - t)^{2} + \left(1 + q_{0}
+\frac{ j_{0}}{6}\right)H_{0}^{3}(t_{0} - t)^{3} + \right. \nonumber \\
&+& \left. \left(1 + \frac{3}{2}q_{0} + \frac{q_{0}^{2}}{4} +
\frac{j_{0}}{3} - \frac{s_{0}}{24}\right)H_{0}^{4}(t_{0} - t)^{4}
+ \left(1 + 2 q_{0} + \frac{3}{4} q_{0}^{2} + \frac{q_{0}
j_{0}}{6} + \frac{j_{0}}{2} - \frac{s}{12} +
l_{0}\right)H_{0}^{5}(t_{0} - t)^{5} + \emph{O}[(t_{0} - t)^{6}]
\right\} = \nonumber \\ &=& \frac{D}{a_{0}} \left\{ 1 +
\frac{1}{2} \frac{H_{0} D}{c} + \left[\frac{2 + q_{0}}{6}\right]
\left(\frac{H_{0} D}{c}\right)^{2} + \left[ \frac{6 + 6 q_{0} +
j_{0}}{24} \right] \left(\frac{H_{0} D}{c}\right)^{3} +
\left[ \frac{24 + 36 q_{0} + 6 q_{0}^{2} + 8 j_{0} - s_{0}}{120} \right] \left(\frac{H_{0} D}{c}\right)^{4} + \right. \nonumber \\
&+& \left. \left[ \frac{12 + 24 q_{0} + 9 q_{0}^{2} + 2 q_{0}
j_{0} + 6 j_{0} - s_{0} + 12 l_{0}}{72} \right] \left(\frac{H_{0}
D}{c}\right)^{5}+ \emph{O}\left[\left(\frac{H_{0}
D}{c}\right)^{6}\right] \right\}
\end{eqnarray}}
To convert from physical distance travelled to \textit{r}
coordinate traversed we have to consider that the Taylor series
expansion of $\sin$-$\sinh$ functions is:
\begin{equation}
r_{0}(D) = \left[\int_{t_{0}-\frac{D}{c}}^{t_{0}} \frac{c \
\mathrm{d}t}{a(t)}\right] - \frac{k}{3!}
\left[\int_{t_{0}-\frac{D}{c}}^{t_{0}} \frac{c \
\mathrm{d}t}{a(t)}\right]^{3} + \emph{O}\left(
\left[\int_{t_{0}-\frac{D}{c}}^{t_{0}} \frac{c \
\mathrm{d}t}{a(t)}\right]^{5}  \right)
\end{equation}
so that Eq.(\ref{eq:a_series}) with curvature $k$ term becomes:
{\setlength\arraycolsep{0.2pt}
\begin{eqnarray}
r_{0}(D) &=& \frac{D}{a_{0}} \left\{ \mathcal{R}_{D}^{0} +
\mathcal{R}_{D}^{1} \frac{H_{0} D}{c} + \mathcal{R}_{D}^{2}
\left(\frac{H_{0} D}{c}\right)^{2} + \mathcal{R}_{D}^{3}
\left(\frac{H_{0} D}{c}\right)^{3} + \right. \nonumber \\
&+&\left. \mathcal{R}_{D}^{4} \left(\frac{H_{0} D}{c}\right)^{4} +
\mathcal{R}_{D}^{5} \left(\frac{H_{0} D}{c}\right)^{5} +
\emph{O}\left[\left(\frac{H_{0} D}{c}\right)^{6}\right] \right\}
\end{eqnarray}}
with: {\setlength\arraycolsep{0.2pt}
\begin{eqnarray}
\mathcal{R}_{D}^{0} &=& 1 \\
\mathcal{R}_{D}^{1} &=& \frac{1}{2} \\
\mathcal{R}_{D}^{2} &=& \frac{1}{6} \left[2 + q_{0} - \frac{k c^{2}}{H_{0}^{2} a_{0}^{2}}\right] \\
\mathcal{R}_{D}^{3} &=& \frac{1}{24} \left[ 6 + 6 q_{0} + j_{0} - 6 \frac{k c^{2}}{H_{0}^{2} a_{0}^{2}}\right]\\
\mathcal{R}_{D}^{4} &=& \frac{1}{120} \left[ 24 + 36 q_{0} + 6 q_{0}^{2} + 8 j_{0} - s_{0} - \frac{5kc^{2}(7 + 2 q_{0})}{a_{0}^{2} H_{0}^{2}}\right] \\
\mathcal{R}_{D}^{5} &=& \frac{1}{144} \left[ 24 + 48 q_{0} + 18
q_{0}^{2} + 4 q_{0} j_{0} + 12 j_{0} - 2 s_{0} + 24 l_{0} -
\frac{3kc^{2}(15 + 10 q_{0} + j_{0})}{a_{0}^{2} H_{0}^{2}} \right]
\end{eqnarray}}
Using these one for luminosity distance we have:
\begin{eqnarray}
d_{L}(z) = \frac{c z}{H_{0}} \left\{ \mathcal{D}_{L}^{0} +
\mathcal{D}_{L}^{1} \ z + \mathcal{D}_{L}^{2} \ z^{2} +
\mathcal{D}_{L}^{3} \ z^{3} + \mathcal{D}_{L}^{4} \ z^{4} +
\emph{O}(z^{5}) \right\}
\end{eqnarray}
with: {\setlength\arraycolsep{0.2pt}
\begin{eqnarray}
\mathcal{D}_{L}^{0} &=& 1 \\
\mathcal{D}_{L}^{1} &=& - \frac{1}{2} \left(-1 + q_{0}\right) \\
\mathcal{D}_{L}^{2} &=& - \frac{1}{6} \left(1 - q_{0} - 3q_{0}^{2} + j_{0} + \frac{k c^{2}}{H_{0}^{2}a_{0}^{2}}\right) \\
\mathcal{D}_{L}^{3} &=& \frac{1}{24} \left(2 - 2 q_{0} - 15
q_{0}^{2} - 15 q_{0}^{3} + 5 j_{0} + 10 q_{0} j_{0} + s_{0} +
\frac{2 k c^{2} (1 + 3 q_{0})}{H_{0}^{2} a_{0}^{2}}\right)\\
\mathcal{D}_{L}^{4} &=& \frac{1}{120} \left[ -6 + 6 q_{0} + 81
q_{0}^{2} + 165 q_{0}^{3} + 105 q_{0}^{4} - 110 q_{0} j_{0} - 105
q_{0}^{2} j_{0} - 15 q_{0} s_{0} + \right. \\
&-& \left.  27 j_{0} + 10 j^{2} - 11 s_{0} - l_{0} -
\frac{5kc^{2}(1 + 8 q_{0} + 9 q_{0}^{2} - 2 j_{0})}{a_{0}^{2}
H_{0}^{2}}\right]
\end{eqnarray}}
While for the angular diameter distance it is:
\begin{eqnarray}
d_{A}(z) = \frac{c z}{H_{0}} \left\{ \mathcal{D}_{A}^{0} +
\mathcal{D}_{A}^{1} \ z + \mathcal{D}_{A}^{2} \ z^{2} +
\mathcal{D}_{A}^{3} \ z^{3} + \mathcal{D}_{A}^{4} \ z^{4} +
\emph{O}(z^{5}) \right\}
\end{eqnarray}
with: {\setlength\arraycolsep{0.2pt}
\begin{eqnarray}
\mathcal{D}_{A}^{0} &=& 1 \\
\mathcal{D}_{A}^{1} &=& - \frac{1}{2} \left(3 + q_{0}\right) \\
\mathcal{D}_{A}^{2} &=& \frac{1}{6} \left[11 + 7 q_{0} + 3q_{0}^{2} - j_{0} - \frac{k c^{2}}{H_{0}^{2}a_{0}^{2}}\right] \\
\mathcal{D}_{A}^{3} &=& - \frac{1}{24} \left[50 + 46 q_{0} + 39
q_{0}^{2} + 15 q_{0}^{3} - 13 j_{0} - 10 q_{0} j_{0} - s_{0}
- \frac{2 k c^{2} (5 + 3 q_{0})}{H_{0}^{2} a_{0}^{2}}\right] \\
\mathcal{D}_{A}^{4} &=& \frac{1}{120} \left[ 274 + 326 q_{0} + 411
q_{0}^{2} + 315 q_{0}^{3} + 105 q_{0}^{4} - 210 q_{0} j_{0} - 105
q_{0}^{2} j_{0} - 15 q_{0} s_{0} + \right. \\
&-& \left. 137 j_{0} + 10 j^{2} - 21 s_{0} - l_{0} -
\frac{5kc^{2}(17 + 20 q_{0} + 9 q_{0}^{2} - 2 j_{0})}{a_{0}^{2}
H_{0}^{2}}\right]
\end{eqnarray}}

If we want to use the same notation of \cite{CV07}, we define
$\Omega_{0} = 1 + \frac{k c^{2}}{H_{0}^{2} a_{0}^{2}}$, which can
be considered a purely cosmographic parameter, or $\Omega_{0} = 1
- \Omega_{k} = \Omega_{m,0} + \Omega_{r,0} + \Omega_{X,0}$ if we
consider the dynamics of the Universe. With this parameter
Eqs.(26)-(28) become: {\setlength\arraycolsep{0.2pt}
\begin{eqnarray}
\mathcal{D}_{L,y}^{0} &=& 1 \\
\mathcal{D}_{L,y}^{1} &=& - \frac{1}{2} \left(- 3 + q_{0}\right) \\
\mathcal{D}_{L,y}^{2} &=& - \frac{1}{6} \left(12 - 5 q_{0} + 3q_{0}^{2} - j_{0} - \Omega_{0} \right) \\
\mathcal{D}_{L,y}^{3} &=& \frac{1}{24} \left[52 - 20 q_{0} + 21
q_{0}^{2} - 15 q_{0}^{3} - 7 j_{0} + 10 q_{0} j_{0} + s_{0}
- 2 \Omega_{0} (1 + 3 q_{0}) \right]\\
\mathcal{D}_{L,y}^{4} &=& \frac{1}{120} \left[359 - 184 q_{0} +
186 q_{0}^{2} - 135 q_{0}^{3} + 105 q_{0}^{4} + 90 q_{0} j_{0} -
105 q_{0}^{2} j_{0}
- 15 q_{0} s_{0} + \right. \\
&-& \left. 57 j_{0} + 10 j^{2} + 9 s_{0} - l_{0} - 5 \Omega_{0}
(17 - 6 q_{0} + 9 q_{0}^{2} - 2 j_{0}) \right]
\end{eqnarray}}
and {\setlength\arraycolsep{0.2pt}
\begin{eqnarray}
\mathcal{D}_{A,y}^{0} &=& 1 \\
\mathcal{D}_{A,y}^{1} &=& - \frac{1}{2} \left(1 + q_{0}\right) \\
\mathcal{D}_{A,y}^{2} &=& - \frac{1}{6} \left[- q_{0} - 3q_{0}^{2} + j_{0} + \Omega_{0} \right] \\
\mathcal{D}_{A,y}^{3} &=& - \frac{1}{24} \left[- 2 q_{0} + 3
q_{0}^{2} + 15 q_{0}^{3} - j_{0} - 10 q_{0} j_{0} - s_{0}
+ 2 \Omega_{0} \right] \\
\mathcal{D}_{A,y}^{4} &=& - \frac{1}{120} \left[ 1 - 6 q_{0} + 9
q_{0}^{2} - 15 q_{0}^{3} - 105 q_{0}^{4} + 10 q_{0} j_{0} + 105
q_{0}^{2} j_{0}+ 15 q_{0} s_{0} + \right. \\
&-& \left. 3 j_{0} - 10 j^{2} + s_{0} + l_{0} + 5 \Omega_{0}
\right]
\end{eqnarray}}

Previous relations  in this section have been derived for any
value of the curvature parameter; but since in the following we
will assume a flat Universe,  we will used the simplified versions
for $k = 0$. Now, since we are going to use supernovae data, it
will be useful to give as well the Taylor series of the expansion
of  the luminosity distance at it enters the modulus distance,
which is the quantity about which those observational data inform.
The final expression for the modulus distance based on the Hubble
free luminosity distance, $\mu(z) = 5 \log_{10} d_{L}(z)$, is:
\begin{equation}\label{eq:museries}
\mu(z) = \frac{5}{\log 10} \cdot \left( \log z + \mathcal{M}^{1} z
+ \mathcal{M}^{2} z^2 + \mathcal{M}^{3} z^{3} + \mathcal{M}^{4}
z^{4} \right)\, ,
\end{equation}
with {\setlength\arraycolsep{0.2pt}
\begin{eqnarray}
\mathcal{M}^{1} &=& - \frac{1}{2} \left[ -1 + q_{0} \right] \, ,\\
\mathcal{M}^{2} &=& - \frac{1}{24} \left[7 - 10 q_{0} - 9q_{0}^{2}
+ 4j_{0} \right] \, ,\\
\mathcal{M}^{3} &=& \frac{1}{24}\left[5 - 9 q_{0} - 16 q_{0}^{2} -
10 q_{0}^{3} + 7 j_{0} + 8 q_{0} j_{0} + s_{0} \right] \, ,\\
\mathcal{M}^{4} &=& \frac{1}{2880}\left[-469 + 1004 q_{0} + 2654
q_{0}^{2} + 3300 q_{0}^{3} + 1575 q_{0}^{4} + 200 j_0^{2} -1148 j_{0} +\right. \nonumber \\
&-& \left. -2620 q_{0} j_{0} - 1800 q_{0}^{2} j_{0} - 300 q_{0}
s_{0} - 324 s_{0} - 24 l_{0} \right]\, .
\end{eqnarray}}

\section{$f(R)$-gravity vs cosmography}

\subsection{$f(R)$ preliminaries}

As discussed in the Introduction, much interest has been recently
devoted to the possibility that dark energy  could be nothing else
but a curvature effect according to which the present Universe is
filled by pressureless dust matter only and the acceleration is
the result of  modified Friedmann equations obtained by replacing
the Ricci curvature scalar $R$ with a generic function $f(R)$ in
the gravity action. Under the assumption of a flat Universe, the
Hubble parameter is therefore determined by\footnote{We use here
natural units such that $8 \pi G = 1$.}\,:

\begin{equation}
H^2 = \frac{1}{3} \left [ \frac{\rho_m}{f'(R)} + \rho_{curv}
\right ] \label{eq: hfr}
\end{equation}
where the prime denotes derivative with respect to $R$ and
$\rho_{curv}$ is the energy density of an {\it effective curvature
fluid}\footnote{Note that the name {\it curvature fluid} does not
refer to the FRW curvature parameter $k$, but only takes into
account that such a term is a geometrical one related to the
scalar curvature $R$.}\,:

\begin{equation}
\rho_{curv} = \frac{1}{f'(R)} \left \{ \frac{1}{2} \left [ f(R)  -
R f'(R) \right ] - 3 H \dot{R} f''(R) \right \} \ . \label{eq:
rhocurv}
\end{equation}
Assuming there is no interaction between the matter and the
curvature terms (we are in the so-called {\it Jordan frame}), the
matter continuity equation gives the usual scaling $\rho_M =
\rho_M(t = t_0) a^{-3} = 3 H_0^2 \Omega_M a^{-3}$, with $\Omega_M$
the present day matter density parameter. The continuity equation
for $\rho_{curv}$ then reads\,:

\begin{equation}
\dot{\rho}_{curv} + 3 H (1 + w_{curv}) \rho_{curv}  = \frac{3
H_0^2 \Omega_M \dot{R} f''(R)}{\left [ f'(R) \right ]^2}  a^{-3}
\label{eq: curvcons}
\end{equation}
with

\begin{equation}
w_{curv} = -1 + \frac{\ddot{R} f''(R) + \dot{R} \left [ \dot{R}
f'''(R) - H f''(R) \right ]} {\left [ f(R) - R f'(R) \right ]/2 -
3 H \dot{R} f''(R)} \label{eq: wcurv}
\end{equation}
the barotropic factor of the curvature fluid. It is worth noticing
that the curvature fluid quantities $\rho_{curv}$ and $w_{curv}$
only depends on $f(R)$ and its derivatives up to the third order.
As a consequence, considering only their present day values (which
may be naively obtained by replacing $R$ with $R_0$ everywhere),
two $f(R)$ theories sharing the same values of $f(R_0)$,
$f'(R_0)$, $f''(R_0)$, $f'''(R_0)$ will be degenerate from this
point of view\footnote{One can argue that this is not strictly
true since different $f(R)$ theories will lead to different
expansion rate $H(t)$ and hence different present day values of
$R$ and its derivatives. However, it is likely that two $f(R)$
functions that exactly match each other up to the third order
derivative today will give rise to the same $H(t)$ at least for $t
\simeq t_0$ so that $(R_0, \dot{R}_0, \ddot{R}_0)$ will be almost
the same.}.

Combining Eq.(\ref{eq: curvcons}) with Eq.(\ref{eq: hfr}), one
finally gets the following {\it master equation} for the Hubble
parameter\,:

\begin{eqnarray}
\dot{H} & = & -\frac{1}{2 f'(R)} \left \{ 3 H_0^2 \Omega_M a^{-3} + \ddot{R} f''(R)+ \right . \nonumber \\
~ & ~ & \left . + \dot{R} \left [ \dot{R} f'''(R) - H f''(R)
\right ] \right \} \ . \label{eq: presingleeq}
\end{eqnarray}
Expressing the scalar curvature $R$ as function of the Hubble
parameter as\,:

\begin{equation}
R = - 6 \left ( \dot{H} + 2 H^2 \right ) \label{eq: rvsh}
\end{equation}
and inserting the result into Eq.(\ref{eq: presingleeq}), one ends
with a fourth order nonlinear differential equation for the scale
factor $a(t)$ that cannot be easily solved also for the simplest
cases (for instance, $f(R) \propto R^n$). Moreover, although
technically feasible, a numerical solution of Eq.(\ref{eq:
presingleeq}) is plagued by the large uncertainties on the
boundary conditions (i.e., the present day values of the scale
factor and its derivatives up to the third order) that have to be
set to find out the scale factor.

\subsection{$f(R)$-derivatives and cosmography}

Motivated by these difficulties, we approach now the problem from
a different viewpoint. Rather than choosing a parameterized
expression for $f(R)$ and then numerically solving Eq.(\ref{eq:
presingleeq}) for given values of the boundary conditions, we try
to relate the present day values of its derivatives to the
cosmographic parameters $(q_0, j_0, s_0, l_0)$ so that
constraining them in a model independent way gives us a hint for
what kind of $f(R)$ theory could be able to fit the observed
Hubble diagram\footnote{Note that a similar analysis, but in the
context of the energy conditions in $f(R)$, has yet been presented
in \cite{Bergliaffa}. However, in that work, the author give an
expression for $f(R)$ and then compute the snap parameter to be
compared to the observed one. On the contrary, our analysis does
not depend on any assumed functional expression for $f(R)$.}.

As a preliminary step, it is worth considering again the
constraint equation (\ref{eq: rvsh}). Differentiating with respect
to $t$, we easily get the following relations\,:

\begin{equation}
\begin{array}{l}
\displaystyle{\dot{R} = -6 \left ( \ddot{H} + 4 H \dot{H} \right )} \\ ~ \\
\displaystyle{\ddot{R} = -6 \left ( d^3H/dt^3 + 4 H \ddot{H} + 4 \dot{H}^2 \right )} \\  ~ \\
\displaystyle{d^3R/dt^3{R} = -6 \left ( d^4H/dt^4 + 4 H d^3H/dt^3 + 12 \dot{H} \ddot{H} \right )} \\
\end{array}
\ . \label{eq: prederr}
\end{equation}
Evaluating these at the present time and using Eqs.(\ref{eq:
hdot})\,-\,(\ref{eq: h4dot}), one finally gets\,:

\begin{equation}
R_0 = -6 H_0^2 (1 - q_0) \ , \label{eq: rz}
\end{equation}

\begin{equation}
\dot{R}_0 = -6 H_0^3 (j_0 - q_0 - 2) \ , \label{eq: rdotz}
\end{equation}

\begin{equation}
\ddot{R}_0 = -6 H_0^4 \left ( s_0 + q_0^2 + 8 q_0 + 6 \right ) \ ,
\label{eq: r2dotz}
\end{equation}

\begin{equation}
d^3R_{0}/dt^3 = -6 H_0^5 \left [ l_0 - s_0 + 2 (q_0 + 4) j_0 - 6
(3q_0 + 8) q_0 - 24 \right ] \ , \label{eq: r3dotz}
\end{equation}
which will turn out to be useful in the following.

Let us now come back to the expansion rate and master equations
(\ref{eq: hfr}) and (\ref{eq: presingleeq}). Since they have to
hold along the full evolutionary history of the Universe, they
naively hold also at the present day. As a consequence, we may
evaluate them in $t = t_0$ thus easily obtaining\,:

\begin{eqnarray}
H_0^2 = \frac{H_0^2 \Omega_M}{f'(R_0)} + \frac{f(R_0) - R_0
f'(R_0) - 6 H_0 \dot{R}_0 f''(R_0)}{6 f'(R_0)} \ , \label{eq:
hfrz}
\end{eqnarray}

\begin{eqnarray}
- \dot{H}_0 = \frac{3 H_0^2 \Omega_M}{2 f'(R_0)} +
\frac{\dot{R}_0^2 f'''(R_0) + \left ( \ddot{R}_0 - H_0 \dot{R_0}
\right ) f''(R_0)}{2 f'(R_0)} \ . \label{eq: hdotfrz}
\end{eqnarray}
Using Eqs.(\ref{eq: hdot})\,-\,(\ref{eq: h4dot}) and (\ref{eq:
rz})\,-\,(\ref{eq: r3dotz}), we can rearrange Eqs.(\ref{eq: hfrz})
and (\ref{eq: hdotfrz}) as two relations among the Hubble constant
$H_0$ and the cosmographic parameters $(q_0, j_0, s_0)$, on one
hand, and the present day values of $f(R)$ and its derivatives up
to third order. However, two further relations are needed in order
to close the system and determine the four unknown quantities
$f(R_0)$, $f'(R_0)$, $f''(R_0)$, $f'''(R_0)$. A first one may be
easily obtained by noting that, inserting back the physical units,
the rate expansion equation reads\,:

\begin{displaymath}
H^2 = \frac{8 \pi G}{3 f'(R)} \left [\rho_m + \rho_{curv} f'(R)
\right ]
\end{displaymath}
which clearly shows that, in $f(R)$ gravity, the Newtonian
gravitational constant $G$ is replaced by an effective (time
dependent) $G_{eff} = G/f'(R)$. On the other hand, it is
reasonable to assume that the present day value of $G_{eff}$ is
the same as the Newtonian one so that we get the simple
constraint\,:

\begin{equation}
G_{eff}(z = 0) = G \rightarrow f'(R_0) = 1 \ . \label{eq: fpz}
\end{equation}
In order to get the fourth relation we need to close the system,
we first differentiate both sides of Eq.(\ref{eq: presingleeq})
with respect to $t$. We thus get\,:

\begin{eqnarray}
\ddot{H} & = & \frac{\dot{R}^2 f'''(R) + \left ( \ddot{R} - H
\dot{R} \right ) f''(R) + 3 H_0^2 \Omega_M a^{-3}}{2 \left [
\dot{R} f''(R) \right ]^{-1} \left [ f'(R) \right ]^2} -
\frac{\dot{R}^3 f^{(iv)}(R) + \left
( 3 \dot{R} \ddot{R} - H \dot{R}^2 \right ) f'''(R)}{2 f'(R)} \nonumber \\
~ & - & \frac{\left ( d^3R/dt^3 - H \ddot{R} + \dot{H} \dot{R}
\right ) f''(R) - 9 H_0^2 \Omega_M H a^{-3}}{2 f'(R)} \ ,
\label{eq: h2dotfr}
\end{eqnarray}
with $f^{(iv)}(R) = d^4f/dR^4$. Let us now suppose that $f(R)$ may
be well approximated by its third order Taylor expansion in $R -
R_0$, i.e. we set\,:

\begin{eqnarray}
f(R) = f(R_0) + f'(R_0) (R - R_0) +  \frac{1}{2} f''(R_0) (R -
R_0)^2 + \frac{1}{6} f'''(R_0) (R - R_0)^3 \ . \label{eq:
frtaylor}
\end{eqnarray}
In such an approximation, it is $f^{(n)}(R) = d^nf/R^n = 0$ for $n
\ge 4$ so that naively $f^{(iv)}(R_0) = 0$. Evaluating then
Eq.(\ref{eq: h2dotfr}) at the present day, we get\,:

\begin{eqnarray}
\ddot{H}_0 & = & \frac{\dot{R}_0^2 f'''(R_0) + \left ( \ddot{R}_0
- H_0 \dot{R}_0 \right ) f''(R_0) + 3 H_0^2 \Omega_M}{2 \left [
\dot{R}_0 f''(R_0) \right ]^{-1} \left [ f'(R_0) \right ]^2} -
\frac{ \left ( 3 \dot{R}_0 \ddot{R}_0 - H \dot{R}_0^2 \right )
f'''(R_0)}{2 f'(R_0)} \nonumber \\ ~ & - & \frac{\left (
d^3R_{0}/dt^3 - H_0 \ddot{R}_0 + \dot{H}_0 \dot{R}_0 \right )
f''(R_0) - 9 H_0^3 \Omega_M}{2 f'(R_0)} \ . \label{eq: h2dotfrz}
\end{eqnarray}
We can now schematically proceed as follows. Evaluate
Eqs.(\ref{eq: hdot})\,-\,(\ref{eq: h4dot}) at $z = 0$ and plug
these relations into the left hand sides of Eqs.(\ref{eq: hfrz}),
(\ref{eq: hdotfrz}), (\ref{eq: h2dotfrz}). Insert Eqs.(\ref{eq:
rz})\,-\,(\ref{eq: r3dotz}) into the right hand sides of these
same equations so that only the cosmographic parameters $(q_0,
j_0, s_0, l_0)$ and the $f(R)$ related quantities enter both sides
of these relations. Finally, solve them under the constraint
(\ref{eq: fpz}) with respect to the present day values of $f(R)$
and its derivatives up to the third order. After some  algebra,
one ends up with the desired result\,:

\begin{equation}
\frac{f(R_0)}{6 H_0^2} = - \frac{{\cal{P}}_0(q_0, j_0, s_0, l_0)
\Omega_M + {\cal{Q}}_0(q_0, j_0, s_0, l_0)}{{\cal{R}}(q_0, j_0,
s_0, l_0)} \ , \label{eq: f0z}
\end{equation}

\begin{equation}
f'(R_0) = 1 \ , \label{eq: f1z}
\end{equation}

\begin{equation}
\frac{f''(R_0)}{\left ( 6 H_0^2 \right )^{-1}} = -
\frac{{\cal{P}}_2(q_0, j_0, s_0) \Omega_M + {\cal{Q}}_2(q_0, j_0,
s_0)}{{\cal{R}}(q_0, j_0, s_0, l_0)} \ , \label{eq: f2z}
\end{equation}

\begin{equation}
\frac{f'''(R_0)}{\left ( 6 H_0^2 \right )^{-2}} = -
\frac{{\cal{P}}_3(q_0, j_0, s_0, l_0) \Omega_M + {\cal{Q}}_3(q_0,
j_0, s_0, l_0)}{(j_0 - q_0 - 2) {\cal{R}}(q_0, j_0, s_0, l_0)} \ ,
\label{eq: f3z}
\end{equation}
where we have defined\,:

\begin{eqnarray}
{\cal{P}}_0 & = & (j_0 - q_0 - 2) l_0 - (3s_0 + 7j_0 + 6q_0^2 +
41q_0 + 22) s_0 - \left [ (3q_0 + 16) j_0 + 20q_0^2 + 64q_0 + 12
\right ] j_0 + \nonumber \\ &-& \left ( 3q_0^4 + 25q_0^3 + 96q_0^2
+ 72q_0 + 20 \right ) \ , \label{eq: defp0}
\end{eqnarray}

\begin{eqnarray}
{\cal{Q}}_0 & = & (q_0^2 - j_0 q_0 + 2q_0) l_0 +
\left [ 3q_0s_0 + (4q_0 + 6) j_0 + 6q_0^3 + 44q_0^2 + 22q_0 - 12 \right ] s_0 \nonumber \\
~ & + & \left [ 2j_0^2 + (3q_0^2 + 10q_0 - 6) j_0 + 17q_0^3 +
52q_0^2 + 54q_0 + 36 \right ] j_0 + 3q_0^5 + 28q_0^4 + 118q_0^3 +
\nonumber \\
&+& 72q_0^2 - 76q_0 -64\ , \label{eq: defq0}
\end{eqnarray}

\begin{equation}
{\cal{P}}_2 = 9 s_0 + 6 j_0 + 9q_0^2 + 66q_0 + 42 \ , \label{eq:
defp2}
\end{equation}

\begin{eqnarray}
{\cal{Q}}_2 & = & - \left \{ 6 (q_0 + 1) s_0 + \left [ 2j_0 - 2 (1
- q_0) \right ] j_0 + 6q_0^3 + 50q_0^2 + 74q_0 + 32 \right \} \ ,
\label{eq: defq2}
\end{eqnarray}

\begin{equation}
{\cal{P}}_3 = 3 l_0  + 3 s_0 - 9(q_0 + 4) j_0 - (45q_0^2 + 78q_0 +
12) \ , \label{eq: defp3}
\end{equation}

\begin{eqnarray}
{\cal{Q}}_3 & = & - \left \{ 2 (1 + q_0) l_0 + 2 (j_0 + q_0) s_0 -
( 2j_0 + 4q_0^2 + 12q_0 + 6 ) j_0 - (30q_0^3 + 84q_0^2 + 78q_0 +
24) \right \} \, \label{eq: defq3}
\end{eqnarray}

\begin{eqnarray}
{\cal{R}} & = &  (j_0 - q_0 - 2) l_0 - (3s_0 - 2j_0 + 6q_0^2 +
50q_0 + 40) s_0 + \left [ (3q_0 + 10) j_0 + 11q_0^2 + 4q_0 +
\right. \nonumber \\ &-& \left. 18 \right ] j_0 - (3q_0^4 +
34q_0^3 + 246q_0 + 104) \ . \label{eq: defr}
\end{eqnarray}
Eqs.(\ref{eq: f0z})\,-\,(\ref{eq: defr}) make it possible to
estimate the present day values of $f(R)$ and its first three
derivatives as function of the Hubble constant $H_0$ and the
cosmographic parameters $(q_0, j_0, s_0, l_0)$ provided a value
for the matter density parameter $\Omega_M$ is given. This is a
somewhat problematic point. Indeed, while the cosmographic
parameters may be estimated in a model independent way, the
fiducial value for $\Omega_M$ is usually the outcome of fitting a
given dataset in the framework of an assumed dark energy scenario.
However, it is worth noting that different models all converge
towards the concordance value $\Omega_M \simeq 0.25$ which is also
in agreement with astrophysical (model independent) estimates from
the gas mass fraction in galaxy clusters. On the other hand, it
has been proposed that $f(R)$ theories may avoid the need for dark
matter in galaxies and galaxy clusters
\cite{noipla,prl,CapCardTro07,Frigerio,sobouti,Mendoza,fogdm}. In
such a case, the total matter content of the Universe is
essentially equal to the baryonic one. According to the primordial
elements abundance and the standard BBN scenario, we therefore get
$\Omega_M \simeq \omega_b/h^2$ with $\omega_b = \Omega_b h^2
\simeq 0.0214$ \cite{Kirk} and $h$ the Hubble constant in units of
$100 {\rm km/s/Mpc}$. Setting $h = 0.72$ in agreement with the
results of the HST Key project \cite{Freedman}, we thus get
$\Omega_M = 0.041$ for a baryons only Universe. We will therefore
consider in the following both cases when numerical estimates are
needed.

It is worth noticing that $H_0$ only plays the role of a scaling
parameter giving the correct physical dimensions to $f(R)$ and its
derivatives. As such, it is not surprising that we need four
cosmographic parameters, namely $(q_0, j_0, s_0, l_0)$, to fix the
four $f(R)$ related quantities $f(R_0)$, $f'(R_0)$, $f''(R_0)$,
$f'''(R_0)$. It is also worth stressing that Eqs.(\ref{eq:
f0z})\,-\,({\ref{eq: f3z}) are linear in the $f(R)$ quantities so
that $(q_0, j_0, s_0, l_0)$ uniquely determine the former ones. On
the contrary, inverting them to get the cosmographic parameters as
function of the $f(R)$ ones, we do not get linear relations.
Indeed, the field equations in $f(R)$ theories are nonlinear
fourth order differential equations in the scale factor $a(t)$ so
that fixing the derivatives of $f(R)$ up to third order makes it
possible to find out a class of solutions, not a single one. Each
one of these solutions will be characterized by a different set of
cosmographic parameters thus explaining why the inversion of
Eqs.(\ref{eq: f0z})\,-\,(\ref{eq: defr}) does not give a unique
result for $(q_0, j_0, s_0, l_0)$.

As a final comment, we reconsider the underlying assumptions
leading to the above derived relations. While Eqs.(\ref{eq: hfrz})
and (\ref{eq: hdotfrz}) are exact relations deriving from a
rigorous application of the field equations, Eq.(\ref{eq:
h2dotfrz}) heavily relies on having approximated $f(R)$ with its
third order Taylor expansion (\ref{eq: frtaylor}). If this
assumption fails, the system should not be closed since a fifth
unknown parameter enters the game, namely $f^{(iv)}(R_0)$.
Actually, replacing $f(R)$ with its Taylor expansion is not
possible for all class of $f(R)$ theories. As such, the above
results only hold in those cases where such an expansion is
possible. Moreover, by truncating the expansion to the third
order, we are implicitly assuming that higher order terms are
negligible over the redshift range probed by the data. That is to
say, we are assuming that\,:

\begin{equation}
f^{(n)}(R_0) (R - R_0)^n << \sum_{m = 0}^{3}{\frac{f^{(m)}(R_0)}{m
!} (R - R_0)^m} \ \ {\rm for} \ n \ge 4 \label{eq: checkcond}
\end{equation}
over the redshift range probed by the data. Checking the validity
of this assumption is not possible without explicitly solving the
field equations, but we can guess an order of magnitude estimate
considering that, for all viable models, the background dynamics
should not differ too much from the $\Lambda$CDM one at least up
to $z \simeq 2$. Using then the expression of $H(z)$ for the
$\Lambda$CDM model, it is easily to see that $R/R_0$ is a quickly
increasing function of the redshift so that, in order Eq.(\ref{eq:
checkcond}) holds, we have to assume that $f^{(n)}(R_0) <<
f'''(R_0)$ for $n \ge 4$. This condition is easier to check for
many analytical $f(R)$ models.

Once such a relation is verified, we have still to worry about
Eq.(\ref{eq: fpz}) relying on the assumption that the {\it
cosmological} gravitational constant is {\it exactly} the same as
the {\it local} one. Although reasonable, this requirement is not
absolutely demonstrated. Actually, the numerical value usually
adopted for the Newton constant $G_N$ is obtained from laboratory
experiments in settings that can hardly be considered homogenous
and isotropic. As such, the spacetime metric in such conditions
has nothing to do with the cosmological one so that matching the
two values of $G$ is strictly speaking an extrapolation. Although
commonly accepted and quite reasonable, the condition $G_{local} =
G_{cosmo}$ could (at least, in principle) be violated so that
Eq.(\ref{eq: fpz}) could be reconsidered. Indeed, as we will see,
the condition $f'(R_0) = 1$ may not be verified for some popular
$f(R)$ models recently proposed in literature. However, it is
reasonable to assume that $G_{eff}(z = 0) = G (1 + \varepsilon)$
with $\varepsilon << 1$. When this be the case, we should repeat
the derivation of Eqs.(\ref{eq: f0z})\,-\,(\ref{eq: f3z}) now
using the condition $f'(R_0) = (1 + \varepsilon)^{-1}$. Taylor
expanding the results in $\varepsilon$ to the first order and
comparing with the above derived equations, we can estimate the
error induced by our assumption $\varepsilon = 0$. The resulting
expressions are too lengthy to be reported and depend in a
complicated way on the values of the matter density parameter
$\Omega_M$, the cosmographic parameters $(q_0, j_0, s_0, l_0)$ and
$\varepsilon$. However, we have numerically checked that the error
induced on $f(R_0)$, $f''(R_0)$, $f'''(R_0)$ are much lower than
$10\%$ for value of $\varepsilon$ as high as an unrealistic
$\varepsilon \sim 0.1$. We are confident that our results are
reliable also for these cases.

\section{$f(R)$-gravity and the CPL model}

A determination of $f(R)$ and its derivatives in terms of the
cosmographic parameters need for an estimate of these latter from
the data in a model independent way. Unfortunately, even in the
nowadays era of {\it precision cosmology}, such a program is still
too ambitious to give useful constraints on the $f(R)$
derivatives, as we will see later. On the other hand, the
cosmographic parameters may also be expressed in terms of the dark
energy density and EoS parameters so that we can work out what are
the present day values of $f(R)$ and its derivatives giving the
same $(q_0, j_0, s_0, l_0)$ of the given dark energy model. To
this aim, it is convenient to adopt a parameterized expression for
the dark energy EoS in order to reduce the dependence of the
results on any underlying theoretical scenario. Following the
prescription of the Dark Energy Task Force \cite{DETF}, we will
use the Chevallier\,-\,Polarski\,-\,Linder (CPL) parameterization
for the EoS setting \cite{CPL,Linder03}\,:

\begin{equation}
w = w_0 + w_a (1 - a) = w_0 + w_a z (1 + z)^{-1} \label{eq:
cpleos}
\end{equation}
so that, in a flat Universe filled by dust matter and dark energy,
the dimensionless Hubble parameter $E(z) = H/H_0$ reads\,:

\begin{equation}
E^2(z) = \Omega_M (1 + z)^3 + \Omega_X (1 + z)^{3(1 + w_0 + w_a)}
{\rm e}^{-\frac{3 w_a z}{1 + z}} \label{eq: ecpl}
\end{equation}
with $\Omega_X = 1 - \Omega_M$ because of the flatness assumption.
In order to determine the cosmographic parameters for such a
model, we avoid integrating $H(z)$ to get $a(t)$ by noting that
$d/dt = -(1 + z) H(z) d/dz$. We can use such a relation to
evaluate $(\dot{H}, \ddot{H}, d^3H/dt^3, d^4H/dt^4)$ and then
solve Eqs.(\ref{eq: hdot})\,-\,(\ref{eq: h4dot}), evaluated in $z
= 0$, with respect to the parameters of interest. Some algebra
finally gives\,:

\begin{equation}
q_0 = \frac{1}{2} + \frac{3}{2} (1 - \Omega_M) w_0 \ , \label{eq:
qzcpl}
\end{equation}

\begin{equation}
j_0 = 1 + \frac{3}{2} (1 - \Omega_M) \left [ 3w_0 (1 + w_0) + w_a
\right ] \ , \label{eq: jzcpl}
\end{equation}

\begin{eqnarray}
s_0 & = & -\frac{7}{2} - \frac{33}{4} (1 - \Omega_M) w_a -
\frac{9}{4} (1 - \Omega_M) \left [ 9 + (7 - \Omega_M) w_a \right ]
w_0 - \frac{9}{4} (1 - \Omega_M) (16 - 3\Omega_M) w_0^2 -
\frac{27}{4} (1 - \Omega_M) (3 - \Omega_M) w_0^3 \ , \label{eq:
szcpl}
\end{eqnarray}

\begin{eqnarray}
l_0 & = & \frac{35}{2} + \frac{1 - \Omega_M}{4} \left [ 213 + (7 -
\Omega_M) w_a \right ] w_a + \frac{1 - \Omega_M)}{4} \left [ 489 +
9(82 - 21 \Omega_M) w_a \right ] w_0 + \nonumber \\ &+&
\frac{9}{2} (1 - \Omega_M) \left [ 67 - 21 \Omega_M + \frac{3}{2}
(23 - 11 \Omega_M) w_a \right ] w_0^2 + \frac{27}{4} (1 -
\Omega_M) (47 - 24 \Omega_M) w_0^3 + \nonumber \\ &+& \frac{81}{2}
(1 - \Omega_M) (3 - 2\Omega_M) w_0^4 \ . \label{eq: lzcpl}
\end{eqnarray}
Inserting Eqs.(\ref{eq: qzcpl})\,-\,(\ref{eq: lzcpl}) into
Eqs.(\ref{eq: f0z})\,-\,(\ref{eq: defr}), we get lengthy
expressions (which we do not report here) giving the present day
values of $f(R)$ and its first three derivatives as function of
$(\Omega_M, w_0, w_a)$. It is worth noting that the $f(R)$ model
thus obtained is not dynamically equivalent to the starting CPL
one. Indeed, the two models have the same cosmographic parameters
only today. As such, for instance, the scale factor is the same
between the two theories only over the time period during which
the fifth order Taylor expansion is a good approximation of the
actual $a(t)$. It is also worth stressing that such a procedure
does not select a unique $f(R)$ model, but rather a class of
fourth order theories all sharing the same third order Taylor
expansion of $f(R)$.

\subsection{The $\Lambda$CDM case}

With these caveats in mind, it is worth considering first the
$\Lambda$CDM model which is obtained by setting $(w_0, w_a) = (-1,
0)$ in the above expressions thus giving\,:

\begin{equation}
\left \{
\begin{array}{lll}
\displaystyle{q_0} & = & \displaystyle{\frac{1}{2} - \frac{3}{2} (1-\Omega_M)} \\
~ & ~ & \\
\displaystyle{j_0} & = & \displaystyle{1} \\
~ & ~ & \\
\displaystyle{s_0} & = & \displaystyle{1 - \frac{9}{2} \Omega_M} \\
~ & ~ & \\
\displaystyle{l_0} & = & \displaystyle{1 + 3 \Omega_M + \frac{27}{2} \Omega_M^2} \\
\end{array}
\right . \ . \label{eq: cplcdm}
\end{equation}
When inserted into the expressions for the $f(R)$ quantities,
these relations give the remarkable result\,:

\begin{equation}
f(R_0) = R_0 + 2\Lambda \ \ , \ \ f''(R_0) = f'''(R_0) = 0 \ ,
\label{eq: frlcdm}
\end{equation}
so that we obviously  conclude that the only $f(R)$ theory having
exactly the same cosmographic parameters as the $\Lambda$CDM model
is just $f(R) \propto R$, i.e. GR. It is worth noticing that such
a result comes out as a consequence of the values of $(q_0, j_0)$
in the $\Lambda$CDM model. Indeed, should we have left $(s_0,
l_0)$ undetermined and only fixed $(q_0, j_0)$ to the values in
(\ref{eq: cplcdm}), we should have got the same result in
(\ref{eq: frlcdm}). Since the $\Lambda$CDM model fits well a large
set of different data, we do expect that the actual values of
$(q_0, j_0, s_0, l_0)$ do not differ too much from the
$\Lambda$CDM ones. Therefore, we plug into Eqs.(\ref{eq:
f0z})\,-\,(\ref{eq: defr}) the following expressions\,:

\begin{displaymath}
q_0 = q_0^{\Lambda} {\times} (1 + \varepsilon_q) \ \ , \ \ j_0 =
j_0^{\Lambda} {\times} (1 + \varepsilon_j) \ \ ,
\end{displaymath}

\begin{displaymath}
s_0 = s_0^{\Lambda} {\times} (1 + \varepsilon_s) \ \ , \ \ l_0 =
l_0^{\Lambda} {\times} (1 + \varepsilon_l) \ \ ,
\end{displaymath}
with $(q_0^{\Lambda}, j_0^{\Lambda}, s_0^{\Lambda},
l_0^{\Lambda})$ given by Eqs.(\ref{eq: cplcdm}) and
$(\varepsilon_q, \varepsilon_j, \varepsilon_s, \varepsilon_l)$
quantifyin the deviations from the $\Lambda$CDM values allowed by
the data. A numerical estimate of these quantities may be
obtained, e.g., from a Markov chain analysis, but this is outside
our aims. Since we are here interested in a theoretical
examination, we prefer to consider an idealized situation where
the four quantities above all share the same value $\varepsilon <<
1$. In such a case, we can easily investigate how much the
corresponding $f(R)$ deviates from the GR one considering the two
ratios $f''(R_0)/f(R_0)$ and $f'''(R_0)/f(R_0)$. Inserting the
above expressions for the cosmographic parameters into the exact
(not reported) formulae for $f(R_0)$, $f''(R_0)$ and $f'''(R_0)$,
taking their ratios and then expanding to first order in
$\varepsilon$, we finally get\,:

\begin{equation}
\eta_{20} = \frac{64 - 6 \Omega_M (9\Omega_M + 8)} {\left [ 3
(9\Omega_M + 74) \Omega_M - 556 \right ] \Omega_M^2 + 16} \
{\times} \ \frac{\varepsilon}{27} \ , \label{eq: e20eps}
\end{equation}

\begin{equation}
\eta_{30} = \frac{6 \left [ (81 \Omega_M - 110) \Omega_M + 40
\right ] \Omega_M + 16} {\left [ 3 (9\Omega_M + 74) \Omega_M - 556
\right ] \Omega_M^2 + 16} \ {\times} \ \frac{\varepsilon}{243
\Omega_M^2} \ , \label{eq: e30eps}
\end{equation}
having defined $\eta_{20} = f''(R_0)/f(R_0) {\times} H_0^4$ and
$\eta_{30} = f'''(R_0)/f(R_0) {\times} H_0^6$ which, being
dimensionless quantities, are more suited to estimate the order of
magnitudes of the different terms. Inserting our fiducial values
for $\Omega_M$, we get\,:

\begin{displaymath}
\left \{
\begin{array}{ll}
\displaystyle{\eta_{20} \simeq 0.15 \ {\times} \ \varepsilon} &
{\rm for}
\ \ \Omega_M = 0.041 \\
~ & ~ \\
\displaystyle{\eta_{20} \simeq -0.12 \ {\times} \ \varepsilon} &
{\rm for}
\ \ \Omega_M = 0.250 \\
\end{array}
\right . \ ,
\end{displaymath}

\begin{displaymath}
\left \{
\begin{array}{ll}
\displaystyle{\eta_{30} \simeq 4 \ {\times} \ \varepsilon} & {\rm
for}
\ \ \Omega_M = 0.041 \\
~ & ~ \\
\displaystyle{\eta_{30} \simeq -0.18 \ {\times} \ \varepsilon} &
{\rm for}
\ \ \Omega_M = 0.250 \\
\end{array}
\right . \ .
\end{displaymath}
For values of $\varepsilon$ up to 0.1, the above relations show
that the second and third derivatives are at most two orders of
magnitude smaller than the zeroth order term $f(R_0)$. Actually,
the values of $\eta_{30}$ for a baryon only model (first row)
seems to argue in favor of a larger importance of the third order
term. However, we have numerically checked that the above
relations approximates very well the exact expressions up to
$\varepsilon \simeq 0.1$ with an accuracy depending on the value
of $\Omega_M$, being smaller for smaller matter density
parameters. Using the exact expressions for $\eta_{20}$ and
$\eta_{30}$, our conclusion on the negligible effect of the second
and third order derivatives are significantly strengthened.

Such a result holds under the hypotheses that the narrower are the
constraints on the validity of the $\Lambda$CDM model, the smaller
are the deviations of the cosmographic parameters from the
$\Lambda$CDM ones. It is possible to show that this indeed the
case for the CPL parametrization we are considering. On the other
hand, we have also assumed that the deviations $(\varepsilon_q,
\varepsilon_j, \varepsilon_s, \varepsilon_l)$ take the same
values. Although such hypothesis is somewhat ad hoc, we argue that
the main results are not affected by giving it away. Indeed,
although different from each other, we can still assume that all
of them are very small so that Taylor expanding to the first order
should lead to additional terms into Eqs.(\ref{eq:
e20eps})\,-\,(\ref{eq: e30eps}) which are likely of the same order
of magnitude. We may therefore conclude that, if the observations
confirm that the values of the cosmographic parameters agree
within $\sim 10\%$ with those predicted for the $\Lambda$CDM
model, we must conclude that the deviations of $f(R)$ from the GR
case, $f(R) \propto R$, should be vanishingly small.

It is worth stressing, however, that such a conclusion only holds
for those $f(R)$ models satisfying the constraint (\ref{eq:
checkcond}). It is indeed possible to work out a model having
$f(R_0) \propto R_0$, $f''(R_0) = f'''(R_0) = 0$, but
$f^{(n)}(R_0) \ne 0$ for some $n$. For such a (somewhat ad hoc)
model, Eq.(\ref{eq: checkcond}) is clearly not satisfied so that
the cosmographic parameters have to be evaluated from the solution
of the field equations. For such a model, the conclusion above
does not hold so that one cannot exclude that the resulting $(q_0,
j_0, s_0, l_0)$ are within $10\%$ of the $\Lambda$CDM ones.

\subsection{The constant EoS model}

Let us now take into account the condition $w = -1$, but still
retains $w_a = 0$ thus obtaining the so called {\it quiessence}
models. In such a case, some problems arise because both the terms
$(j_0 - q_0 - 2)$ and ${\cal{R}}$ may vanish for some combinations
of the two model parameters $(\Omega_M, w_0)$. For instance, we
find that $j_0 - q_0 - 2 = 0$  for $w_0 = (w_1, w_2)$ with\,:

\begin{displaymath}
w_1 = \frac{1}{1 - \Omega_M + \sqrt{(1 - \Omega_M) (4 -
\Omega_M)}} \ ,
\end{displaymath}

\begin{displaymath}
w_2 = - \frac{1}{3} \left [ 1 + \frac{4 - \Omega_M}{\sqrt{(1 -
\Omega_M) (4 - \Omega_M)}} \right ] \ .
\end{displaymath}
On the other hand, the equation ${\cal{R}}(\Omega_M, w_0) = 0$ may
have different real roots for $w$ depending on the adopted value
of $\Omega_M$. Denoting collectively with ${\bf w}_{null}$ the
values of $w_0$ that, for a given $\Omega_M$, make $(j_0 - q_0 -
2) {\cal{R}}(\Omega_M, w_0)$ taking the null value, we individuate
a set of quiessence models whose cosmographic parameters give rise
to divergent values of $f(R_0$, $f''(R_0)$ and $f'''(R_0)$. For
such models, $f(R)$ is clearly not defined so that we have to
exclude these cases from further consideration. We only note that
it is still possible to work out a $f(R)$ theory reproducing the
same background dynamics of such models, but a different route has
to be used.

\begin{figure}
\centering
\resizebox{8.5cm}{!}{\includegraphics{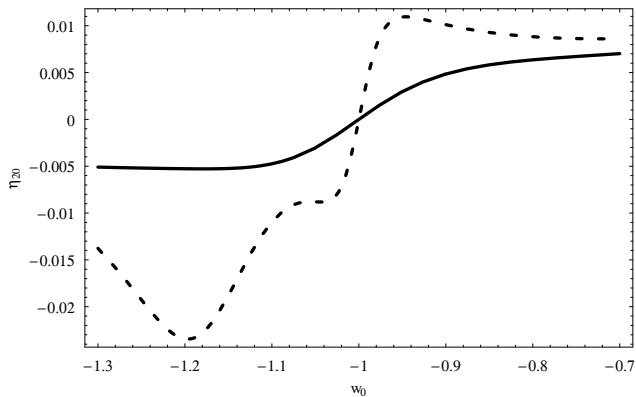}}
\caption{The dimensionless ratio between the present day values of
$f''(R)$ and $f(R)$ as function of the constant EoS $w_0$ of the
corresponding quiessence model. Short dashed and solid lines refer
to models with $\Omega_M = 0.041$ and $0.250$ respectively.}
\label{fig: r20}
\end{figure}

Since both $q_0$ and $j_0$ now deviate from the $\Lambda$CDM
values, it is not surprising that both $f''(R_0)$ and $f'''(R_0)$
take finite non null values. However, it is more interesting to
study the two quantities $\eta_{20}$ and $\eta_{30}$ defined above
to investigate the deviations of $f(R)$ from the GR case. These
are plotted in Figs.\,\ref{fig: r20} and \ref{fig: r30} for the
two fiducial $\Omega_M$ values. Note that the range of $w_0$ in
these plots have been chosen in order to avoid divergences, but
the lessons we will draw also hold for the other $w_0$ values.

As a general comment, it is clear that, even in this case,
$f''(R_0)$ and $f'''(R_0)$ are from two to three orders of
magnitude smaller that the zeroth order term $f(R_0)$. Such a
result could be yet guessed from the previous discussion for the
$\Lambda$CDM case. Actually, relaxing the hypothesis $w_0 = -1$ is
the same as allowing the cosmographic parameters to deviate from
the $\Lambda$CDM values. Although a direct mapping between the two
cases cannot be established, it is nonetheless evident that such a
relation can be argued thus making the outcome of the above plots
not fully surprising. It is nevertheless worth noting that, while
in the $\Lambda$CDM case, $\eta_{20}$ and $\eta_{30}$ always have
opposite signs, this is not the case for quiessence models with $w
> -1$. Indeed, depending on the value of $\Omega_M$, we can have
$f(R)$ theories with both $\eta_{20}$ and $\eta_{30}$ positive.
Moreover, the lower is $\Omega_M$, the higher are the ratios
$\eta_{20}$ and $\eta_{30}$ for a given value of $w_0$. This can
be explained qualitatively noticing that, for a lower $\Omega_M$,
the density parameter of the curvature fluid (playing the role of
an effective dark energy) must be larger thus claiming for higher
values of the second and third derivatives (see also \cite{emilio}
for a different approach to the problem).

\begin{figure}
\centering
\resizebox{8.5cm}{!}{\includegraphics{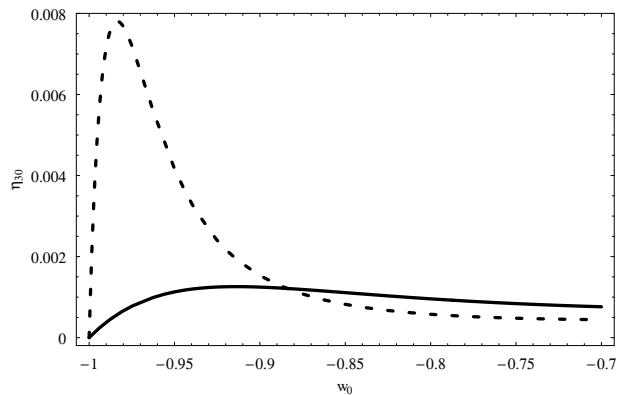}}
\caption{The dimensionless ratio between the present day values of
$f'''(R)$ and $f(R)$ as function of the constant EoS $w_0$ of the
corresponding quiessence model. Short dashed and solid lines refer
to models with $\Omega_M = 0.041$ and $0.250$ respectively.}
\label{fig: r30}
\end{figure}

\subsection{The general case}

Finally, we consider evolving dark energy models with $w_a \ne 0$.
Needless to say, varying three parameters allows to get a wide
range of models that cannot be discussed in detail. Therefore, we
only concentrate on evolving dark energy models with $w_0 = -1$ in
agreement with some most recent analysis. The results on
$\eta_{20}$ and $\eta_{30}$ are plotted in Figs.\,\ref{fig:
r20cpl} and \ref{fig: r30cpl} where these quantities as functions
of $w_a$. Note that we are considering models with positive $w_a$
so that $w(z)$ tends to $w_0 + w_a > w_0$ for $z \rightarrow
\infty$ so that the EoS dark energy can eventually approach the
dust value $w = 0$. Actually, this is also the range favored by
the data. We have, however, excluded values where $\eta_{20}$ or
$\eta_{30}$ diverge. Considering how they are defined, it is clear
that these two quantities diverge when $f(R_0) = 0$ so that the
values of $(w_0, w_a)$ making $(\eta_{20}, \eta_{30})$ to diverge
may be found solving\,:

\begin{displaymath}
{\cal{P}}_0(w_0, w_a) \Omega_M + {\cal{Q}}_0(w_0, w_a) = 0
\end{displaymath}
where ${\cal{P}}_0(w_0, w_a)$ and ${\cal{Q}}_0(w_0, w_a)$ are
obtained by inserting Eqs.(\ref{eq: qzcpl})\,-\,(\ref{eq: lzcpl})
into the defintions (\ref{eq: defp0})\,-\,(\ref{eq: defq0}). For
such CPL models, there is no any $f(R)$ model having the same
cosmographic parameters and, at the same time, satisfying all the
criteria needed for the validity of our procedure. Actually, if
$f(R_0) = 0$, the condition (\ref{eq: checkcond}) is likely to be
violated so that higher than third order must be included in the
Taylor expansion of $f(R)$ thus invalidating the derivation of
Eqs.(\ref{eq: f0z})\,-\,(\ref{eq: f3z}).

Under these caveats, Figs.\,\ref{fig: r20cpl} and \ref{fig:
r30cpl} demonstrate that allowing the dark energy EoS to evolve
does not change significantly our conclusions. Indeed, the second
and third derivatives, although being not null, are nevertheless
negligible with respect to the zeroth order term thus arguing in
favour of a GR\,-\,like $f(R)$ with only very small corrections.
Such a result is, however, not fully unexpected. From
Eqs.(\ref{eq: qzcpl}) and (\ref{eq: jzcpl}), we see that, having
setted $w_0 = -1$, the $q_0$ parameter is the same as for the
$\Lambda$CDM model, while $j_0$ reads $j_0^{\Lambda} + (3/2)(1 -
\Omega_M) w_a$. As we have stressed above, the Hilbert - Einstein
Lagrangian $f(R) = R + 2 \Lambda$ is recovered when $(q_0, j_0) =
(q_0^{\Lambda}, j_0^{\Lambda})$ whatever the values of $(s_0,
l_0)$ are. Introducing a $w_a \ne 0$ makes $(s_0, l_0)$ to differ
from the $\Lambda$CDM values, but the first two cosmographic
parameters are only mildly affected. Such deviations are then
partially washed out by the complicated way they enter in the
determination of the present day values of $f(R)$ and its first
three derivatives.

\begin{figure}
\centering \resizebox{8.5cm}{!}{\includegraphics{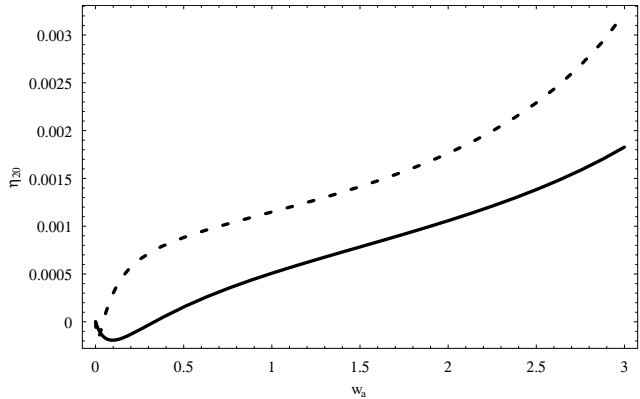}}
\caption{The dimensionless ratio between the present day values of
$f''(R)$ and $f(R)$ as function of the $w_a$ parameter for models
with $w_0 = -1$. Short dashed and solid lines refer to models with
$\Omega_M = 0.041$ and $0.250$ respectively.} \label{fig: r20cpl}
\end{figure}

\section{Constraining $f(R)$ parameters}

In the previous section, we have worked an alternative method to
estimate $f(R_0)$, $f''(R_0)$, $f'''(R_0)$ resorting to a model
independent parameterization of the dark energy EoS. However, in
the ideal case, the cosmographic parameters are directly estimated
from the data so that Eqs.(\ref{eq: f0z})\,-\,(\ref{eq: defr}) can
be used to infer the values of the $f(R)$ related quantities.
These latter can then be used to put constraints on the parameters
entering an assumed fourth order theory assigned by a $f(R)$
function characterized by a set of parameters ${\bf p} = (p_1,
\ldots, p_n)$ provided that the hypotheses underlying the
derivation of Eqs.(\ref{eq: f0z})\,-\,(\ref{eq: defr}) are indeed
satisfied. We show below two interesting cases which clearly
highlight the potentiality and the limitations of such an
analysis.

\subsection{Double power law Lagrangian}

As a first interesting example, we set\,:

\begin{equation}
f(R) = R \left (1 + \alpha R^{n} + \beta R^{-m} \right )
\label{eq: frdpl}
\end{equation}
with $n$ and $m$ two positive real numbers (see, for example,
\cite{double} for some physical motivations). The following
expressions are immediately obtained\,:

\begin{displaymath}
\left \{
\begin{array}{lll}
f(R_0) & = & R_0 \left (1 + \alpha R_0^{n} + \beta R_0^{-m} \right
) \\ ~ & ~ & ~ \\ f'(R_0) & = & 1 + \alpha (n + 1) R_0^n - \beta
(m - 1) R_0^{-m} \\  ~ & ~ & ~ \\ f''(R_0) & = & \alpha n (n + 1)
R_0^{n - 1} + \beta m (m - 1) R_0^{-(1 + m)} \\ ~ & ~ & ~ \\
f'''(R_0) & = & \alpha n (n + 1) (n - 1) R_0^{n - 2} \\ ~ & - &
\beta m (m + 1) (m - 1) R_0^{-(2 + m)}
\end{array}
\right . \ . \label{eq: derdpl}
\end{displaymath}
Denoting by $\phi_i$ (with $i = 0, \ldots, 3$) the values of
$f^{(i)}(R_0)$ determined through Eqs.(\ref{eq:
f0z})\,-\,(\ref{eq: defr}), we can solve\,:

\begin{figure}
\centering \resizebox{8.5cm}{!}{\includegraphics{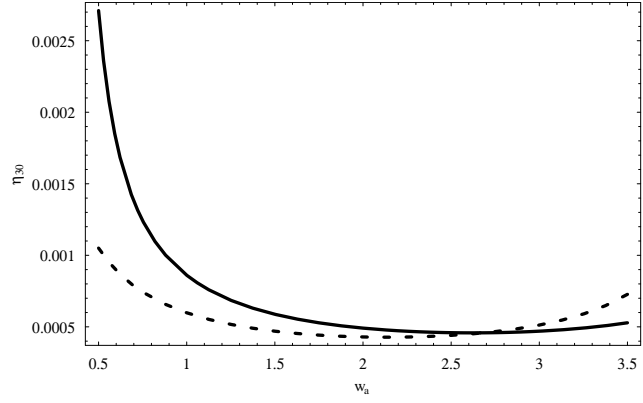}}
\caption{The dimensionless ratio between the present day values of
$f'''(R)$ and $f(R)$ as function of the $w_a$ parameter for models
with $w_0 = -1$. Short dashed and solid lines refer to models with
$\Omega_M = 0.041$ and $0.250$ respectively. } \label{fig: r30cpl}
\end{figure}

\begin{displaymath}
\left \{
\begin{array}{lll}
f(R_0) & = & \phi_0 \\ f'(R_0) & = & \phi_1 \\ f''(R_0) & = & \phi_2 \\
f'''(R_0) & = & \phi_3 \\
\end{array}
\right .
\end{displaymath}
which is a system of four equations in the four unknowns $(\alpha,
\beta, n, m)$ that can be analytically solved proceeding as
follows. First, we solve the first and second equation with
respect to $(\alpha, \beta)$ obtaining\,:

\begin{equation}
\left \{
\begin{array}{lll}
\alpha & = & \displaystyle{\frac{1 - m}{n + m} \left ( 1 - \frac{\phi_0}{R_0} \right ) R_0^{-n}} \\
~ & ~ & ~ \\ \beta & = & \displaystyle{- \frac{1 + n}{n + m} \left
( 1 -
\frac{\phi_0}{R_0} \right ) R_0^{m}} \\
\end{array}
\right . \ , \label{eq: ab12}
\end{equation}
while, solving the third and fourth equations, we get\,:

\begin{equation}
\left \{
\begin{array}{lll}
\alpha & = & \displaystyle{\frac{\phi_2 R_0^{1 - n} \left [ 1 + m + (\phi_3/\phi_2) R_0 \right ]}{n (n + 1) (n + m)}} \\
~ & ~ & ~ \\
\beta & = & \displaystyle{\frac{\phi_2 R_0^{1 + n} \left [ 1 - n + (\phi_3/\phi_2) R_0 \right ]}{m (1 - m) (n + m)}} \\
\end{array}
\right . \ . \label{eq: ab34}
\end{equation}
Equating the two solutions, we get a systems of two equations in
the two unknowns $(n, m)$, namely\,:

\begin{equation}
\left \{
\begin{array}{lll}
\displaystyle{\frac{n (n + 1) (1 - m) \left ( 1 - \phi_0/R_0
\right )} {\phi_2 R_0 \left [ 1 + m + (\phi_3/\phi_2) R_0 \right
]}} & = & 1 \\ ~ & ~ & ~ \\ \displaystyle{\frac{m (n + 1) (m - 1)
\left ( 1 - \phi_0/R_0 \right )} {\phi_2 R_0 \left [ 1 - n +
(\phi_3/\phi_2) R_0 \right ]}} & = & 1
\end{array}
\right . \ .
\end{equation}
Solving with respect to $m$, we get two solutions, the first one
being $m = -n$ which has to be discarded since makes $(\alpha,
\beta)$ goes to infinity. The only acceptable solution is\,:

\begin{equation}
m = - \left [ 1 - n + (\phi_3/\phi_2) R_0 \right ] \label{eq:
msol}
\end{equation}
which, inserted back into the above system, leads to a second
order polynomial equation for $n$ with solutions\,:

\begin{equation}
n = \frac{1}{2} \left [1 + \frac{\phi_3}{\phi_2} R_0 {\pm}
\frac{\sqrt{{\cal{N}}(\phi_0, \phi_2, \phi_3)}}{\phi_2 R_0 (1 +
\phi_0/R_0)} \right ] \label{eq: nsol}
\end{equation}
where we have defined\,:

\begin{eqnarray}
{\cal{N}}(\phi_0, \phi_2, \phi_3) & = & \left ( R_0^2 \phi_0^2 - 2
R_0^3
\phi_0 + R_0^4 \right ) \phi_3^2 \nonumber \\
~ & + & 6 \left ( R_0 \phi_0^2 - 2 R_0^2 \phi_0 + R_0^3 \right )
\phi_2
\phi_3 \nonumber \\
~ & + & 9 \left ( \phi_0^2 - 2 R_0 \phi_0 + R_0^2 \right )
\phi_2^2 \nonumber \\ ~ & + & 4 \left ( R_0^2 \phi_0 - R_0^3
\right ) \phi_2^3 \ . \label{eq: defn}
\end{eqnarray}
Depending on the values of $(q_0, j_0, s_0, l_0)$, Eq.(\ref{eq:
nsol}) may lead to one, two or any acceptable solution, i.e. real
positive values of $n$. This solution has then to be inserted back
into Eq.(\ref{eq: msol}) to determine $m$ and then into
Eqs.(\ref{eq: ab12}) or (\ref{eq: ab34}) to estimate $(\alpha,
\beta)$. If the final values of $(\alpha, \beta, n, m)$  are
physically viable, we can conclude that the model in Eq.(\ref{eq:
frdpl}) is in agreement with the data giving the same cosmographic
parameters inferred from the data themselves. Exploring
analytically what is the region of the $(q_0, j_0, s_0, l_0)$
parameter space which leads to acceptable $(\alpha, \beta, n, m)$
solutions is a daunting task far outside the aim of the present
work.

\subsection{The Hu and Sawicki model}

One of the most pressing problems of $f(R)$ theories is the need
to escape the severe constraints imposed by the Solar System
tests. A successful model has been recently proposed by Hu and
Sawicki \cite{Hu} (HS) setting\footnote{Note that such a model
does not pass the matter instability test so that some viable
generalizations \cite{Odi,Cogno08,Nodi07} have been proposed.}\,:

\begin{equation}
f(R) = R - R_c \frac{\alpha (R/R_c)^n}{1 + \beta (R/R_c)^n} \ .
\label{eq: frhs}
\end{equation}
As for the double power law model discussed above, there are four
parameters which we can be expressed in terms of the cosmographic
parameters $(q_0, j_0, s_0, l_0)$.

As a first step, it is trivial to get\,:

\begin{equation}
\left \{
\begin{array}{lll}
f(R_0) & = & \displaystyle{R_0 - R_c \frac{\alpha R_{0c}^n}{1 +
\beta R_{0c}^n}} \\  ~ & ~ & ~ \\  f'(R_0) & = & \displaystyle{1
- \frac{\alpha n R_c R_{0c}^{n}}{R_0 (1 + \beta R_{0c}^n)^2}} \\
~ & ~ & ~ \\ f''(R_0) & = & \displaystyle{\frac{\alpha n R_c
R_{0c}^n \left [ (1 - n) + \beta (1 + n) R_{0c}^n \right ]}{R_0^2
(1 + \beta R_{0c}^n)^3}} \\ ~ & ~ & ~ \\ f'''(R_0) & = &
\displaystyle{\frac{\alpha n R_c R_{0c}^n (A n^2 + B n + C)}{R_0^3
(1 + \beta R_{0c}^n)^4}}
\end{array}
\right . \label{eq: derhs}
\end{equation}
with $R_{0c} = R_0/R_c$ and\,:

\begin{equation}
\left \{
\begin{array}{lll}
A & = & - \beta^2 R_{0c}^{2n} + 4 \beta R_{0c}^n - 1 \\  ~ & ~ & ~ \\
B & = & 3 (1 - \beta^2  R_{0c}^{2n}) \\ ~ & ~ & ~ \\ C & = &  -2
(1 - \beta  R_{0c}^{n})^2
\end{array}
\right . \ .
\end{equation}
Equating Eqs.(\ref{eq: derhs}) to the four quantities $(\phi_0,
\phi_1, \phi_2, \phi_3)$ defined as above, we could, in principle,
solve this system of four equations in four unknowns to get
$(\alpha, \beta, R_c, n)$ in terms of $(\phi_0, \phi_1, \phi_2,
\phi_3)$ and then, using Eqs.(\ref{eq: f0z})\,-\,(\ref{eq: defr})
as functions of the cosmographic parameters. However, setting
$\phi_1 = 1$ as required by Eq.(\ref{eq: f1z}) gives the only
trivial solution $\alpha n R_c = 0$ so that the HS model reduces
to the Einstein\,-\,Hilbert Lagrangian $f(R) = R$. In order to
escape this problem, we can relax the condition $f'(R_0) = 1$ to
$f'(R_0) = (1 + \varepsilon)^{-1}$.  As we have discussed in
Sect.\,IV, this is the same as assuming that the present day
effective gravitational constant $G_{eff, 0} = G_N/f'(R_0)$ only
slightly differs from the usual Newtonian one which seems to be a
quite reasonable assumption. Under this hypothesis, we can
analytically solve for $(\alpha, \beta, R_c, n)$ in terms of
$(\phi_0, \varepsilon, \phi_2, \phi_3)$. The actual values of
$(\phi_0, \phi_2, \phi_3)$ will be no more given by Eqs.(\ref{eq:
f0z})\,-\,(\ref{eq: f3z}), but we have checked that they deviate
from those expressions\footnote{Note that the correct expressions
for $(phi_0, \phi_2, \phi_3)$ may still formally be written as
Eqs.(\ref{eq: f0z})\,-\,(\ref{eq: f3z}), but the polynomials
entering them are now different and also depend on powers of
$\varepsilon$.} much less than $10\%$ for $\varepsilon$ up to
$10\%$ well below any realistic expectation.

With this caveat in mind, we first solve

\begin{displaymath}
f(R_0) = \phi_0 \ \ , \ \ f''(R_0) = (1 + \varepsilon)^{-1}
\end{displaymath}
to get\,:

\begin{eqnarray}
\alpha & = & \frac{n (1 + \varepsilon)}{\varepsilon} \left (
\frac{R_0}{R_c} \right )^{1 - n} \left ( 1 - \frac{\phi_0}{R_0} \right )^2 \ , \nonumber \\
\beta & = & \frac{n (1 + \varepsilon)}{\varepsilon} \left (
\frac{R_0}{R_c} \right )^{-n} \left [ 1  - \frac{\phi_0}{R_0}  -
\frac{\varepsilon}{n (1 + \varepsilon)}\right ] \ . \nonumber
\end{eqnarray}
Inserting these expressions in Eqs.(\ref{eq: derhs}), it is easy
to check that $R_c$ cancels out so that we can no more determine
its value. Such a result is, however, not unexpected. Indeed,
Eq.(\ref{eq: frhs}) can trivially be rewritten as\,:

\begin{displaymath}
f(R) = R - \frac{\tilde{\alpha} R^n}{1 + \tilde{\beta} R^n}
\end{displaymath}
with $\tilde{\alpha} = \alpha R_c^{1 - n}$ and $\tilde{\beta} =
\beta R_c^{-n}$ which are indeed the quantities that are
determined by the above expressions for $(\alpha, \beta)$.
Reversing the discussion, the present day values of $f^{(i)}(R)$
depend on $(\alpha, \beta, R_c)$ only through the two parameters
$(\tilde{\alpha}, \tilde{\beta})$. As such, the use of
cosmographic parameters is unable to break this degeneracy.
However, since $R_c$ only plays the role of a scaling parameter,
we can arbitrarily set its value without loss of generality.

On the other hand, this degeneracy allows us to get a consistency
relation to immediately check whether the HS model is viable or
not. Indeed, solving the equation $f''(R_0) = \phi_2$, we get\,:

\begin{displaymath}
n = \frac{(\phi_0/R_0) + [(1 + \varepsilon)/\varepsilon](1 -
\phi_2 R_0) - (1 - \varepsilon)/(1 + \varepsilon)}{1 - \phi_0/R_0}
\ ,
\end{displaymath}
which can then be inserted into the equations $f'''(R_0) = \phi_3$
to obtain a complicated relation among $(\phi_0, \phi_2, \phi_3)$
which we do not report for sake of shortness. Solving such a
relation with respect to $\phi_3/\phi_0$ and Taylor expanding to
first order in $\varepsilon$, the constraint we get reads\,:

\begin{displaymath}
\frac{\phi_3}{\phi_0} \simeq - \frac{1 + \varepsilon}{\varepsilon}
\frac{\phi_2}{R_0} \left [ R_0 \left ( \frac{\phi_2}{\phi_0}
\right ) + \frac{\varepsilon \phi_0^{-1}}{1 + \varepsilon} \left (
1 - \frac{2 \varepsilon}{1 - \phi_0/R_0} \right ) \right ] \ .
\end{displaymath}
If the cosmographic parameters $(q_0, j_0, s_0, l_0)$ are known
with sufficient accuracy, one could compute the values of $(R_0,
\phi_0, \phi_2. \phi_3)$ for a given $\varepsilon$ (eventually
using the expressions obtained for $\varepsilon = 0$) and then
check if they satisfied this relation. If this is not  the case,
one can immediately give off the HS model also without the need of
solving the field equations and fitting the data. Actually, given
the still large errors on the cosmographic parameters, such a test
only remains in the realm of (quite distant) future applications.
However, the HS model works for other tests as shown in \cite{Hu}
and so a consistent cosmography analysis has to be combined with
them.

\section{Constraints on $f(R)$-derivatives from the data}

Eqs.(\ref{eq: f0z})\,-\,(\ref{eq: defr}) relate the present day
values of $f(R)$ and its first three derivatives to the
cosmographic parameters $(q_0, j_0, s_0, l_0)$ and the matter
density $\Omega_M$. In principle, therefore, a measurement of
these latter quantities makes it possible to put constraints on
$f^{(i)}(R_0)$, with $i = \{0, \ldots, 3\}$, and hence on the
parameters of a given fourth order theory through the method shown
in the previous section. Actually, the cosmographic parameters are
affected by errors which obviously propagate onto the $f(R)$
quantities. Actually, the covariance matrix for the cosmographic
parameters is not diagonal so that one has also take care of this
to estimate the final errors on $f^{(i)}(R_0)$. A similar
discussion also holds for the errors on the dimensionless ratios
$\eta_{20}$ and $\eta_{30}$ introduced above. As a general rule,
indicating with $g(\Omega_M, {\bf p})$ a generic $f(R)$ related
quantity depending on $\Omega_M$ and the set of cosmographic
parameters ${\bf p}$, its uncertainty reads\,:

\begin{equation}
\sigma_{g}^2 = \left | \frac{\partial g}{\partial \Omega_M} \right
|^2 \sigma_{M}^2 + \sum_{i = 1}^{i = 4}{ \left | \frac{\partial
g}{\partial p_i} \right |^2 \sigma_{p_i}^2} + \sum_{i \neq j}{2
\frac{\partial g}{\partial p_i} \frac{\partial g}{\partial p_j}
C_{ij}} \label{eq: error}
\end{equation}
where $C_{ij}$ are the elements of the covariance matrix (being
$C_{ii} = \sigma_{p_i}^2$), we have set $(p_1, p_2, p_3, p_4) =
(q_0, j_0, s_0, l_0)$. and assumed that the error $\sigma_M$ on
$\Omega_M$ is uncorrelated with those on ${\bf p}$. Note that this
latter assumption strictly holds if the matter density parameter
is estimated from an astrophysical method (such as estimating the
total matter in the Universe from the estimated halo mass
function). Alternatively, we will assume that $\Omega_M$ is
constrained by the CMBR related experiments. Since these latter
mainly probes the very high redshift Universe ($z \simeq z_{lss}
\simeq 1089$), while the cosmographic parameters are concerned
with the present day cosmo, one can argue that the determination
of $\Omega_M$ is not affected by the details of the model adopted
for describing the late Universe. Indeed, we can reasonably assume
that, whatever is the dark energy candidate or $f(R)$ theory, the
CMBR era is well approximated by the standard GR with a model
comprising only dust matter. As such, we will make the simplifying
(but well motivated) assumption that $\sigma_M$ may be reduced to
very small values and is uncorrelated with the cosmographic
parameters.

Under this assumption, the problem of estimating the errors on
$g(\Omega_M, {\bf p})$ reduces to estimating the covariance matrix
for the cosmographic parameters given the details of the data set
used as observational constraints. We address this issue by
computing the Fisher information matrix (see, e.g., \cite{Teg97}
and references therein) defined as\,:

\begin{equation}
F_{ij} = \left \langle \frac{\partial^2 L}{\partial \theta_i
\partial \theta_j} \right \rangle \label{eq: deffij}
\end{equation}
with $L = -2 \ln{{\cal{L}}(\theta_1, \ldots, \theta_n)}$,
${\cal{L}}(\theta_1, \ldots, \theta_n)$ the likelihood of the
experiment, $(\theta_1, \ldots, \theta_n)$ the set of parameters
to be constrained, and $\langle \ldots \rangle$ denotes the
expectation value. Actually, the expectation value is computed by
evaluating the Fisher matrix elements for fiducial values of the
model parameters $(\theta_1, \ldots, \theta_n)$, while the
covariance matrix ${\bf C}$ is finally obtained as the inverse of
${\bf F}$.

A key ingredient in the computation of ${\bf F}$ is the definition
of the likelihood which depends, of course, of what experimental
constraint one is using. To this aim, it is worth remembering that
our analysis is based on fifth order Taylor expansion of the scale
factor $a(t)$ so that we can only rely on observational tests
probing quantities that are well described by this truncated
series. Moreover, since we do not assume any particular model, we
can only characterize the background evolution of the Universe,
but not its dynamics which, being related to the evolution of
perturbations, unavoidably need the specification of a physical
model. As a result, the SNeIa Hubble diagram is the ideal
test\footnote{See the conclusions for further discussion on this
issue.} to constrain the cosmographic parameters. We therefore
defined the likelihood as\,:

\begin{equation}
\begin{array}{l}
{\cal{L}}(H_0, {\bf p}) \propto \exp{-\chi^2(H_0, {\bf p})/2} \\ ~ \\
\chi^2(H_0, {\bf p}) = \sum_{n =
1}^{{\cal{N}}_{SNeIa}}{\displaystyle{\left [ \frac{\mu_{obs}(z_i)
- \mu_{th}(z_n, H_0, {\bf p})}{\sigma_i(z_i)} \right ]^2}}
\end{array} \ ,
\label{eq: deflike}
\end{equation}
where the distance modulus to redshift $z$ reads\,:

\begin{equation}
\mu_{th}(z, H_0, {\bf p}) = 25 + 5 \log{(c/H_0)} + 5 \log{d_L(z,
{\bf p})} \ , \label{eq: defmuth}
\end{equation}
and $d_L(z)$ is the Hubble free luminosity distance\,:

\begin{equation}
d_L(z) = (1 + z) \int_{0}^{z}{\frac{dz}{H(z)/H_0}} \ . \label{eq:
defdlhf}
\end{equation}
Using the fifth order Taylor expansion of the scale factor, we get
for $d_L(z, {\bf p})$ an analytical expression (reported in
Appendix A) so that the computation of $F_{ij}$ does not need any
numerical integration (which makes the estimate faster). As a last
ingredient, we need to specify the details of the SNeIa survey
giving the redshift distribution of the sample and the error on
each measurement. Following \cite{Kim}, we adopt\footnote{Note
that, in \cite{Kim}, the authors assume the data are separated in
redshift bins so that the error becomes $\sigma^2 =
\sigma_{sys}^2/{\cal{N}}_{bin} + {\cal{N}}_{bin} (z/z_{max})^2
\sigma_m^2$ with ${\cal{N}}_{bin}$ the number of SNeIa in a bin.
However, we prefer to not bin the data so that ${\cal{N}}_{bin} =
1$.}\,:

\begin{displaymath}
\sigma_(z) = \sqrt{\sigma_{sys}^2 + \left ( \frac{z}{z_{max}}
\right )^2 \sigma_m^2}
\end{displaymath}
with $z_{max}$ the maximum redshift of the survey, $\sigma_{sys}$
an irreducible scatter in the SNeIa distance modulus and
$\sigma_m$ to be assigned depending on the photometric accuracy.

In order to run the Fisher matrix calculation, we have to set a
fiducial model which we set according to the $\Lambda$CDM
predictions for the cosmographic parameters. For $\Omega_M = 0.3$
and $h = 0.72$ (with $h$ the Hubble constant in units of $100 {\rm
km/s/Mpc}$), we get\,:

\begin{displaymath}
(q_0, j_0, s_0, l_0) = (-0.55, 1.0, -0.35, 3.11) \ .
\end{displaymath}
As a first consistency check, we compute the Fisher matrix for a
survey mimicking the recent database in \cite{D07} thus setting
$({\cal{N}}_{SNeIa}, \sigma_m) = (192, 0.33)$. After marginalizing
over $h$ (which, as well known, is fully degenerate with the SNeIa
absolute magnitude ${\cal{M}}$), we get for the uncertainties\,:

\begin{displaymath}
(\sigma_1, \sigma_2, \sigma_3, \sigma_4) = (0.38, 5.4, 28.1, 74.0)
\end{displaymath}
where we are still using the indexing introduced above for the
cosmographic parameters. These values compare reasonably well with
those obtained from a cosmographic fitting of the Gold SNeIa
dataset\footnote{Actually, such estimates have been obtained
computing the mean and the standard deviation from the
marginalized likelihoods of the cosmographic parameters. As such,
the central values do not represent exactly the best fit model,
while the standard deviations do not give a rigorous description
of the error because the marginalized likelihoods are manifestly
non - Gaussian. Nevertheless, we are mainly interested in an order
of magnitude estimate so that we do not care about such
statistical details.} \cite{John04,John05}\,:

\begin{displaymath}
q_0 = -0.90 {\pm} 0.65 \ \ , \ \ j_0 = 2.7 {\pm} 6.7 \ \ ,
\end{displaymath}

\begin{displaymath}
s_0 = 36.5 {\pm} 52.9 \ \ , \ \ l_0 = 142.7 {\pm} 320 \ \ .
\end{displaymath}
Because of the Gaussian assumptions it relies on, the Fisher
matrix forecasts are known to be lower limits to the accuracy a
given experiment can attain on the determination of a set of
parameters. This is indeed the case with the comparison suggesting
that our predictions are quite optimistic. It is worth stressing,
however, that the analysis in \cite{John04,John05} used the Gold
SNeIa dataset which is poorer in high redshift SNeIa than the
\cite{D07} one we are mimicking so that larger errors on the
higher order parameters $(s_0, l_0)$ are expected.

Rather than computing the errors on $f(R_0)$ and its first three
derivatives, it is more interesting to look at the precision
attainable on the dimensionless ratios $(\eta_{20}, \eta_{30}$
introduced above since they quantify how much deviations from the
linear order are present. For the fiducial model we are
considering, both $\eta_{20}$ and $\eta_{30}$ vanish, while, using
the covariance matrix for a present day survey and setting
$\sigma_M/\Omega_M \simeq 10\%$, their uncertainties read\,:

\begin{displaymath}
(\sigma_{20}, \sigma_{30}) = (0.04, 0.04) \ .
\end{displaymath}
As an application, we can look at Figs.\,\ref{fig: r20} and
\ref{fig: r30} showing how $(\eta_{20}, \eta_{30})$ depend on the
present day EoS $w_0$ for $f(R)$ models sharing the same
cosmographic parameters of a dark energy model with constant EoS.
As it is clear, also considering only the $1 \sigma$ range, the
full region plotted is allowed by such large constraints on
$(\eta_{20}, \eta_{30})$ thus meaning that the full class of
corresponding $f(R)$ theories is viable. As a consequence, we may
conclude that the present day SNeIa data are unable to
discriminate between a $\Lambda$ dominated Universe and this class
of fourth order gravity theories.

As a next step, we consider a SNAP\,-\,like survey \cite{SNAP}
thus setting $({\cal{N}}_{SNeIa}, \sigma_m) = (2000, 0.02)$. We
use the same redshift distribution in Table 1 of \cite{Kim} and
add 300 nearby SNeIa in the redshift range $(0.03, 0.08)$. The
Fisher matrix calculation gives for the uncertainties on the
cosmographic parameters\,:

\begin{displaymath}
(\sigma_1, \sigma_2, \sigma_3, \sigma_4) = (0.08, 1.0, 4.8, 13.7)
\ .
\end{displaymath}
The significant improvement of the accuracy in the determination
of $(q_0, j_0, s_0, l_0)$ translates in a reduction of the errors
on $(\eta_{20}, \eta_{30})$ which now read\,:

\begin{displaymath}
(\sigma_{20}, \sigma_{30}) = (0.007, 0.008)
\end{displaymath}
having assumed that, when SNAP data will be available, the matter
density parameter $\Omega_M$ has been determined with a precision
$\sigma_M/\Omega_M \sim 1\%$. Looking again at Figs.\,\ref{fig:
r20} and \ref{fig: r30}, it is clear that the situation is
improved. Indeed, the constraints on $\eta_{20}$ makes it possible
to narrow the range of allowed models with low matter content (the
dashed line), while models with typical values of $\Omega_M$ are
still viable for $w_0$ covering almost the full horizontal axis.
On the other hand, the constraint on $\eta_{30}$ is still too weak
so that almost the full region plotted is allowed.

Finally, we consider an hypothetical future SNeIa survey working
at the same photometric accuracy as SNAP and with the same
redshift distribution, but increasing the number of SNeIa up to
${\cal{N}}_{SNeIa} = 6 {\times} 10^4$ as expected from, e.g., DES
\cite{DES}, PanSTARRS \cite{PanSTARRS}, SKYMAPPER \cite{SKY},
while still larger numbers may potentially be achieved by ALPACA
\cite{ALPACA} and LSST \cite{LSST}. Such a survey can achieve\,:

\begin{displaymath}
(\sigma_1, \sigma_2, \sigma_3, \sigma_4) = (0.02, 0.2, 0.9, 2.7)
\end{displaymath}
so that, with $\sigma_M/\Omega_M \sim 0.1\%$, we get\,:

\begin{displaymath}
(\sigma_{20}, \sigma_{30}) = (0.0015, 0.0016) \ .
\end{displaymath}
Fig.\,\ref{fig: r20} shows that, with such a precision on
$\eta_{20}$, the region of $w_0$ values allowed essentially
reduces to the $\Lambda$CDM value, while, from Fig.\,\ref{fig:
r30}, it is clear that the constraint on $\eta_{30}$ definitively
excludes models with low matter content further reducing the range
of $w_0$ values to quite small deviations from the $w_0 = -1$. We
can therefore conclude that such a survey will be able to
discriminate between the concordance $\Lambda$CDM model and all
the $f(R)$ theories giving the same cosmographic parameters as
quiessence models other than the $\Lambda$CDM itself.

A similar discussion may be repeated for $f(R)$ models sharing the
same $(q_0, j_0, s_0, l_0)$ values as the CPL model even if it is
less intuitive to grasp the efficacy of the survey being the
parameter space multivalued. For the same reason, we have not
explored what is the accuracy on the double power\,-\,law or HS
models, even if this is technically possible. Actually, one should
first estimate the errors on the present day value of $f(R)$ and
its three time derivatives and then propagate them on the model
parameters using the expressions obtained in Sect. VI. The
multiparameter space to be explored makes this exercise quite
cumbersome so that we leave it for a furthcoming work where we
will explore in detail how these models compare to the present and
future data.

\section{What we have  learnt from Cosmography}

The recent amount of good quality data have given a new input to
the observational cosmology. As often in science, new and better
data lead to unexpected discoveries as in the case of the nowadays
accepted evidence for cosmic acceleration. However, a fierce and
strong debate is still open on what this cosmic speed up implies
for theoretical cosmology. The equally impressive amount of
different (more or less) viable candidates have also generated a
great confusion so that model independent analyses are welcome. A
possible solution could come from   cosmography  rather than
assuming {\it ad hoc} solutions of the cosmological Friedmann
equations. Present day and future SNeIa surveys have renewed the
interest in the determination of the cosmographic parameters so
that it is worth investigating how these quantities can constrain
cosmological models.

Motivated by this consideration, in the framework of metric
formulation of $f(R)$ gravity, we have here derived the
expressions of the present day values of $f(R)$ and its first
three derivatives as function of the matter density parameter
$\Omega_M$, the Hubble constant $H_0$ and the cosmographic
parameters $(q_0, j_0, s_0, l_0)$. Although based on a third order
Taylor expansion of $f(R)$, we have shown that such relations hold
for a quite large class of models so that they are valid tools to
look for viable $f(R)$ models without the need of solving the
mathematically difficult nonlinear fourth order differential field
equations.

Notwithstanding the common claim that we live in the era of {\it
precision cosmology}, the constraints on $(q_0, j_0, s_0, l_0)$
are still too weak to efficiently apply the program we have
outlined above. As such, we have shown how it is possible to
establish a link between the popular CPL parameterization of the
dark energy equation of state and the derivatives of $f(R)$,
imposing that they share the same values of the cosmographic
parameters. This analysis has lead to the quite interesting
conclusion that the only $f(R)$ function able to give the same
values of $(q_0, j_0, s_0, l_0)$ as the $\Lambda$CDM model is
indeed $f(R) = R + 2 \Lambda$. If future observations will tell us
that the cosmographic parameters are those of the $\Lambda$CDM
model, we can therefore rule out all $f(R)$ theories satisfying
the hypotheses underlying our derivation of Eqs.(\ref{eq:
f0z})\,-\,(\ref{eq: f3z}). Actually, such a result should not be
considered as a no way out for higher order gravity. Indeed, one
could still work out a model with null values of $f''(R_0)$ and
$f'''(R_0)$ as required by the above constraints, but
non\,-\,vanishing higher order derivatives. One could well argue
that such a contrived model could be rejected on the basis of the
Occam razor, but nothing prevents from still taking it into
account if it turns out to be both in agreement with the data and
theoretically well founded.

If new SNeIa surveys will determine the cosmographic parameters
with good accuracy, acceptable constraints on the two
dimensionless ratios $\eta_{20} \propto f''(R_0)/f(R_0)$ and
$\eta_{30} \propto f'''(R_0)/f(R_0)$ could be obtained thus
allowing to discriminate among rival $f(R)$ theories. To
investigate whether such a program is feasible, we have pursued a
Fisher matrix based forecasts of the accuracy future SNeIa surveys
can achieve on the cosmographic parameters and hence on
$(\eta_{20}, \eta_{30})$. It turns out that a SNAP\,-\,like survey
can start giving interesting (yet still weak) constraints allowing
to reject $f(R)$ models with low matter content, while a
definitive improvement is achievable with future SNeIa survey
observing $\sim 10^4$ objects thus making it possible to
discriminate between $\Lambda$CDM and a large class of fourth
order theories. It is worth stressing, however, that the
measurement of $\Omega_M$ should come out as the result of a model
independent probe such as the gas mass fraction in galaxy clusters
which, at present, is still far from the $1\%$ requested
precision. On the other hand, one can also rely on the $\Omega_M$
estimate from the CMBR anisotropy and polarization spectra even if
this comes to the price of assuming that the physics at
recombination is strictly described by GR so that one has to limit
its attention to $f(R)$ models reducing to $f(R) \propto R$ during
that epoch. However, such an assumption is quite common in many
$f(R)$ models available in literature so that it is not a too
restrictive limitation.

A further remark is in order concerning what kind of data can be
used to constrain the cosmographic parameters. The use of the
fifth order Taylor expansion of the scale factor makes it possible
to not specify any underlying physical model thus relying on the
minimalist assumption that the Universe is described by the flat
Robertson\,-\,Walker metric. While useful from a theoretical
perspective, such a generality puts severe limitations to the
dataset one can use. Actually, we can only resort to observational
tests depending only on the background evolution so that the range
of astrophysical probes reduces to standard candles (such as SNeIa
and possibly Gamma Ray Bursts \cite{izzo}) and standard rods (such
as the angular size\,-\,redshift relation for compact
radiosources). Moreover, pushing the Hubble diagram to $z \sim 2$
may rise the question of the impact of gravitational lensing
amplification on the apparent magnitude of the adopted standard
candle. The magnification probability distribution function
depends on the growth of perturbations
\cite{Holz,Holz05,Hui,Friem,Coor} so that one should worry about
the underlying physical model in order to estimate whether this
effect biases the estimate of the cosmographic parameters.
However, it has been shown \cite{R06,Jon,Gunna,Nordin,Sark} that
the gravitational lensing amplification does not alter
significantly the measured distance modulus for $z \sim 1$ SNeIa.
Although such an analysis has been done for GR based models, we
can argue that, whatever is the $f(R)$ model, the growth of
perturbations finally leads to a distribution of structures along
the line of sight that is as similar as possible to the observed
one so that the lensing amplification is approximately the same.
We can therefore argue that the systematic error made by
neglecting lensing magnification is lower than the statistical
ones expected by the future SNeIa surveys. On the other hand, one
can also try further reducing this possible bias using the method
of flux averaging \cite{WangFlux} even if, in such a case, our
Fisher matrix calculation should be repeated accordingly. It is
also worth noting that the constraints on the cosmographic
parameters may be tightened by imposing some physically motivated
priors in the parameter space. For instance, we can impose that
the Hubble parameter $H(z)$ stays always positive over the full
range probed by the data or that the transition from past
deceleration to present acceleration takes place over the range
probed by the data (so that we can detect it). Such priors should
be included in the likelihood definition so that the Fisher matrix
should be recomputed which is left for a forthcoming work.

Although the present day data are still too limited to efficiently
discriminate among rival $f(R)$ models, we are confident that an
aggressive strategy aiming at a very precise determination of the
cosmographic parameters could offer stringent constraints on
higher order gravity without the need of solving the field
equations or addressing the complicated problems related to the
growth of perturbations. Almost 80 years after the pioneering
distance\,-\,redshift diagram by Hubble, the old cosmographic
approach appears nowadays as a precious observational tool to
investigate the new developments of cosmology.

\section{The weak-field limit of $f(R)$-gravity}
\label{sec:ext_theories}

Before facing the problem of galaxy clusters by $f(R)$-gravity, a
discussion is due on the weak-field limit of such a theory which,
being of fourth order in metric formalism, could lead to results
radically different with respect to the case $f(R)=R$, the
standard second order General Relativity.

Let us consider the general action\,:

\begin{equation}\label{actfR}
{\cal A}\, = \,\int d^4x\sqrt{-g}\left[f(R)+{\cal X}{\cal
L}_m\right]\,,
\end{equation}
where $f(R)$ is an  analytic function of the Ricci scalar $R$, $g$
is the determinant of the metric $g_{\mu\nu}$, ${\displaystyle
{\cal X}=\frac{16\pi G}{c^4}}$ is the coupling constant and ${\cal
L}_m$ is the standard perfect-fluid matter Lagrangian. Such an
action is the straightforward generalization of the
Hilbert-Einstein action of GR obtained for $f(R)=R$. Since we are
considering the  metric approach, field equations are obtained by
varying (\ref{actfR}) with respect to the metric\,:

\begin{eqnarray}\label{fe}
f'R_{\mu\nu}-\frac{1}{2}fg_{\mu\nu}-f'_{;\mu\nu}+g_{\mu\nu}\Box
f'=\frac{\mathcal{X}}{2}T_{\mu\nu}\,.
\end{eqnarray}
where $T_{\mu\nu}=\frac{-2}{\sqrt{-g}}\frac{\delta(\sqrt{-g}{\cal
L}_m)}{\delta g^{\mu\nu}}$ is the energy momentum tensor of
matter, the prime indicates the derivative with respect to $R$ and
$\Box={{}_{;\sigma}}^{;\sigma}$. We adopt the signature
$(+,-,-,-)$.

As discussed in details in  \cite{noi-prd}, we deal with the
Newtonian and the post-Newtonian limit of $f(R)$ - gravity on a
spherically symmetric background. Solutions for the field
equations can be obtained by imposing the spherical symmetry
\cite{arturocqg}:

\begin{eqnarray}\label{me}
ds^2\,=\,g_{00}(x^0,r)d{x^0}^2+g_{rr}(x^0,r)dr^2-r^2d\Omega
\end{eqnarray}
where $x^0\,=\,ct$ and $d\Omega$ is the angular element.

To develop the  post-Newtonian limit of the theory, one can
consider a perturbed metric with respect to a Minkowski background
$g_{\mu\nu}\,=\,\eta_{\mu\nu}+h_{\mu\nu}$. The metric coefficients
can be developed as:

\begin{eqnarray}\label{definexpans}
\left\{\begin{array}{ll} g_{tt}(t,
r)\simeq1+g^{(2)}_{tt}(t,r)+g^{(4)}_{tt}(t,r)
\\\\
g_{rr}(t,r)\simeq-1+g^{(2)}_{rr}(t,r)\\\\
g_{\theta\theta}(t,r)=-r^2\\\\
g_{\phi\phi}(t,r)=-r^2\sin^2\theta
\end{array}\right.\,,
\end{eqnarray}
where we put, for the sake of simplicity, $c\,=\,1$\,,\
$x^0=ct\rightarrow t$. We want to obtain the most general result
without imposing particular forms for the $f(R)$-Lagrangian. We
only consider analytic Taylor expandable functions
\begin{eqnarray}\label{sertay}
f(R)\simeq f_0+f_1R+f_2R^2+f_3R^3+...\,.
\end{eqnarray}
To obtain the post-Newtonian approximation of $f(R)$ - gravity,
one has to plug the expansions (\ref{definexpans}) and
(\ref{sertay}) into the field equations (\ref{fe}) and then expand
the system up to the orders $O(0),\, O(2)$ and $O(4)$ . This
approach provides general results and specific (analytic)
Lagrangians are selected by  the coefficients $f_i$ in
(\ref{sertay}) \cite{noi-prd}.

If we now consider the $O(2)$ - order of approximation, the field
equations (\ref{fe}),  in the vacuum case, results to be

\begin{eqnarray}\label{eq2}
\left\{\begin{array}{ll}
f_1rR^{(2)}-2f_1g^{(2)}_{tt,r}+8f_2R^{(2)}_{,r}-f_1rg^{(2)}_{tt,rr}+4f_2rR^{(2)}=0
\\\\
f_1rR^{(2)}-2f_1g^{(2)}_{rr,r}+8f_2R^{(2)}_{,r}-f_1rg^{(2)}_{tt,rr}=0
\\\\
2f_1g^{(2)}_{rr}-r[f_1rR^{(2)}
\\\\
-f_1g^{(2)}_{tt,r}-f_1g^{(2)}_{rr,r}+4f_2R^{(2)}_{,r}+4f_2rR^{(2)}_{,rr}]=0
\\\\
f_1rR^{(2)}+6f_2[2R^{(2)}_{,r}+rR^{(2)}_{,rr}]=0
\\\\
2g^{(2)}_{rr}+r[2g^{(2)}_{tt,r}-rR^{(2)}+2g^{(2)}_{rr,r}+rg^{(2)}_{tt,rr}]=0
\end{array} \right.\end{eqnarray}
It is evident that the trace equation (the fourth in the system
(\ref{eq2})), provides a differential equation with respect to the
Ricci scalar which allows to solve  the system  at $O(2)$ - order.
One obtains the general solution\,:
\begin{eqnarray}\label{sol}
\left\{\begin{array}{ll}
g^{(2)}_{tt}=\delta_0-\frac{2GM}{f_1r}-\frac{\delta_1(t)e^{-r\sqrt{-\xi}}}{3\xi
r}+\frac{\delta_2(t)e^{r\sqrt{-\xi}}}{6({-\xi)}^{3/2}r}
\\\\
g^{(2)}_{rr}=-\frac{2GM}{f_1r}+\frac{\delta_1(t)[r\sqrt{-\xi}+1]e^{-r\sqrt{-\xi}}}{3\xi
r}-\frac{\delta_2(t)[\xi r+\sqrt{-\xi}]e^{r\sqrt{-\xi}}}{6\xi^2r}
\\\\
R^{(2)}=\frac{\delta_1(t)e^{-r\sqrt{-\xi}}}{r}-\frac{\delta_2(t)\sqrt{-\xi}e^{r\sqrt{-\xi}}}{2\xi
r}\end{array} \right.
\end{eqnarray}
where $\xi\doteq\displaystyle\frac{f_1}{6f_2}$, $f_1$ and $f_2$
are the expansion coefficients obtained by the $f(R)$-Taylor
series.
 In the limit $f\rightarrow R$, for a point-like source of mass $M$ we recover the standard
Schwarzschild solution. Let us notice that the integration
constant $\delta_0$ is  dimensionless, while the two arbitrary
time-functions  $\delta_1(t)$ and $\delta_2(t)$ have respectively
the dimensions of $lenght^{-1}$ and $lenght^{-2}$; $\xi$ has the
dimension $lenght^{-2}$. As extensively discussed in
\cite{noi-prd}, the functions $\delta_i(t)$ ($i=1,2$) are
completely arbitrary since the differential equation system
(\ref{eq2}) depends only on spatial derivatives. Besides, the
integration constant $\delta_0$ can be set to zero, as in the
standard theory of  potential,  since it represents an unessential
additive quantity.
 In order to obtain the physical prescription of the asymptotic flatness
at infinity, we can discard the Yukawa growing mode in (\ref{sol})
and then the metric is \,:

\begin{eqnarray}\label{mesol}
ds^2&=&\biggl[1-\frac{2GM}{f_1r}-\frac{\delta_1(t)e^{-r\sqrt{-\xi}}}{3\xi
r}\biggr]dt^2\nonumber\\
&-&\biggl[1+\frac{2GM}{f_1r}-\frac{\delta_1(t)(r\sqrt{-\xi}+1)e^{-r\sqrt{-\xi}}}{3\xi
r}\biggr]dr^2\nonumber\\
&-&r^2d\Omega\,.
\end{eqnarray}
The Ricci scalar curvature is
\begin{equation}
R\,=\,\frac{\delta_1(t)e^{-r\sqrt{-\xi}}}{r}\,.
\end{equation}

The solution can be given also in terms of gravitational
potential. In particular, we have an explicit Newtonian-like term
into the definition. The first of (\ref{sol}) provides the second
order solution in term of the metric expansion (see the definition
(\ref{definexpans})). In particular, it is
$g_{tt}\,=\,1+2\phi_{grav}\,=\,1+g_{tt}^{(2)}$ and then the
gravitational potential of an analytic $f(R)$-theory is

\begin{eqnarray}\label{gravpot}
\phi_{grav}\,=\,-\frac{GM}{f_1r}-\frac{\delta_1(t)e^{-r\sqrt{-\xi}}}{6\xi
r}\,.
\end{eqnarray}
Among the possible analytic $f(R)$-models, let us consider the
Taylor expansion where the cosmological term (the above $f_0$) and
terms higher than second have been discarded. For the sake of
simplicity, we rewrite the Lagrangian (\ref{sertay}) as
\begin{equation}
f(R) \sim a_1 R + a_2 R^2 + ...
\end{equation}
and specify the above gravitational potential (\ref{gravpot}),
generated by a point-like matter distribution, as:
\begin{equation}
\label{gravpot1} \phi(r) = -\frac{3 G M}{4 a_1
r}\left(1+\frac{1}{3}e^{-\frac{r}{L}}\right)\,,
\end{equation}
where
\begin{equation}\label{lengths model}
L \equiv L(a_{1},a_{2}) =  \left( -\frac{6 a_2}{a_1}
\right)^{1/2}\,.
\end{equation}
$L$ can be defined as  the {\it interaction length} of the
problem\footnote{Such a length is function of the series
coefficients, $a_{1}$ and $a_{2}$, and it is not a free
independent parameter in the following fit procedure.} due to the
correction to the Newtonian potential. We have changed the
notation to remark that we are doing only a specific choice in the
wide class of potentials (\ref{gravpot}), but the following
considerations are completely general.

\section{Extended systems}
\label{sec:ext_systems}

The gravitational potential (\ref{gravpot1}) is a point-like one.
Now we have to generalize this solution for
 extended systems. Let us describe galaxy clusters as
spherically symmetric systems and then we have to extend the above
considerations
 to this geometrical configuration. We simply consider
the system composed by many infinitesimal mass elements $dm$ each
one contributing with a point-like gravitational potential. Then,
summing up all terms, namely integrating them on a spherical
volume, we obtain a suitable potential. Specifically, we have to
solve
 the integral:
\begin{equation}
\Phi(r) = \int_{0}^{\infty} r'^{2} dr' \int_{0}^{\pi} \sin \theta'
d\theta' \int_{0}^{2\pi} d\omega' \hspace{0.1cm} \phi(r')\,.
\end{equation}
The point-like potential (\ref{gravpot1})can be split in two
terms. The \textit{Newtonian} component is
\begin{equation}
\phi_{N}(r) = -\frac{3 G M}{4 a_1 r}
\end{equation}
The extended integral of such a part is the well-known (apart from
the numerical constant $\frac{3}{4 a_{1}}$) expression. It is
\begin{equation}
\Phi_{N}(r) = -\frac{3}{4 a_1}\frac{G M(<r)}{r} \\
\end{equation}
where $M(<r)$ is the mass enclosed in a sphere with radius $r$.
The \textit{correction} term:
\begin{equation}
\phi_{C}(r) = -\frac{G M}{4 a_1}\frac{e^{-\frac{r}{L}}}{r}
\end{equation}
considering some analytical steps in the integration of the
angular part, gives the expression:
\begin{equation}
\Phi_{C}(r) = - \frac{2 \pi G}{4} \cdot L \int_{0}^{\infty} dr' r'
\rho(r') \cdot \frac{e^{-\frac{|r-r'|}{L}} -
e^{-\frac{|r+r'|}{L}}}{r}
\end{equation}
The radial integral is  numerically estimated once the mass
density is given. We underline a fundamental difference between
such a term and the Newtonian one: while in the latter, the matter
outside the spherical shell of radius $r$ does not contribute to
the potential, in the former external matter takes part to the
integration procedure. For this reason we split the corrective
potential in two terms:
\begin{itemize}
    \item if $r' < r$:
      {\setlength\arraycolsep{0.2pt}
      \begin{eqnarray}
      \Phi_{C,int}(r) &=& - \frac{2 \pi G}{4} \cdot L \int_{0}^{r} dr' r'
      \rho(r') \cdot \frac{e^{-\frac{|r-r'|}{L}} -
      e^{-\frac{|r+r'|}{L}}}{r}  \nonumber \\
     & =& - \frac{2 \pi G}{4} \cdot L \int_{0}^{r} dr' r' \rho(r') \cdot
      e^{-\frac{r+r'}{L}} \left( \frac{-1 + e^{\frac{2r'}{L}}}{r}
      \right)\nonumber
      \end{eqnarray}}
   \item if $r' > r$:
      {\setlength\arraycolsep{0.2pt}
      \begin{eqnarray}
      \Phi_{C,ext}(r) &=& - \frac{2 \pi G}{4} \cdot L \int_{r}^{\infty}
      dr' r' \rho(r') \cdot \frac{e^{-\frac{|r-r'|}{L}} -
      e^{-\frac{|r+r'|}{L}}}{r} = \nonumber \\
      &=& - \frac{2 \pi G}{4} \cdot L \int_{r}^{\infty} dr' r' \rho(r')
      \cdot e^{-\frac{r+r'}{L}} \left( \frac{-1 + e^{\frac{2r}{L}}}{r}
      \right)\nonumber
      \end{eqnarray}}
\end{itemize}
The total potential of the spherical mass distribution will be
\begin{equation}\label{eq:total corrected potential}
\Phi(r) = \Phi_{N}(r) + \Phi_{C,int}(r) + \Phi_{C,ext}(r)
\end{equation}
As we will show below, for our purpose, we need the gravitational
potential derivative with respect to the variable $r$; the two
derivatives may not be evaluated analytically so we estimate them
numerically, once we have given an expression for the
\textit{total} mass density $\rho(r)$. While the Newtonian term
gives the simple expression:
\begin{equation}
- \frac{\mathrm{d}\Phi_{N}}{\mathrm{d}r}(r) = - \frac{3}{4 a_{1}}
\frac{G M(<r)}{r^{2}}
\end{equation}
The internal and external derivatives of the corrective potential
terms are much longer. We do not give them explicitly for sake of
brevity, but  they are integral-functions of the form
\begin{equation}
{\mathcal{F}}(r, r') = \int_{\alpha(r)}^{\beta(r)} dr' \ f(r,r')
\end{equation}
from which one has: {\setlength\arraycolsep{0.2pt}
\begin{eqnarray}
\frac{\mathrm{d}\mathcal{F}(r, r')}{\mathrm{d}r} &=&
\int_{\alpha(r)}^{\beta(r)} dr'
\frac{\mathrm{d}f(r,r')}{\mathrm{d}r} + \nonumber \\
&-& f(r,\alpha(r)) \frac{\mathrm{d}\alpha}{\mathrm{d}r}(r) +
f(r,\beta(r)) \frac{\mathrm{d}\beta}{\mathrm{d}r}(r)
\end{eqnarray}}
Such an expression is numerically derived once the integration
extremes are given. A general consideration is in order at this
point. Clearly, the Gauss theorem holds only for the Newtonian
part  since, for this term, the force law scales as $1/r^2$. For
the total potential (\ref{gravpot1}), it does not hold anymore due
to the correction. From a physical point of view, this is not a
problem because the full conservation laws are determined, for
$f(R)$-gravity,  by the contracted Bianchi identities  which
assure the self-consistency. For a detailed discussion, see
\cite{CapCardTro07,capfra,faraoni}.

\section{The cluster mass profiles}
\label{sec:mass_profiles}

Clusters of galaxies are generally considered self-bound
gravitational systems with spherical symmetry and in hydrostatic
equilibrium if virialized. The last two hypothesis are still
widely used, despite of the fact that it has been widely proved
that most clusters show more complex morphologies and/or signs of
strong interactions or dynamical
activity, especially in their innermost regions (\cite{Chak08,DeFil05}). \\
Under the hypothesis of  spherical symmetry in hydrostatic
equilibrium, the structure equation can be derived from the
collisionless Boltzmann equation
\begin{equation}\label{Boltzmann equation}
\frac{d}{dr}(\rho_{gas}(r) \hspace{0.1cm} \sigma^{2}_{r}) +
\frac{2\rho_{gas}(r)}{r}(\sigma^{2}_{r}-\sigma^{2}_{\theta,\omega})
= -\rho_{gas}(r) \cdot \frac{d\Phi(r)}{dr}
\end{equation}
where $\Phi$ is the gravitational potential of the cluster,
$\sigma_{r}$ and $\sigma_{\theta,\omega}$ are the mass-weighted
velocity dispersions in the radial and tangential directions,
respectively, and $\rho$ is gas mass-density. For an isotropic
system, it is
\begin{equation}\label{velocity dispersion}
\sigma_{r} = \sigma_{\theta,\omega}
\end{equation}
The pressure profile can be related to these quantities by
\begin{equation}\label{pressure}
P(r) = \sigma^{2}_{r} \rho_{gas}(r)
\end{equation}
Substituting Eqs. (\ref{velocity dispersion}) and (\ref{pressure})
into Eq. (\ref{Boltzmann equation}), we have, for an isotropic
sphere,
\begin{equation}\label{isotropic sphere}
\frac{d P(r)}{dr} = - \rho_{gas}(r) \frac{d\Phi(r)}{dr}
\end{equation}
For a gas sphere with temperature profile $T(r)$, the velocity
dispersion becomes
\begin{equation}\label{temperature}
\sigma^{2}_{r} = \frac{k T(r)}{\mu m_{p}}
\end{equation}
where $k$ is the Boltzmann constant, $\mu \approx 0.609$ is the
mean mass particle and $m_{p}$ is the proton mass. Substituting
Eqs. (\ref{pressure}) and (\ref{temperature}) into Eq.
(\ref{isotropic sphere}), we obtain
\[
\frac{d}{dr} \left( \frac{k T(r)}{\mu m_{p}} \rho_{gas}(r) \right)
= -\rho_{gas}(r) \frac{d \Phi}{dr}
\]
or, equivalently,
\begin{equation}\label{eq:Boltzmann potential}
-\frac{d\Phi}{dr} = \frac{k T(r)}{\mu m_{p}
r}\left[\frac{d\ln\rho_{gas}(r)}{d\ln r} + \frac{d\ln T(r)}{d\ln
r}\right]
\end{equation}
Now the total gravitational potential of the cluster is:
\begin{equation}\label{eq:total corrected potential1}
\Phi(r) = \Phi_{N}(r) + \Phi_{C}(r)
\end{equation}
with
\begin{equation}
\Phi_{C}(r) = \Phi_{C,int}(r) + \Phi_{C,ext}(r)
\end{equation}
It is worth underlining that if we consider \textit{only} the
standard Newtonian  potential, the \textit{total} cluster mass
$M_{cl,N}(r)$ is composed by gas mass $+$ mass of galaxies $+$
cD-galaxy mass $+$ dark matter and it is given by the expression:
{\setlength\arraycolsep{0.2pt}
\begin{eqnarray}
\label{eq:M_tot} M_{cl,N}(r) &=& M_{gas}(r) + M_{gal}(r) +
M_{CDgal}(r) + M_{DM}(r)  \nonumber
\\
&=& - \frac{k T(r)}{\mu m_{p} G} r
\left[\frac{d\ln\rho_{gas}(r)}{d\ln r}+\frac{d\ln T(r)}{d\ln
r}\right]
\end{eqnarray}}
$M_{cl,N}$  means the standard estimated \textit{Newtonian} mass.
Generally the galaxy part contribution  is considered negligible
with respect to the other two components so we have:
\[
M_{cl,N}(r) \approx M_{gas}(r) + M_{DM}(r) \approx
\]
\[
\hspace{1.35cm} \approx - \frac{k T(r)}{\mu m_{p}} r
\left[\frac{d\ln\rho_{gas}(r)}{d\ln r}+\frac{d\ln T(r)}{d\ln
r}\right]
\]
Since the gas-mass estimates are provided by X-ray observations,
the equilibrium equation can be used to derive the amount of dark
matter present in a cluster of galaxies and its spatial
distribution.

Inserting the previously defined \textit{extended-corrected}
potential of Eq.~(\ref{eq:total corrected potential1}) into
Eq.~(\ref{eq:Boltzmann potential}), we obtain:
\begin{equation}
\label{eq:corrected_mass} -\frac{\mathrm{d}\Phi_{N}}{\mathrm{d}r}
-\frac{\mathrm{d}\Phi_{C}}{\mathrm{d}r} =\frac{k T(r)}{\mu m_{p}
r}\left[\frac{\mathrm{d}\ln\rho_{gas}(r)}{\mathrm{d}\ln r} +
\frac{\mathrm{d}\ln T(r)}{\mathrm{d}\ln r}\right]
\end{equation}
from which the \textit{extended-corrected} mass estimate follows:
{\setlength\arraycolsep{0.2pt}
\begin{eqnarray}\label{eq:fit relation}
M_{cl,EC}(r) &+& \frac{4 a_{1}}{3G} r^{2}
\frac{\mathrm{d}\Phi_{C}}{\mathrm{d}r}(r) = \nonumber \\ &=&
\frac{4 a_{1}}{3} \left[ - \frac{k T(r)}{\mu m_{p}G} r
\left(\frac{d\ln\rho_{gas}(r)}{d\ln r}+\frac{d\ln T(r)}{d\ln
r}\right) \right]
\end{eqnarray}}
Since the use of a corrected potential avoids, in principle, the
additional requirement of dark matter, the total cluster mass, in
this case, is given by:
\begin{equation}
M_{cl,EC}(r) = M_{gas}(r) + M_{gal}(r) + M_{CDgal}(r)
\end{equation}
and the mass density in the $\Phi_{C}$ term is
\begin{equation}
\rho_{cl,EC}(r) = \rho_{gas}(r) + \rho_{gal}(r) + \rho_{CDgal}(r)
\end{equation}
with the  density components derived from observations.

In this work, we will use  Eq.~(\ref{eq:fit relation}) to compare
the baryonic mass profile $M_{cl,EC}(r)$, estimated from
observations, with the theoretical deviation from the Newtonian
gravitational potential, given by the expression ${\displaystyle
-\frac{4 a_{1}}{3G} r^{2}
\frac{\mathrm{d}\Phi_{C}}{\mathrm{d}r}(r)}$. Our goal is to
reproduce the observed mass profiles for a sample of galaxy
clusters.

\section{The Galaxy Cluster Sample}
\label{sec:sample}

The formalism described in \S~\ref{sec:mass_profiles} can be
applied to a sample of $12$ galaxy clusters. We shall use the
cluster sample studied in ~\cite{Vik05,Vik06} which consists of
$13$ low-redshift clusters spanning a temperature range $0.7\div
9.0\ {\rm keV}$ derived from high quality {\it Chandra} archival
data. In all these clusters, the surface brightness and the gas
temperature profiles are measured out to large radii, so that mass
estimates can be extended up to r$_{500}$ or beyond.

\subsection{The Gas Density Model}
\label{sec:gas_model}

The gas density distribution of the clusters in the sample is
described by the analytic model proposed in~\cite{Vik06}. Such a
model modifies the classical $\beta-$model to represent the
characteristic properties of the observed X-ray surface brightness
profiles, i.e. the power-law-type cusps of gas density in the
cluster center, instead of a flat core and the steepening of the
brightness profiles at large radii. Eventually, a second
$\beta-$model, with a small core radius, is added to improve  the
model close to the cluster cores. The  analytical form for the
particle emission  is given by: {\setlength\arraycolsep{0.2pt}
\begin{eqnarray}
\label{gas density vik} n_{p}n_{e} = n_{0}^{2} \cdot
\frac{(r/r_{c})^{-\alpha}}{(1+r^{2}/r_{c}^{2})^{3\beta-\alpha/2}}
& \cdot&\frac{1}{(1+r^{\gamma}/r_{s}^{\gamma})^{\epsilon/\gamma}}+
\nonumber \\
&+& \frac{n_{02}^{2}}{(1+r^{2}/r_{c2}^{2})^{3\beta_{2}}}
\end{eqnarray}}
which can be easily converted to a mass density using the
relation:
\begin{equation}
\label{eq:gas_density} \rho_{gas} = n_T \cdot \mu m_p =
\frac{1.4}{1.2} n_e m_p
\end{equation}
where $n_T$ is the total number density of particles in the gas.
The resulting model has a large number of parameters, some of
which do not have a direct physical interpretation. While this can
often be inappropriate and computationally inconvenient, it suits
well our case, where the main requirement is a detailed
qualitative
description of the cluster profiles.\\
In \cite{Vik06},  Eq.~(\ref{gas density vik}) is applied to a
restricted range of distances from the cluster center, i.e.
between an inner cutoff $r_{min}$, chosen to exclude the central
temperature bin ($\approx 10\div 20\ {\rm kpc}$) where the ICM is
likely to be multi-phase, and $r_{det}$, where the X-ray surface
brightness is at least $3 \sigma$ significant. We have
extrapolated the above function to values outside this restricted
range using the following criteria:
\begin{itemize}
  \item for $r < r_{min}$, we have performed a linear extrapolation
  of the first three terms out to $r = 0$ kpc;
  \item for $r > r_{det}$, we have performed a linear extrapolation
  of the last three terms out to a distance $\bar{r}$ for which
  $\rho_{gas}(\bar{r})=\rho_{c}$, $\rho_{c}$ being the critical
  density of the Universe at the cluster redshift:
  $\rho_{c} = \rho_{c,0} \cdot (1 + z)^{3}$. For radii larger than $\bar{r}$,
  the gas density is assumed constant at $\rho_{gas}(\bar{r})$.
\end{itemize}
We point out that, in Table~\ref{tab1}, the radius limit $r_{min}$
is almost the same as given in the previous definition. When the
value given by \cite{Vik06} is less than the cD-galaxy radius,
which is defined in the next section, we  choose  this last one as
the lower limit. On the contrary, $r_{max}$ is quite different
from $r_{det}$: it is fixed by considering the higher value of
temperature profile
and not by imaging methods. \\
We then compute the gas mass $M_{gas}(r)$ and the total mass
$M_{cl,N}(r)$, respectively, for all clusters in our sample,
substituting Eq.~(\ref{gas density vik}) into
Eqs.~(\ref{eq:gas_density}) and (\ref{eq:M_tot}), respectively;
the gas temperature profile has been described in details in
\S~\ref{sec:T_prof}. The resulting mass values, estimated at
$r=r_{max}$, are listed in Table~\ref{tab1}.

\begin{table}
\begin{minipage}{\textwidth}
 \caption{Column 1: Cluster name. Column2: Richness. Column 3: cluster total mass.
 Column 4: gas mass. Column 5: galaxy mass. Column 6: cD-galaxy mass. All mass values are estimated at $r=r_{max}$. Column 7:
 ratio of total galaxy mass to gas mass. Column 8: minimum radius. Column 9: maximum
radius.}
 \label{tab1}
 \resizebox*{0.75\textwidth}{!}{
 \begin{tabular}{ccccccccc}
    \hline
  name & R &$M_{cl,N}$ & $M_{gas}$ & $M_{gal}$  & $M_{cDgal}$  & $\frac{gal}{gas}$ & $r_{min}$ & $r_{max}$  \\
  &  & $(M_{\odot})$ &($M_{\odot}$)& ($M_{\odot}$) & ($M_{\odot}$)  &  & (kpc) & (kpc) \\
  \hline
  \hline
  A133    & 0 & $4.35874\cdot10^{14}$ & $2.73866\cdot10^{13}$ & $5.20269\cdot10^{12}$ & $1.10568\cdot10^{12}$ & $0.23$ &  $86$ & $1060$ \\
  A262    & 0 & $4.45081\cdot10^{13}$ & $2.76659\cdot10^{12}$ & $1.71305\cdot10^{11}$ & $5.16382\cdot10^{12}$ & $0.25$ &  $61$ & $ 316$ \\
  A383    & 2 & $2.79785\cdot10^{14}$ & $2.82467\cdot10^{13}$ & $5.88048\cdot10^{12}$ & $1.09217\cdot10^{12}$ & $0.25$ &  $52$ & $ 751$ \\
  A478    & 2 & $8.51832\cdot10^{14}$ & $1.05583\cdot10^{14}$ & $2.15567\cdot10^{13}$ & $1.67513\cdot10^{12}$ & $0.22$ &  $59$ & $1580$ \\
  A907    & 1 & $4.87657\cdot10^{14}$ & $6.38070\cdot10^{13}$ & $1.34129\cdot10^{13}$ & $1.66533\cdot10^{12}$ & $0.24$ & $563$ & $1226$ \\
  A1413   & 3 & $1.09598\cdot10^{15}$ & $9.32466\cdot10^{13}$ & $2.30728\cdot10^{13}$ & $1.67345\cdot10^{12}$ & $0.26$ &  $57$ & $1506$ \\
  A1795   & 2 & $5.44761\cdot10^{14}$ & $5.56245\cdot10^{13}$ & $4.23211\cdot10^{12}$ & $1.93957\cdot10^{12}$ & $0.11$ &  $79$ & $1151$ \\
  A1991   & 1 & $1.24313\cdot10^{14}$ & $1.00530\cdot10^{13}$ & $1.24608\cdot10^{12}$ & $1.08241\cdot10^{12}$ & $0.23$ &  $55$ & $ 618$ \\
  A2029   & 2 & $8.92392\cdot10^{14}$ & $1.24129\cdot10^{14}$ & $3.21543\cdot10^{13}$ & $1.11921\cdot10^{12}$ & $0.27$ &  $62$ & $1771$ \\
  A2390   & 1 & $2.09710\cdot10^{15}$ & $2.15726\cdot10^{14}$ & $4.91580\cdot10^{13}$ & $1.12141\cdot10^{12}$ & $0.23$ &  $83$ & $1984$ \\
  MKW4    & - & $4.69503\cdot10^{13}$ & $2.83207\cdot10^{12}$ & $1.71153\cdot10^{11}$ & $5.29855\cdot10^{11}$ & $0.25$ &  $60$ & $ 434$ \\
  RXJ1159 & - & $8.97997\cdot10^{13}$ & $4.33256\cdot10^{12}$ & $7.34414\cdot10^{11}$ & $5.38799\cdot10^{11}$ & $0.29$ &  $64$ & $ 568$ \\
  \hline
 \end{tabular}}
 \end{minipage}
\end{table}

\subsection{The Temperature Profiles}
\label{sec:T_prof}

As  stressed in \S~\ref{sec:gas_model}, for the purpose of this
work, we need an accurate qualitative description of the radial
behavior of the gas properties. Standard isothermal or polytropic
models, or even the more complex one proposed in \cite{Vik06}, do
not provide a good description of the data at all radii and for
all clusters in the present sample. We hence describe the gas
temperature profiles using the straightforward X-ray spectral
analysis
results, without the introduction of any analytic model.\\
X-ray spectral values have been provided by A. Vikhlinin (private
communication). A detailed description of the relative spectral
analysis can be found in \cite{Vik05}.

\subsection{The Galaxy Distribution Model}
\label{sec:gal_model}

The galaxy density can be modelled as proposed by \cite{Bah96}.
Even if the galaxy distribution is a \textit{point}-distribution
instead of
 a continuous function, assuming that galaxies are in
equilibrium with gas, we can use a $\beta-$model, $\propto
r^{-3}$, for $r < R_{c}$ from the  cluster center, and a steeper
one, $\propto r^{-2.6}$, for $r > R_{c}$, where $R_{c}$ is the
cluster core radius (its value is taken from Vikhlinin 2006). Its
final expression is:
\begin{equation}\label{gal density bahcall}
\rho_{gal}(r) = \left\{%
\begin{array}{ll}
    \rho_{gal,1} \cdot \left[1+
\left(\frac{r}{R_{c}}\right)^{2} \right]^{-\frac{3}{2}} & \hbox{$r < R_{c}$} \\
    \rho_{gal,2} \cdot \left[1+
\left(\frac{r}{R_{c}}\right)^{2} \right]^{-\frac{2.6}{2}} & \hbox{$r > R_{c}$} \\
\end{array}%
\right.
\end{equation}
where the constants $\rho_{gal,1}$ and $\rho_{gal,2}$ are chosen
in the following way:
\begin{itemize}
 \item \cite{Bah96} provides the central number density of galaxies in
 rich compact clusters for  galaxies located within a
 $1.5$ h$^{-1}$Mpc radius from the cluster center and brighter than $m_3+2^m$
 (where $m_3$ is the magnitude of the third brightest galaxy):
 $n_{gal,0} \sim 10^{3} h^{3}$ galaxies Mpc$^{-3}$. Then we  fix
 $\rho_{gal,1}$ in the range $\sim 10^{34}\div 10^{36}$ kg/kpc$^{3}$.
 For any cluster obeying the condition chosen for the mass
 ratio gal-to-gas, we  assume a typical elliptical and
 cD galaxy mass in the range $10^{12}\div 10^{13} M_{\odot}$.
 \item the constant $\rho_{gal,2}$ has been fixed with the only
 requirement that the galaxy density function has to be continuous at
 $R_{c}$.
\end{itemize}
We have tested the effect of varying galaxy density in the above
range
 $\sim 10^{34}\div 10^{36}$ kg/kpc$^{3}$ on the cluster with the lowest mass, namely A262.
 In this case, we would
expect  great variations with respect to  other clusters; the
result is that the contribution due to galaxies and  cD-galaxy
gives a variation $\leq 1\%$ to the final estimate of  fit
parameters. \\
The cD galaxy density has been modelled as described in
\cite{SA06}; they use a Jaffe model of the form:
\begin{equation}\label{jaffe cd galaxy}
\rho_{CDgal} = \frac{\rho_{0,J}}{\left(\frac{r}{r_{c}}\right)^{2}
\left(1+\frac{r}{r_{c}}\right)^{2}}
\end{equation}
where $r_{c}$ is the core radius while the central density is
obtained from ${\displaystyle M_{J} = \frac{4}{3} \pi R_{c}^{3}
\rho_{0,J}}$. The mass of the cD galaxy has been fixed at $1.14
\times 10^{12}$ $M_{\odot}$, with $r_{c} = R_{e}/0.76$, with
$R_{e} = 25$ kpc being the effective radius of the galaxy. The
central galaxy for each cluster in the sample is assumed to have
approximately this stellar mass.

We have assumed that the total galaxy-component mass (galaxies
plus cD galaxy masses) is $\approx 20\div25\%$ of the gas mass: in
\cite{Schindler02}, the mean fraction of gas versus the total mass
(with dark matter) for a cluster is estimated to be $15\div 20\%$,
while the same quantity for galaxies is $3\div 5\%$. This means
that the relative mean mass ratio gal-to-gas in a cluster is
$\approx 20\div 25\%$. We have varied the parameters
$\rho_{gal,1}$, $\rho_{gal,2}$ and $M_{J}$ in their previous
defined ranges to obtain a mass ratio between total galaxy mass
and total gas mass which lies in this range. Resulting galaxy mass
values and ratios ${\displaystyle
\frac{\mathrm{gal}}{\mathrm{gas}}}$, estimated at $r=r_{max}$, are
listed in Table~\ref{tab1}.

In Fig.~(1), we show how each component is spatially distributed.
The CD-galaxy is dominant with respect to the other galaxies only
in the inner region (below $100$ kpc). As already stated in
\S~\ref{sec:gas_model}, cluster innermost regions have been
excluded from our analysis and so the contribution due to the
cD-galaxy is practically negligible in our analysis. The gas is,
as a consequence, clearly the dominant visible component, starting
from innermost regions out to large radii, being galaxy mass only
$20\div 25\%$ of gas mass. A similar behavior is shown by all the
clusters considered in our sample.

\begin{table}
\begin{minipage}{\textwidth}
 \caption{\footnotesize{Column 1: Cluster name. Column 2: first derivative coefficient, $a_{1}$, of f(R) series.
          Column3: $1\sigma$ confidence interval for $a_{1}$. Column 4:
          second derivative coefficient, $a_{2}$, of f(R) series. Column 5: $1\sigma$ confidence interval for $a_{2}$.
          Column 6: characteristic length, L, of the modified gravitational potential, derived from $a_{1}$ and $a_{2}$.
          Column 7 : $1\sigma$ confidence interval for $L$.}}
 \label{tab2}
 \resizebox*{0.75\textwidth}{!}{
 \begin{tabular}{ccccccc}
  \hline
  name & $a_{1}$ & [$a_{1}-1\sigma$, $a_{1}+1\sigma$] & $a_{2}$ & [$a_{2}-1\sigma$, $a_{2}+1\sigma$] & $L$ & [$L-1\sigma$, $L+1\sigma$]\\
  &  &  & $({\mathrm{kpc}}^{2})$ & $({\mathrm{kpc}}^{2})$ &  (kpc) & (kpc)\\
  \hline
  \hline
  A133    &     $0.085$   &       [$0.078$, $0.091$]    &           $-4.98 \cdot 10^{3}$   &    [$-2.38 \cdot 10^{4}$, $-1.38  \cdot 10^{3}$]   &      $591.78$   &   [$323.34$, $1259.50$]  \\
  A262    &     $0.065$   &       [$0.061$, $0.071$]    &                       $-10.63$   &                              [$-57.65$, $-3.17$]   &       $31.40$   &      [$17.28$, $71.10$]  \\
  A383    &     $0.099$   &       [$0.093$, $0.108$]    &           $-9.01 \cdot 10^{2}$   &     [$-4.10 \cdot 10^{3}$, $-3.14 \cdot 10^{2}$]   &      $234.13$   &    [$142.10$, $478.06$]  \\
  A478    &     $0.117$   &       [$0.114$, $0.122$]    &           $-4.61 \cdot 10^{3}$   &     [$-1.01 \cdot 10^{4}$, $-2.51 \cdot 10^{3}$]   &      $484.83$   &    [$363.29$, $707.73$]  \\
  A907    &     $0.129$   &       [$0.125$, $0.136$]    &           $-5.77 \cdot 10^{3}$   &     [$-1.54 \cdot 10^{4}$, $-2.83 \cdot 10^{3}$]   &      $517.30$   &    [$368.84$, $825.00$]  \\
  A1413   &     $0.115$   &       [$0.110$, $0.119$]    &           $-9.45 \cdot 10^{4}$   &     [$-4.26 \cdot 10^{5}$, $-3.46 \cdot 10^{4}$]   &     $2224.57$   &  [$1365.40$, $4681.21$]  \\
  A1795   &     $0.093$   &       [$0.084$, $0.103$]    &           $-1.54 \cdot 10^{3}$   &     [$-1.01 \cdot 10^{4}$, $-2.49 \cdot 10^{2}$]   &      $315.44$   &    [$133.31$, $769.17$]  \\
  A1991   &     $0.074$   &       [$0.072$, $0.081$]    &                       $-50.69$   &                    [$-3.42 \cdot 10^{2}$, $-13$]   &       $64.00$   &     [$32.63$, $159.40$]  \\
  A2029   &     $0.129$   &       [$0.123$, $0.134$]    &           $-2.10 \cdot 10^{4}$   &     [$-7.95 \cdot 10^{4}$, $-8.44 \cdot 10^{3}$]   &      $988.85$   &   [$637.71$, $1890.07$]  \\
  A2390   &     $0.149$   &       [$0.146$, $0.152$]    &           $-1.40 \cdot 10^{6}$   &     [$-5.71 \cdot 10^{6}$, $-4.46 \cdot 10^{5}$]   &     $7490.80$   & [$4245.74$, $15715.60$]  \\
  MKW4    &     $0.054$   &       [$0.049$, $0.060$]    &                       $-23.63$   &                  [$-1.15 \cdot 10^{2}$, $-8.13$]   &       $51.31$   &     [$30.44$, $110.68$]  \\
  RXJ1159 &     $0.048$   &       [$0.047$, $0.052$]    &                       $-18.33$   &                  [$-1.35 \cdot 10^{2}$, $-4.18$]   &       $47.72$   &     [$22.86$, $125.96$]  \\
  \hline
 \end{tabular}}
 \end{minipage}
\end{table}

\subsection{Uncertainties on mass profiles}
\label{sec:uncertainties}

Uncertainties on the cluster total mass profiles have been
estimated performing Monte-Carlo simulations~\cite{NeuBoh95}. We
proceed to simulate temperature profiles and choose random
radius-temperature values couples for each bin which we have in
our temperature data given by \cite{Vik05}. Random temperature
values have been extracted from a Gaussian distribution centered
on the spectral values, and with a dispersion fixed to its $68\%$
confidence level. For the radius, we choose a random value inside
each bin. We have performed 2000 simulations for each cluster and
perform two cuts on the simulated profile. First, we exclude those
profiles that give an unphysical negative estimate of the mass:
this is possible when our simulated couples of quantities give
rise to too high temperature-gradient. After this cut, we have
$\approx1500$ simulations for any cluster. Then we have ordered
the resulting mass values for increasing radius values. Extreme
mass estimates (outside the $10\div90\%$ range) are excluded from
the obtained distribution, in order to avoid other high mass
gradients which give rise to masses too different from real data.
The resulting limits provide the errors on the total mass.
Uncertainties on the electron-density profiles has not been
included in the simulations, being them negligible with respect to
those of the gas-temperature profiles.

\subsection{Fitting the mass profiles}
\label{sec:mass_profiles_fit}

In the above sections, we have shown that, with the aid of X-ray
observations, modelling theoretically the galaxy distribution and
using Eq.~(\ref{eq:fit relation}), we obtain an
estimate of the baryonic content of clusters.\\
We have hence performed a best-fit analysis of the theoretical
 Eq.~(\ref{eq:fit relation})
{\setlength\arraycolsep{0.2pt}
\begin{eqnarray}\label{eq:theo_bar}
M_{bar,th}(r) &=& \frac{4 a_{1}}{3} \left[ - \frac{k T(r)}{\mu
m_{p}G} r \left(\frac{d\ln\rho_{gas}(r)}{d\ln r}+\frac{d\ln
T(r)}{d\ln r}\right) \right] + \nonumber \\ &-& \frac{4 a_{1}}{3G}
r^{2} \frac{\mathrm{d}\Phi_{C}}{\mathrm{d}r}(r)
\end{eqnarray}}
versus the observed mass contributions
\begin{equation}\label{eq:obs_bar}
M_{bar,obs}(r) = M_{gas}(r) + M_{gal}(r) + M_{CDgal}(r)
\end{equation}
Since not all the data involved in the above estimate have
measurable errors, we cannot perform an \textit{exact}
$\chi$-square minimization: Actually, we can minimize the
quantity:
\begin{equation}
\chi^{2} = \frac{1}{N-n_{p}-1} \cdot \sum_{i=1}^{N}
\frac{(M_{bar,obs}-M_{bar,theo})^{2}}{M_{bar,theo}}
\end{equation}
where $N$ is the number of data and $n_{p} = 2$ the  free
 parameters of the model. We minimize the $\chi$-square using the Markov
Chain Monte Carlo Method (MCMC). For each cluster, we have run
various chains to set the best parameters of the used algorithm,
the Metropolis-Hastings one: starting from an initial parameter
vector $\mathbf{p}$ (in our case ${\mathbf{p}} = (a_{1},a_{2})$),
we generate a new trial point $\mathbf{p'}$ from a tested proposal
density $q(\mathbf{p'},\mathbf{p})$, which represents the
conditional probability to get $\mathbf{p'}$, given $\mathbf{p}$.
This new point is accepted with probability
\[
\alpha({\mathbf{p}}, {\mathbf{p'}}) = min \left\{1,
\frac{L({\mathbf{d}}|{\mathbf{p'}}) P({\mathbf{p'}})
q({\mathbf{p'}},{\mathbf{p}})}{L({\mathbf{d}}|{\mathbf{p}})
P({\mathbf{p}}) q({\mathbf{p}},{\mathbf{p'}})}\ \right\}
\]
where ${\mathbf{d}}$ are the data, $L({\mathbf{d}}|{\mathbf{p'}})
\propto \exp(-\chi^{2}/2)$ is the likelihood function,
$P({\mathbf{p}})$ is the prior on the parameters. In our case, the
prior on the fit parameters is related to Eq.~(\ref{lengths
model}): being $L$ a length, we need to force the ratio $a_1/a_2$
to  be positive. The proposal density is  Gaussian symmetric with
respect of the two vectors $\mathbf{p}$ and $\mathbf{p'}$, namely
$q({\mathbf{p}},{\mathbf{p'}}) \propto \exp(-\Delta p^{2} / 2
\sigma^{2})$, with $\Delta {\mathbf{p}} = {\mathbf{p}} -
{\mathbf{p'}}$; we decide to fix the dispersion $\sigma$ of any
trial distribution of parameters equal to $20\%$ of trial $a_{1}$
and $a_{2}$ at any step. This means that the parameter $\alpha$
reduces to the ratio between the likelihood functions.

\begin{figure}[ht]
\centering \resizebox{13.5cm}{!}
  {\includegraphics[width=85mm]{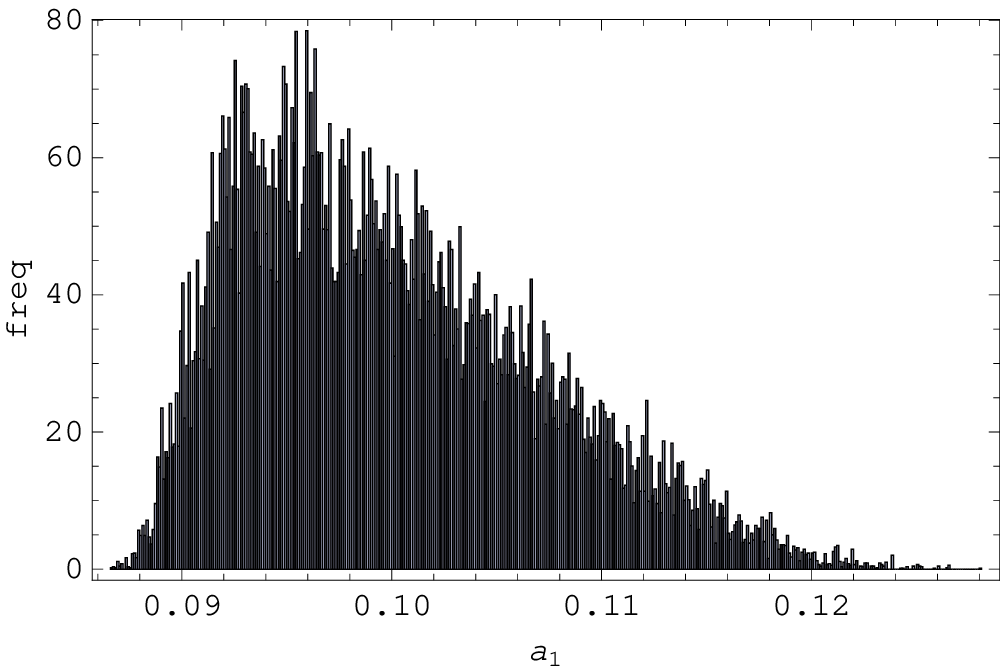}
  \includegraphics[width=85mm]{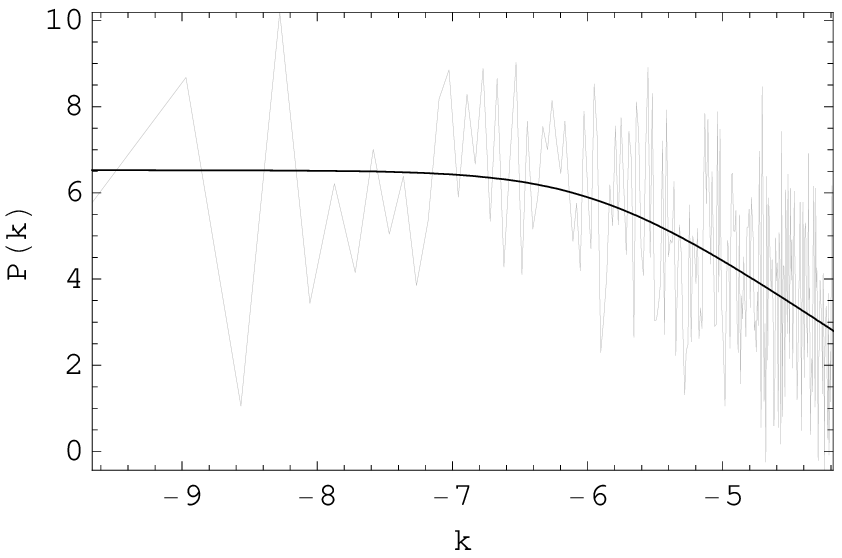}}
  \caption{\footnotesize{Left panel: histogram of the sample points for parameter $a_{1}$
  in Abell 383 coming out the MCMC implementation
  used to estimate best fit values and errors for our fitting procedure as described in
  \S~\ref{sec:mass_profiles_fit}. Binning (horizontal axis) and relative
  frequencies (vertical axis) are given by automatic procedure
  from Mathematica6.0. Right panel: power spectrum test on sample chain for
  parameter $a_{1}$ using the method described in \S~\ref{sec:mass_profiles_fit}. Black line
  is the logarithm of the analytical template Eq.~(\ref{eq:dunk_template}) for power spectrum; gray line is the discrete power spectrum
  obtained using Eq.~(\ref{eq:disc_power_def})
  -~(\ref{eq:disc_power_coeff}).}}
\end{figure}

We have run one chain of $10^{5}$ points for every cluster; the
convergence of the chains has been tested using the power spectrum
analysis from \cite{D05}. The key idea of this method is, at the
same time, simple and powerful: if we take the \textit{power
spectra} of the MCMC samples, we will have a great correlation on
small scales but, when the chain reaches convergence, the spectrum
becomes flat (like a white noise spectrum); so that, by checking
the spectrum of just one chain (instead of many parallel chains as
in Gelmann-Rubin test) will be sufficient to assess the reached
convergence. Remanding to \cite{D05} for a detailed discussion of
all the mathematical steps. Here  we calculate the discrete power
spectrum of the chains:
\begin{equation}\label{eq:disc_power_def}
P_{j} = |a_{N}^{j}|^{2}
\end{equation}
with
\begin{equation}\label{eq:disc_power_coeff}
a_{N}^{j} = \frac{1}{\sqrt{N}}\sum_{n=0}^{N-1} x_{n}
\exp{\left[i\frac{2\pi j}{N}n\right]}
\end{equation}
where $N$ and $x_{n}$ are the length and the element of the sample
from the MCMC, respectively, $j=1,\ldots,\frac{N}{2}-1$. The
wavenumber $k_{j}$ of the spectrum is related to the index $j$ by
the relation $k_{j}=\frac{2\pi j}{N}$. Then we fit it with the
analytical template:
\begin{equation}\label{eq:dunk_template}
P(k) = P_{0} \frac{(k^{*}/k)^{\alpha}}{1+(k^{*}/k)^{\alpha}}
\end{equation}
or in the equivalent logarithmic form:
\begin{equation}
\ln P_{j} = \ln P_{0} + \ln
\left[\frac{(k^{*}/k_{j})^{\alpha}}{1+(k^{*}/k_{j})^{\alpha}}\right]-\gamma
+ r_{j}
\end{equation}
where $\gamma=0.57216$ is the Euler-Mascheroni number and $r_{j}$
are random measurement errors with $<r_{j}> = 0$ and $<r_{i}r_{j}>
= \delta_{ij} \pi^{2}/6$. From the fit, we estimate the two
fundamental parameters, $P_{0}$ and $j^{*}$ (the index
corresponding to $k^{*}$). The first one is the value of the power
spectrum extrapolated for $k \rightarrow 0$ and,  from it, we can
derive the convergence ratio from ${\displaystyle r \approx
\frac{P_{0}}{N}}$; if $r < 0.01$, we can assume that the
convergence is reached. The second parameter is related to the
turning point from a power-law to a flat spectrum. It has to be
$>20$ in order to be sure that the number of points in the sample,
coming from the convergence region, are more than the noise
points. If these two conditions are verified for all the
parameters, then the chain has reached the convergence and the
statistics derived from  MCMC well describes  the underlying
probability distribution (typical results are shown in Figs.
(2)-(3)). Following \cite{D05} prescriptions, we perform the fit
over the range $1 \leq j \leq j_{max}$, with $j_{max} \sim 10
j^{*}$, where a first estimation of $j^{*}$ can be obtained from a
fit with $j_{max} = 1000$, and then performing a second iteration
in order to have a better estimation of it. Even if the
convergence is achieved after few thousand steps of the chain, we
have decided to run longer chains of $10^{5}$ points to reduce the
noise from the histograms and avoid under- or over-  estimations
of errors on the parameters. The $i-\sigma$ confidence levels are
easily estimated deriving them from the final sample the
$15.87$-th and $84.13$-th quantiles (which define the $68\%$
confidence interval) for $i=1$, the $2.28$-th and $97.72$-th
quantiles (which define the $95\%$ confidence interval) for $i=2$
and the $0.13$-th and $99.87$-th quantiles (which define the
$99\%$ confidence interval) for $i=3$.

After the description of the method, let us now comment on the
achieved results.

\section{Results}
\label{sec:results}

The numerical results of our fitting analysis are summarized in
Table 2; we give the best fit values of the independent fitting
parameters $a_{1}$ and $a_{2}$, and of the gravitational length
$L$, considered as a function of the previous two quantities. In
Figs.~(3)-~(5), we give the typical results of fitting, with
histograms and power spectrum of samples derived by the MCMC, to
assess the reached convergence (flat spectrum at  large scales).

The goodness and the properties of the fits are shown in
Figs.~(6)-~(17). The main property of our results is the presence
of a \textit{typical scale} for each cluster above which our model
works really good (typical relative differences are less than
$5\%$), while for lower scale there is a great difference. It is
possible to see, by a rapid inspection, that this turning-point is
located at a radius $\approx 150$ kpc. Except for very large
clusters, it is  clear that this value is independent of the
cluster, being approximately the same for any member of the
considered sample.

There are two main independent explanations that could justify
this trend: limits due to  a break in the state of hydrostatic
equilibrium or limits in the series expansion of the
$f(R)$-models.

If the hypothesis of hydrostatic equilibrium is not correct, then
we are in a regime where the fundamental relations
Eqs.~(\ref{Boltzmann equation})-~(\ref{eq:Boltzmann potential}),
are not working. As discussed in \cite{Vik05},  the central (70
kpc) region of every cluster is strongly affected by radiative
cooling and thus it cannot directly be related to the depth of the
cluster potential well. This means that, in this region, the gas
is not in hydrostatic equilibrium but in a multi-phase, turbulent
state, mainly driven by some astrophysical, non-gravitational
interaction.  In this case, the gas cannot be used as a good
standard tracer.

We have also to consider another limit of our modelling: the
requirement that the $f(R)$-function is Taylor expandable. The
corrected gravitational potential which we have considered is
derived in the weak field limit, which means
\begin{equation}
R - R_0 << \frac{a_{1}}{a_{2}}
\end{equation}
where $R_0$ is the background value of the curvature. If this
condition is not satisfied, the approach does not work (see
\cite{noi-prd} for a detailed discussion of this point).
Considering that $a_{1}/a_{2}$ has the dimension of $length^{-2}$
this condition  defines the \textit{length scale} where our series
approximation can work. In other words, this indicates the limit
in which the model can be compared with  data.

For the considered sample, the fit of the parameters $a_{1}$ and
$a_{2}$, spans the length range $\{19; 200\}$ kpc (except for the
biggest cluster). It is evident that every galaxy cluster has a
\textit{proper} gravitational length scale. It is worth noticing
that a similar situation, but at completely different scales, has
been found out for low surface brightness galaxies modelled by
$f(R)$-gravity \cite{CapCardTro07}.

Considering the  data at our disposal and the analysis which we
have performed, it is not possible to quantify exactly the
quantitative amount of  these two different phenomena (i.e. the
radiative cooling and the validity of the weak field limit).
However, they are not mutually exclusive but should be considered
in details in view of a more refined modelling \footnote{Other
secondary phenomena as cooling flows, merger and asymmetric shapes
have to be considered in view of a detailed modelling of clusters.
However, in this work, we are only interested  to show that
extended gravity could be a valid alternative to dark matter in
order to explain the  cluster dynamics. }.

Similar issues are present also in \cite{Brown06B}: they use the
the Metric - Skew - Tensor - Gravity (MSTG) as a generalization of
the Einstein General Relativity and derive the gas mass profile of
a sample of clusters with gas being the only baryonic component of
the clusters. They consider some clusters  included in our sample
 (in particular, A133, A262, A478, A1413, A1795, A2029, MKW4) and they find the
same different trend for $r \leq 200$ kpc, even if with a
different behavior with respect of us: our model gives lower
values than X-ray gas mass data while their model gives higher
values with respect to X-ray gas mass data. This stresses the need
for a more accurate modelling of the gravitational potential.

However, our goal is to show that potential (\ref{gravpot1}) is
suitable to fit the mass profile of galaxy clusters and that it
comes from a self-consistent theory.

In general, it can be shown  that the weak field limit of extended
theories of gravity has Yukawa-like corrections
\cite{b10,kenmoku}. Specifically, given theory of gravity of order
$(2n+2)$, the Yukawa corrections to the Newtonian potential are
$n$ \cite{Qua91}. This means that if the effective Lagrangian of
the theory is
\begin{equation}
{\cal L}=f(R,\Box R,..\Box^k R,..\Box^n R)\sqrt{-g}
\end{equation}
we have
  \begin{equation} \label{yukawa} \phi(r)=-\frac{G
M}{r}\left[1+\sum_{k=1}^{n}\alpha_k e^{-r/L_k}\right]\,.
\end{equation}
Standard General Relativity, where Yukawa corrections are not
present, is recovered for $n=0$ (second order theory) while the
 $f(R)$-gravity is obtained for $n=1$ (fourth-order theory). Any $\Box$ operator introduces two
further derivation orders in the field equations. This kind of
Lagrangian comes out when  quantum field theory is formulated on
curved spacetime \cite{birrell}. In the series (\ref{yukawa}), $G$
is the value of the gravitational constant considered at infinity,
$L_k$ is the interaction length of the $k$-th component of the
non-Newtonian corrections. The amplitude $\alpha_k$ of each
component is normalized to the standard Newtonian term; the sign
of $\alpha_k$ tells us if the corrections are attractive or
repulsive (see \cite{will} for details). Moreover, the variation
of the gravitational coupling is involved. In our case, we are
taking into account only the first term of the series. It is the
the leading term. Let us rewrite (\ref{gravpot1}) as
\begin{equation}
\label{yukawa1} \phi(r)=-\frac{G M}{r}\left[1+\alpha_1
e^{-r/L_1}\right]\,.
\end{equation}
The effect of non-Newtonian term can be parameterized by
$\{\alpha_1,\,L_1\}$ which could be a useful parameterisation
which respect to our previous
 $\{a_1,\,a_2\}$ or $\{G_{eff},\,L\}$ with
$G_{eff}=3G/(4a_1)$. For large distances, where $r\gg L_1$, the
exponential term vanishes and the gravitational coupling is $G$.
If $r\ll L_1$, the exponential becomes 1 and, by differentiating
Eq.(\ref{yukawa1}) and comparing with the gravitational force
measured in laboratory, we get
\begin{equation} \label{yukawa2}
G_{lab}=G\left[1+\alpha_1\left(1+\frac{r}{L_1}\right)e^{-r/L_1}\right]
\simeq G(1+\alpha_1)\,,
\end{equation}
where $G_{lab}=6.67\times 10^{-8}$ g$^{-1}$cm$^3$s$^{-2}$ is the
usual Newton constant measured by Cavendish-like experiments. Of
course, $G$ and $G_{lab}$ coincide in the standard Newtonian
gravity. It is worth noticing that, asymptotically, the inverse
square law holds but the measured coupling constant differs by a
factor $(1+\alpha_1)$. In general, any  correction introduces a
characteristic length that acts at a certain scale for the
self-gravitating systems as in the case of galaxy cluster which we
are examining here. The range of $L_k$ of the $k$th-component of
non-Newtonian force can be identified with the mass $m_k$ of a
pseudo-particle whose effective Compton's length can be defined as
\begin{equation} \label{27}
L_k=\frac{\hbar}{m_k c}\,.
\end{equation}
The interpretation of this fact is that, in the weak energy limit,
fundamental theories which attempt to unify gravity with the other
forces introduce, in addition to the massless graviton, particles
{\it with mass} which also  carry the gravitational interaction
\cite{gibbons}. See, in particular, \cite{PPNfelix} for
$f(R)$-gravity. These masses are related to effective length
scales which can be parameterized as
\begin{equation}\label{28}
L_k=2\times 10^{-5}\left(\frac{1\,
\mbox{eV}}{m_k}\right)\mbox{cm}\,.
\end{equation}
There have been several attempts to experimentally constrain $L_k$
and $\alpha_k$ (and then $m_k$) by experiments on scales in the
range $1 \,\mbox{cm}<r< 1000\,\mbox{km}$, using different
techniques \cite{fischbach,speake,eckhardt}. In this case, the
expected masses of particles which should carry the additional
gravitational force are in the range $10^{-13} \mbox{eV}<m_k<
10^{-5}\, \mbox{eV}$. The general outcome of these experiments,
even retaining only the term $k=1$, is that {\it geophysical
window} between the laboratory and the astronomical scales has to
be taken into account. In fact, the range
\begin{equation}
|\alpha_1|\sim 10^{-2}\,,\qquad L_1\sim 10^2\div
10^3\,\,\mbox{m}\,,
\end{equation}
is not excluded at all in this window. An interesting suggestion
has been given by Fujii \cite{fujii1}, which proposed that the
exponential deviation from the Newtonian standard potential  could
arise from the microscopic interaction which couples  the nuclear
isospin and the baryon number.

The astrophysical counterparts of these non-Newtonian corrections
seemed ruled out till some years ago due to the fact that
experimental tests of General Relativity seemed to predict  the
Newtonian potential in the weak energy limit, ''inside" the Solar
System. However, as it has been shown, several alternative
theories seem to evade the Solar System constraints (see
\cite{PPNfelix,lecian} and the references therein for recent
results) and, furthermore, indications of an anomalous,
long--range acceleration revealed from the data analysis of
Pioneer 10/11, Galileo, and Ulysses spacecrafts (which are now
almost outside the Solar System) makes these Yukawa--like
corrections come again into play \cite{anderson}. Besides, it is
possible to reproduce phenomenologically the flat rotation curves
of spiral galaxies considering the values
\begin{equation}\label{sand}
\alpha_1=-0.92\,,\qquad L_1\sim 40\,\,\mbox{kpc}\,.
\end{equation}
The main hypothesis of this approach is that the additional
gravitational interaction is carried by some ultra-soft boson
whose range of mass is $m_1\sim 10^{-27}\div 10^{-28}$eV. The
action of this boson becomes efficient at galactic scales without
the request of enormous amounts of dark matter to stabilize the
systems \cite{sanders}.

Furthermore, it is possible to  use a combination of two
exponential correction terms and give a detailed explanation of
the kinematics of galaxies and galaxy clusters, again without dark
matter model \cite{eckhardt}.

It is worthwhile to note that both the spacecrafts measurements
and galactic rotation curves indications come from ''outside" the
usual Solar System boundaries used up to now to test General
Relativity. However, the above results {\it do not come} from any
fundamental theory  to explain the outcome of Yukawa corrections.
In their contexts, these terms are phenomenological.

Another important remark in this direction deserves the fact that
some authors \cite{mcgaugh} interpret also the  experiments on
cosmic microwave background like the experiment BOOMERANG and WMAP
\cite{Boom,WMAP} in the framework of {\it modified Newtonian
dynamics} again without invoking any dark matter model.

All these facts point towards the line of thinking that also
corrections to the standard gravity have to be seriously taken
into account beside dark matter searches.

In our case, the parameters $a_{1,2}$, which determine the
gravitational correction and the gravitational coupling, come out
"directly" from a field theory with the only requirement that the
effective action of gravity could be  more general than the
Hilbert-Einstein theory $f(R)=R$. This main hypothesis comes from
fundamental physics motivations due to the fact that any
unification scheme or quantum field theory on curved space have to
take into account higher order terms in curvature invariants
\cite{birrell}. Besides, several recent results point out that
such corrections have a main role also at astrophysical and
cosmological scales. For a detailed discussion, see
\cite{odi,capfra,faraoni}.

\begin{figure}
\centering
\resizebox{9.5cm}{!}{\includegraphics[width=75mm]{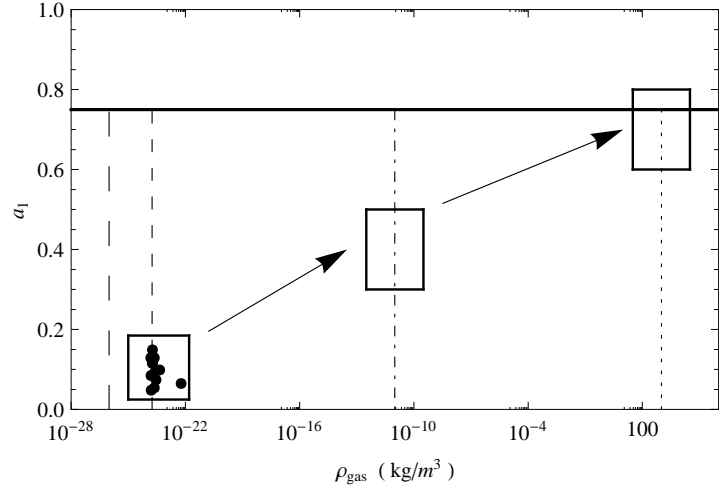}}
  \caption{\footnotesize{Density vs $a_{1}$: predictions on the behavior of
  $a_{1}$. The horizontal black bold line indicates the
  Newtonian-limit, $a_{1} \rightarrow 3/4$ which we expect to be realized on
  scales comparable with Solar System. Vertical lines indicate typical
  approximated values of matter density (without dark matter) for different
  gravitational structures: Universe (large dashed) with critical
  density $\rho_{crit} \approx 10^{-26}$ $\mathrm{kg/m^{3}}$; galaxy clusters (short
  dashed) with $\rho_{cl} \approx 10^{-23}$ $\mathrm{kg/m^{3}}$; galaxies (dot-dashed)
  with $\rho_{gal} \approx 10^{-11}$ $\mathrm{kg/m^{3}}$; sun (dotted) with
  $\rho_{sun} \approx 10^{3}$ $\mathrm{kg/m^{3}}$. Arrows and boxes show the predicted
  trend for $a_{1}$.}}
\label{fig:plots_finali1}
\end{figure}

With this philosophy in mind,  we have plotted the trend of
$a_{1}$ as a function of the density in
Fig.\ref{fig:plots_finali1}. As one can see, its values are
strongly constrained in a narrow region of the parameter space, so
that $a_{1}$ can be considered a "tracer" for the size of
gravitational structures. The value of $a_{1}$  range between
$\{0.8 \div 0.12 \}$ for larger clusters and $\{0.4\div 0.6 \}$
for  poorer structures (i.e. galaxy groups like MKW4 and RXJ1159).
We expect a particular trend when applying the model to different
gravitational structures. In Fig.~\ref{fig:plots_finali1},  we
give characteristic values of density which range from the biggest
 structure, the observed Universe (large dashed vertical line), to the
smallest one, the Sun (vertical dotted line), through intermediate
steps like clusters (vertical short dashed line) and galaxies
(vertical dot-dashed line). The bold black horizontal line
represents the Newtonian limit  $a_{1}=3/4$ and the  boxes
indicate the possible values of $a_{1}$ that we obtain by applying
our theoretical model to different structures.

\begin{figure}
\centering\resizebox{9.5cm}{!}{\includegraphics[width=75mm,height=60mm]{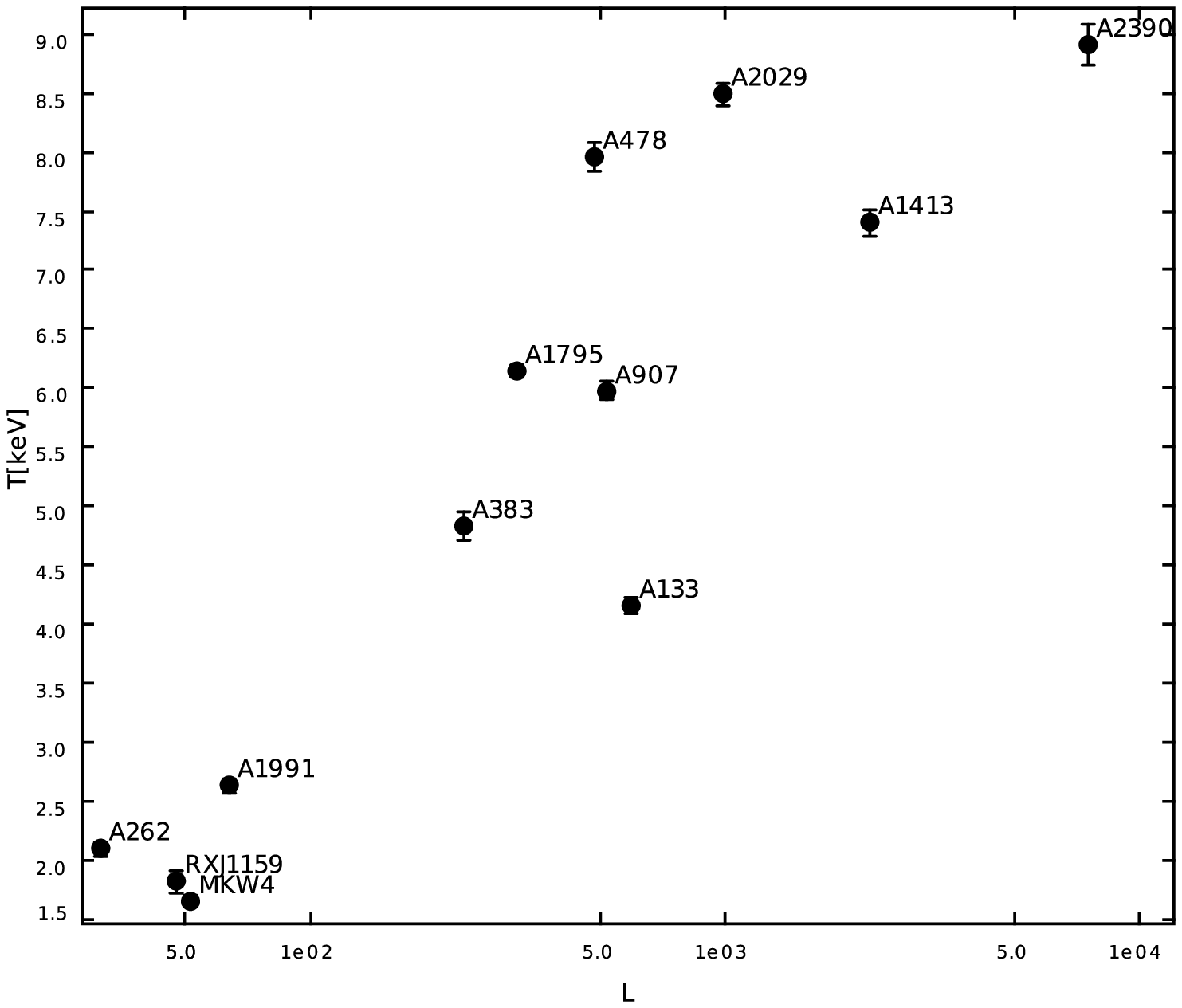}}
\centering\resizebox{9.5cm}{!}{\includegraphics[width=75mm]{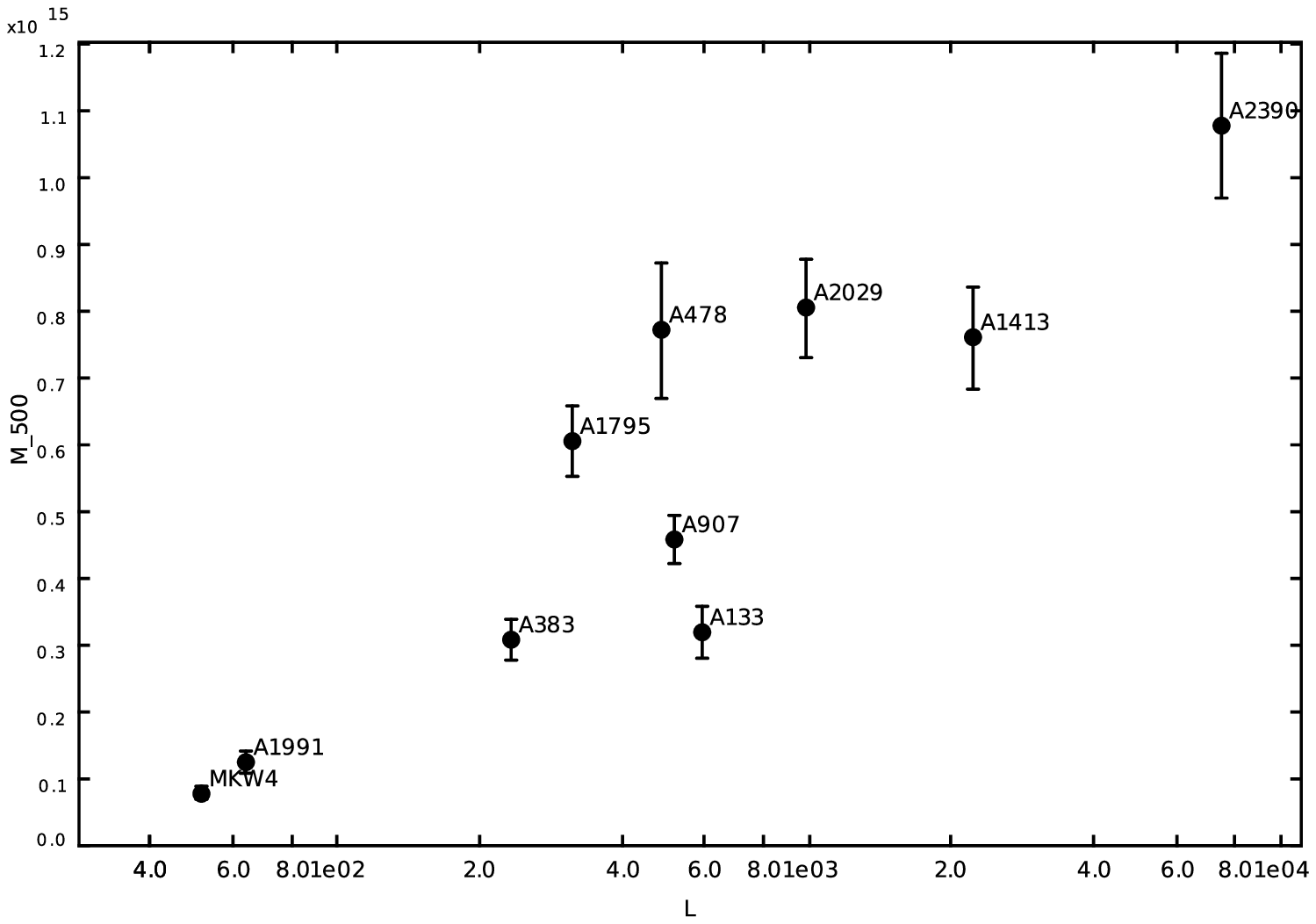}}
  \caption{\footnotesize{Single temperature fit to the total cluster spectrum (upper panel) and total cluster mass within r$_{500}$
  (given as a function of M$_{\odot}$) (lower panel) are plotted as a function of the characteristic
  gravitational length $L$. Temperature and mass values are from \cite{Vik06}.}}
\label{fig:PropPhysVSL}
\end{figure}

Similar considerations hold also for the characteristic
gravitational length $L$ directly related to both $a_{1}$ and
$a_{2}$. The parameter $a_{2}$ shows a very large range of
variation $\{-10^{6} \div -10 \}$ with respect to the density (and
the mass)  of the clusters. The value of $L$ changes with the
sizes of gravitational structure (see Fig.~\ref{fig:PropPhysVSL}),
so it can be considered, beside the Schwarzschild radius, a sort
of additional gravitational radius. Particular care must be taken
when considering Abell~2390, which shows large cavities in the
X-ray surface brightness distribution, and whose central region,
highly asymmetric, is not expected to be in hydrostatic
equilibrium. All results at small and medium radii for this
cluster could hence be strongly biased by these
effects~\cite{Vik06}; the same will hold for the resulting
exceptionally high value of $L$.
 Fig.~\ref{fig:PropPhysVSL} shows how observational properties of the cluster, which well characterize its
gravitational potential (such as the average temperature and the
total cluster mass within r$_{500}$, plotted in the left and right
panel, respectively), well correlate with the characteristic
gravitational length $L$.

For clusters, we can define a gas-density-weighted  and a
gas-mass-weighted mean, both depending on the series parameters
$a_{1,2}$. We have:
\begin{eqnarray}
<L>_{\rho} &=& 318 \; \mathrm{kpc} \qquad <a_{2}>_{\rho} = -3.40
\cdot 10^{4} \nonumber \\
<L>_{M} &=& 2738 \; \mathrm{kpc} \qquad <a_{2}>_{M} = -4.15 \cdot
10^{5}
\end{eqnarray}
It is straightforward to note the correlation with the sizes of
the cluster cD-dominated-central region and the "gravitational"
interaction length of the whole cluster. In other words, the
parameters $a_{1,2}$, directly related to the first and second
derivative of a given analytic $f(R)$-model determine the
characteristic sizes of the self gravitating structures.

\section{What we have  learnt from clusters}
\label{sec:discussion}

We have investigated the possibility that the high observational
mass-to-light ratio of galaxy clusters  could be addressed by
$f(R)$- gravity  without assuming huge amounts of dark matter. We
point out  that this proposal comes out from the fact that, up to
now, no definitive candidate for dark-matter has been observed at
fundamental level and then  alternative solutions to the problem
should be viable. Furthermore, several results in $f(R)$-gravity
seem to confirm that valid alternatives to $\Lambda$CDM can be
achieved in cosmology. Besides, as discussed in the  Introduction,
the rotation curves of spiral galaxies can be explained in the
weak field limit of $f(R)$-gravity.  Results of our analysis go in
this direction.

We have chosen a sample of  relaxed galaxy clusters  for which
accurate spectroscopic temperature measurements and gas mass
profiles are available. For the sake of simplicity, and considered
the sample at our disposal, every cluster has been modelled as a
self-bound gravitational system with spherical symmetry and in
hydrostatic equilibrium.  The mass distribution has been described
by a corrected  gravitational  potential obtained from a generic
analytic $f(R)$-theory. In fact, as soon as $f(R)\neq R$,
Yukawa-like exponential corrections emerge in the weak field limit
while the standard Newtonian potential is recovered only for
$f(R)=R$, the Hilbert-Einstein theory.

Our goal has been to analyze if the dark-matter content of
clusters can be addressed by these  correction potential terms. As
discussed in detail in the previous sections and how it is
possible to see by a rapid inspection of figures,  the clusters of
the sample are consistent with the proposed model at $1\sigma$
confidence level.  This shows, at least \textit{qualitatively},
that the high mass-to-light ratio of clusters can be explained by
using a modified gravitational potential. The good agreement is
achieved on distance scales starting from $150$ kpc up to $1000$
kpc. The differences observed at smaller scales can be ascribed to
non-gravitational phenomena, such as cooling flows, or to the fact
that the gas mass is not a good tracer at this scales. The
remarkable result is that we have obtained a consistent agreement
with data only using the corrected gravitational potential in a
large range of radii. In order to put in evidence this trend, we
have plotted the baryonic mass vs radii considering, for each
cluster,  the scale where the trend is clearly evident.

In our knowledge, the fact that $f(R)$-gravity could work at these
scales  has been only supposed but never achieved by a direct
fitting with data (see \cite{fogdm,lobo} for a review). Starting
from the series coefficients $a_{1}$ and $a_{2}$, it is possible
to state that, at cluster scales, two characteristic sizes emerge
from the weak field limit of the theory.  However, at smaller
scales, e.g. Solar System scales, standard Newtonian gravity has
to be dominant in agreement with observations and experiments.

In summary, if our considerations are right, gravitational
interaction depends on the scale and the {\it infrared limit} is
led by the series coefficient of the considered effective
gravitational Lagrangian. Roughly speaking, we expect that
starting from cluster scale to galaxy scale, and then down to
smaller  scales as Solar System or Earth, the terms of the series
lead the clustering of self-gravitating systems beside other
non-gravitational phenomena. In our case, the Newtonian limit is
recovered for $a_{1} \rightarrow 3/4$ and $L(a_{1},a_{2})\gg r$ at
small scales and for $L(a_{1},a_{2})\ll r$ at large scales. In the
first case, the gravitational coupling has to be redefined, in the
second $G_{\infty}\simeq G$. In these limits, the linear Ricci
term is dominant in the gravitational Lagrangian and the Newtonian
gravity is restored \cite{Qua91}. Reversing the argument, this
could be the starting point to achieve a theory  capable of
explaining the strong segregation in masses and sizes of
gravitationally-bound systems.

\begin{figure}[ht]
\centering\resizebox{9.5cm}{!}{\includegraphics[width=75mm]{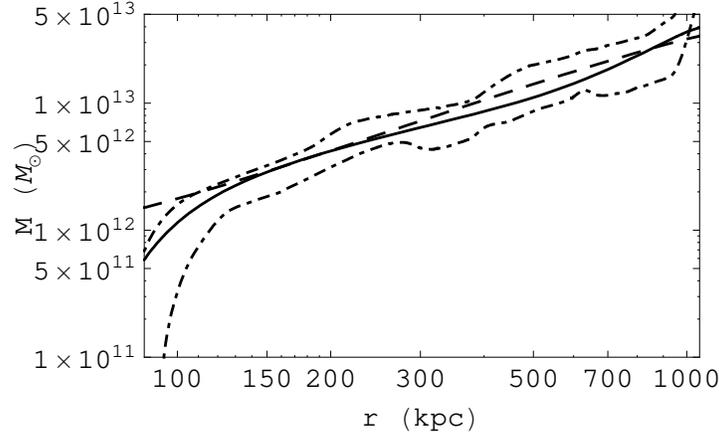}}
  \caption{\footnotesize{As an example of the above results, we have plotted the baryonic mass vs radii for Abell A133. Dashed line is the experimental-observed estimation Eq.~(\ref{eq:obs_bar}) of
  baryonic matter component (i.e. gas, galaxies and cD-galaxy); solid line is the theoretical estimation Eq.~(\ref{eq:theo_bar})
  for baryonic matter component. Dotted lines are the 1-$\sigma$ confidence levels given by errors on fitting
  parameters plus statistical errors on mass profiles as discussed in \S~\ref{sec:uncertainties} in the right panel.}}
\end{figure}

\begin{figure}[ht]
\centering\resizebox{9.5cm}{!}{\includegraphics[width=75mm]{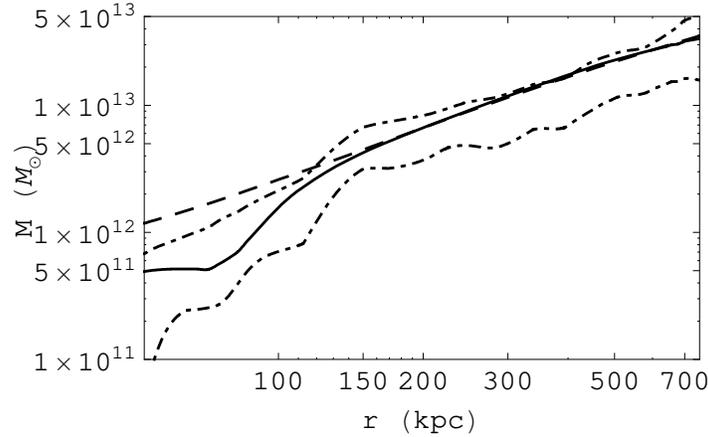}}
  \caption{\footnotesize{As the above case, for cluster Abell 383.}}
\end{figure}

\begin{figure}[ht]
\centering\resizebox{9.5cm}{!}{\includegraphics[width=75mm]{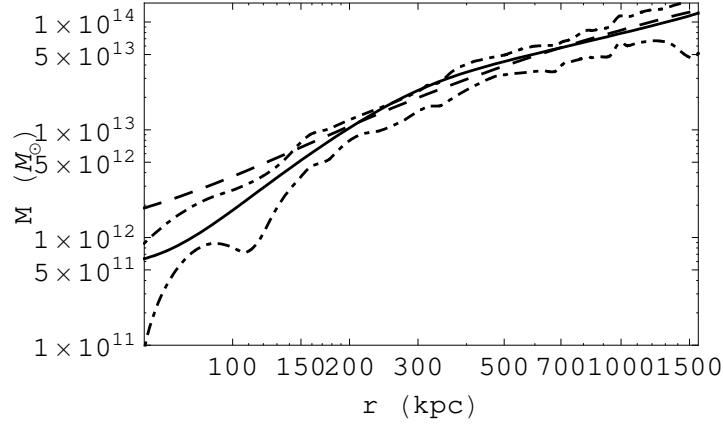}}
  \caption{\footnotesize{As the above cases, for cluster Abell 478.}}
\end{figure}

\begin{figure}[ht]
\centering\resizebox{9.5cm}{!}{\includegraphics[width=75mm]{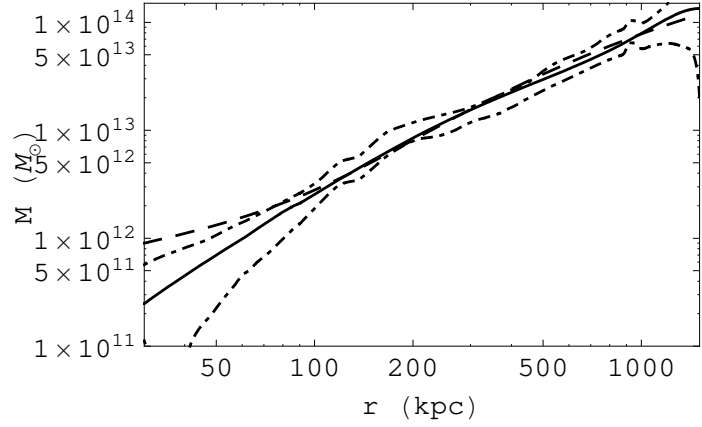}}
  \caption{\footnotesize{As the above cases, for cluster Abell 1413.}}
\end{figure}

\begin{figure}[ht]
\centering\resizebox{9.5cm}{!}{\includegraphics[width=75mm]{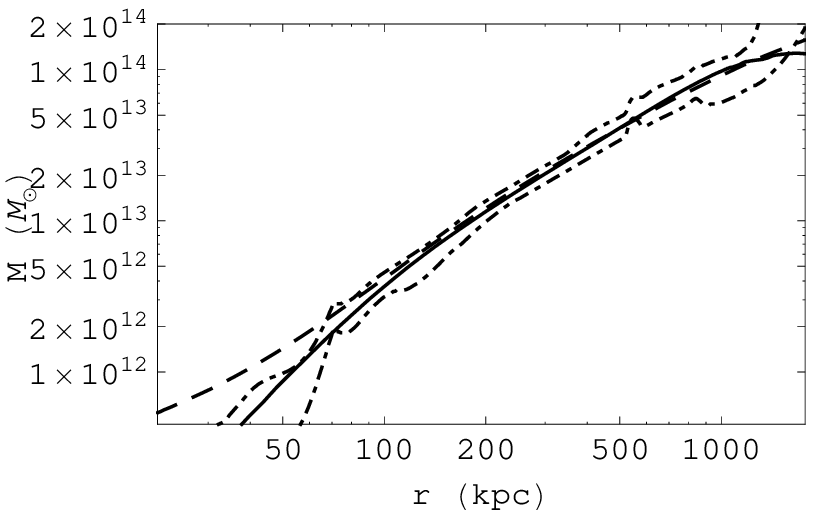}}
  \caption{\footnotesize{As the above cases, for cluster Abell 2029.}}
\end{figure}

\section{Conclusions}
The present status of art of  cosmology shows that the Standard
Cosmological Model, based on General Relativity, nucleosynthesis,
cosmic abundances and large scale structure, has some evident
difficulties. These ones, first of all, rely on some lack of a
self-consistent formulation of missing matter and cosmic
acceleration issues; such  shortcomings give rise to further
difficulties in interpreting observational data. With an aphorism,
one can say that \textit{we have a book, but not the alphabet to
read it}.

Nowadays there two main philosophical approaches aimed to solve
this problem. From one side,  there are researchers  which try to
solve shortcomings of  Standard Cosmological Model assuming that
General Relativity is  right but  we need some  exotic, invisible
kinds of energy and matter to explain cosmic dynamics and  large
scale structure. On the other side, there are people  which
believe that General Relativity is not the definitive and
comprehensive theory of gravity, and that it should be revised at
ultraviolet scales (quantum gravity) and infrared scales
(extragalactic and cosmic scales). In the latter case, dark energy
and dark matter could be nothing else but the signals that we need
a more general theory at large scales, also if General Relativity
works very well up to Solar System scales. To some extent, this
could be seen as a sort of philosophical  debate without solution,
but there are possibilities to move the question toward a
physical viewpoint.

The $f(R)$-gravity is strictly related to the second point of
view. It is a fruitful approach to generalize General Relativity
towards the solution also if, most of the models in literature are
nothing else but phenomenological models. It is interesting to
note that as soon as Einstein formulated his General Relativity,
many authors (and Einstein himself) started to explore
\textit{other possibilities} (see \cite{capfra} for a review). At
the beginning, these researches were mainly devoted to check the
mathematical consistency of General Relativity but the issues to
achieve the unification of gravity with the other interactions
(e.g. electromagnetism) pushed several authors to develop
alternative gravity theories. Today, one of the goals of
alternative gravity is to understand the effective content and
dynamics  of the Universe. This question is recently become
dramatic since assuming that more than 95\% of cosmic
matter-energy is {\it unknown at fundamental level} is highly
disturbing. Alternative gravity could be a way out to this
situation.  The present status of observations, also if we are
living in the era of {\it Precision Cosmology},   does not allow
in discriminating between alternative gravity, from one side, and
the presence of dark energy and dark matter, from the other side
(the forthcoming LHC experiments should aid in this sense if new
fundamental particles will be detected).

However, as discussed in this review, cosmography may be a useful
tool to discriminate among different cosmological models being, by
definition,  a model-independent approach: any cosmographic
parameter can be estimated without assigning an {\it a priori}
cosmological model. So cosmography can be used in two ways:
\begin{itemize}
   \item One can use it to discriminate between General
         Relativity and alternative theories. This issue
         strictly depends on the possibility to have
         good quality data at disposal. We need some minimum sensibility and error
         requirements on data surveys  to
         solve this question. At the moment, we have not them and we are not
         able to do this since  standard candles are not available
         at very high red shifts \cite{izzo}.
   \item We can use the cosmographic parameters to constraint
          cosmological models as we have done in this paper for $f(R)$-gravity.
          Being these parameters model-independent, they results natural "priors" to any theory.
          As above, the accuracy in estimating them is a crucial issue.
\end{itemize}
We have used, essentially, SNeIa but other classes of objects have
to be considered in order to improve such an accuracy (e.g. CMBR,
bright galaxies, GRBs, BAOs, weak lensing and so on). Forthcoming
space missions will be extremely useful in this sense.

Beside cosmography, we have discussed also if $f(R)$-gravity could
be useful to address the problem of mass profile and dynamics of
galaxy clusters. This issue is crucial in view of achieving any
correct model for large scale structure.

Taking into account the weak-field limit of a generic analytic
$f(R)$-function, it is possible to obtain  a
\textit{scale-dependent} gravity, where  scales of
self-gravitating systems could naturally emerge. In this way, one
could successfully explain  dark matter profiles   ranging from
galaxies to clusters of galaxies. The results are preliminary but
seems to indicate a way in which the dark matter puzzle could be
completely solved.

In conclusion, the main lesson of this work is that since it is
very difficult to discriminate among the huge amount of
cosmological models which try to explain the data (deductive
approach), it could be greatly fruitful to "reconstruct" the final
cosmological  model by an inductive approach, that is without
imposing it a priori but adopting the philosophy to use the
minimum number of parameters \footnote{Following the Occam razor
prescriptions: \textit{"Entia non sunt multiplicanda praeter
necessitatem."}}. This "inverse scattering approach" could be not
fully satisfactory but could lead to self-consistent results.


\addcontentsline{toc}{chapter}{Bibliography}

\begin{thebibliography}{999}

\bibitem{Alam03} Alam, U., Sahni, V., Saini, T.D., Starobinsky, A.A., 2003, MNRAS, 344, 1057
\bibitem{DETF} Albrecht et al., {\it Dark energy task force final report}, FERMILAB-FN-0793-A, astro-ph/0609591, 2006
\bibitem{SNAP} Aldering G. et al., astro\,-\,ph/0405232, 2004; see also {\tt snap.lbl.gov}
\bibitem{alle04} Allemandi, G., Borowiec, A., Francaviglia, M., 2004, Phys. Rev.  D 70, 103503
\bibitem{allrugg}Allemandi, G., Francaviglia, M., Ruggiero, M., Tartaglia, A. 2005, Gen. Rel. Grav., 37, 1891
\bibitem{anderson}  Anderson J.D., et al., 1998,  Phys. Rev. Lett. {\bf 81}, 2858
\bibitem{Appleby} Appleby, S.A., Battye, R.A., 2007, Phys. Lett. B, 654, 7
\bibitem{SNLS}{ast05} Astier, P. et al. 2006, A\&A, 447, 31

\bibitem{Bah96} Bahcall N.~A., 1996, in "Formation of structure in the Universe, 1995, Jerusalem Winter School", astro-ph/9611148,
\bibitem{bach03} Bahcall, N.A. et al. 2003, ApJ, 585, 182
\bibitem{bb03} Bahcall, N.A., Bode, P. 2003, ApJ, 588, 1
\bibitem{Barris04}Barris, B.J., et al., 2004, ApJ, 602, 571
\bibitem{Bau70} Bautz, L.~P., Morgan, W.~W., 1970, ApJ, 162, L149
\bibitem{wmap} Bennett C.L. et al., 2003, ApJS, 148, 1
\bibitem{birrell} Birrell, N.D. and Davies, P.C.W., 1982, Quantum Fields in Curved Space, Cambridge Univ. Press, Cambridge
\bibitem{fogdm} Boehmer, C.G., Harko, T., Lobo, F.S.N., 2008, Astropart. Phys. 29, 386
\bibitem{borowiec} Borowiec, A., Godlowski,  W., Szydlowski, M. 2006, Phys. Rev. D, 74, 043502
\bibitem{Brown06A} Brownstein, J.~R., Moffat, J.~W., 2006, ApJ, 636, 721
\bibitem{Brown06B} Brownstein, J.~R., Moffat, J.~W., 2006, MNRAS, 367, 527
\bibitem{Brown07} Brownstein, J.~R., Moffat, astro-ph/0702146

\bibitem{capozzcurv} Capozziello, S. 2002, Int. J. Mod. Phys. D, 11, 483
\bibitem{noiijmpd} Capozziello, S., Cardone, V.F., Carloni, S., Troisi, A. 2003, Int. J. Mod. Phys. D, 12, 1969
\bibitem{ccf} Capozziello, S., Cardone, V.F., Francaviglia, M. 2006, Gen. Rel. Grav., 38, 711
\bibitem{cct} Capozziello, S., Cardone, V.F., Troisi, A. 2005, Phys. Rev. D, 71, 043503
\bibitem{prl} Capozziello, S., Cardone, V.F., Troisi, A. 2006, JCAP, 0608, 001
\bibitem{CapCardTro07} Capozziello, S., Cardone, V.~F., Troisi, A., 2007, MNRAS, 375, 1423
\bibitem{noipla} Capozziello, S., Cardone, V.F., Carloni, S., Troisi, A. 2004, Phys. Lett. A, 326, 292
\bibitem{emilio} Capozziello, S., Cardone, V.F., Elizalde, E., Nojiri, S., Odintsov, S.D., 2006, Phys. Rev. D 73, 043512
\bibitem{arturocqg} Capozziello, S., Stabile A., Troisi, A., 2007, Class. quant. grav., 24, 2153
\bibitem{noi-prd} Capozziello, S., Stabile, A., Troisi, A., 2008, Physical Review D, 76, 104019
\bibitem{PPNfelix} Capozziello, S.,  De Laurentis, M., Nojiri, S., Odintsov, S.D., 2008, arXiv:0808.1335 [hep-th]
\bibitem{noirev} Capozziello, S., Carloni, S., Troisi, A. 2003, Rec. Res. Devel. Astronomy. \& Astrophys., 1, 625, astro\,-\,ph/0303041
\bibitem{capfra} Capozziello S. and Francaviglia M., 2008, Gen. Rel. Grav.: Special Issue on Dark Energy   40,  357.
\bibitem{izzo} Capozziello S. and Izzo L., 2008, A\&A, 490, 31.
\bibitem{ppnantro} Capozziello, S., Troisi, A. 2005, Phys. Rev. D, 72, 044022
\bibitem{Care04} Caresia, P., Matarrese, S., Moscardini, L., 2004, ApJ, 605, 21
\bibitem{sante} Carloni, S., Dunsby, P.K.S., Capozziello, S., Troisi, A. 2005, Class. Quantum Grav. 22, 2839
\bibitem{cdtt} Carroll, S.M., Duvvuri, V., Trodden, M., Turner, M.S. 2004, Phys. Rev. D, 70, 043528
\bibitem{CarLam} Carroll, S.M., Press, W.H., Turner, E.L. 1992, ARA\&A, 30, 499
\bibitem{CV07} Cattoen, C., Visser, M., 2007, Class. Quant. Grav., 24, 5985
\bibitem{cembranos} Cembranos, J.A.R. 2006, Phys. Rev. D, 73, 064029
\bibitem{Chak08} Chakrabarty, D., de Filippis, E.,\& Russell, H.\ 2008, A\&A, 487, 75
\bibitem{CPL} Chevallier, M., Polarski, D., 2001, Int. J. Mod. Phys. D, 10, 213
\bibitem{chiba} Chiba, T. 2003, Phys. Lett. B, 575, 1
\bibitem{cb05} Clifton, T., Barrow, J.D. 2005, Phys. Rev. D, 72, 103005
\bibitem{clo05} Clocchiati, A. et al. 2006, APJ, 642, 1
\bibitem{Cogno08} Cognola, G., Elizalde, E., Nojiri, S., Odintsov, S.D., Sebastiani, L., Zerbini, S., 2008, Phys. Rev. D, 77, 046009
\bibitem{cole05} Cole, S. et al. 2005, MNRAS, 362, 505
\bibitem{Coor} Cooray, A., Huterer, D., Holz, D.E., 2006, Phys. Rev. Lett., 96, 021301
\bibitem{copeland} Copeland E.J., Sami M.,  Tsujikawa S., 2006, Int. Jou. Mod. Phys. D 15, 1753
\bibitem{ALPACA} Corasaniti, P.S., LoVerde, M., Crotts, A. et al., 2006, MNRAS, 369, 798
\bibitem{chd99} Croft, R.A.C., Hu, W., Dave, R. 1999, Phys. Rev. Lett., 83, 1092

\bibitem{Boom} de Bernardis, P. et al. 2000, Nature, 404, 955
\bibitem{dbb02} de Blok, W.J.G., Bosma, A. 2002, A\&A, 385, 816
\bibitem{db05} de Blok, W.J.G. 2005, ApJ, 634, 227
\bibitem{DeFil05} De Filippis, E., Sereno, M., Bautz, Longo G., M.~W., 2005, ApJ, 625, 108
\bibitem{Dabro05} Dabrowski, M.P., 2005, Phys. Lett. B, 625, 184
\bibitem{Dabro061} Dabrowski, M.P., Stachowiak, T., 2006, Annals of Physics, 321, 771
\bibitem{Dabro062} Dabrowski, M.P., 2006, Annalen der Physik, 15, 352
\bibitem{D07} Davis, T., et al., 2007, ApJ, 666, 716
\bibitem{Demia06} Demianski, M., Piedipalumbo, E., Rubano, C., Tortora, C., 2006, A\&A, 454, 55
\bibitem{dick} Dick, R. 2004, Gen. Rel. Grav., 36, 217
\bibitem{Dode02} Dodelson S. et al., 2002, ApJ, 572, 140
\bibitem{dolgov} Dolgov, A.D., Kawasaki, M. 2003, Phys. Lett. B, 573, 1
\bibitem{D05}Dunkley, J., Bucher, M., Ferreira, P.~G., Moodley, K., Skordis, C., 2005, MNRAS, 356, 925
\bibitem{DGP} Dvali, G.R., Gabadadze G., Porrati M., 2000, Phys. Lett. B, 485, 208
\bibitem{DGP2} Dvali, G.R., Gabadadze, G., Kolanovic, M., Nitti, F., 2001, Phys. Rev. D, 64, 084004
\bibitem{DGP3} Dvali, G.R., Gabadadze, G., Kolanovic, M., Nitti, F., 2002, Phys. Rev. D, 64, 024031

\bibitem{eckhardt}  Eckhardt, D.H., 1993 Phys. Rev. {\bf 48 D}, 3762.
\bibitem{eis05} Eisenstein, D. et al. 2005, ApJ, 633, 560
\bibitem{eke98} Eke, V.R., Cole, S., Frenk, C.S., Petrick, H.J. 1998, MNRAS, 298, 1145

\bibitem{faraoni} Sotiriou, T.P. and Faraoni, V., 2008, arXiv:0805.1726 [gr-qc]
\bibitem{fischbach} E. Fischbach, E.,  Sudarsky, D.,  Szafer, A.,   Talmadge, C. and Aroson, S.H., 1986, Phys. Rev. Lett. {\bf 56}, 3
\bibitem{Freedman} Freedman, W.L. et al., 2001, ApJ, 553, 47
\bibitem{Friem} Frieman, J.A., 1997, Comments Astrophys., 18, 323
\bibitem{Frigerio} Frigerio Martins C. and Salucci P., 2007, Mon.Not.Roy.Astron.Soc. 381, 1103
\bibitem{fujii1} Y. Fujii, Y., 1988, Phys. Lett. {\bf B 202}, 246
\bibitem{stbook} Fujii, I., Maeda, K., {\it The scalar\,-\,tensor theory of gravity}, Cambridge University Press, Cambridge (UK), 2003

\bibitem{Garn98} Garnavich P.M. et al., 1998, ApJ, 509, 74
\bibitem{gibbons} G.W. Gibbons, G.W. and  Whiting, B.F., 1981,  {\it Nature} {\bf 291}, 636
\bibitem{Gunna} Gunnarsson, C., Dahlen, T., Goobar, A., J\"{o}nsson, J., M\"{o}rtsell, E., 2006, ApJ, 640, 417

\bibitem{John04} John, M.V., 2004, ApJ, 614, 1
\bibitem{John05} John, M.V., 2005, ApJ, 630, 667
\bibitem{Jon} J\"{o}nsson, J., Dahlen, T., Goobar, A., Gunnarsson, C., M\"{o}rtsell, E., Lee, K., 2006, ApJ, 639, 991

\bibitem{Hawk03} Hawkins E. et al., 2003, MNRAS, 346, 78
\bibitem{Holz} Holz, D.E., Wald, R.M., 1998, Phys. Rev. D, 58, 063501
\bibitem{Holz05} Holz, D.E., Linder, E.V., 2005, ApJ, 631, 678
\bibitem{Hu} Hu, W., Sawicki, I., 2007, Phys. Rev. D, 76, 064004
\bibitem{Hui} Hui, L., Greene, P.B., 2006, Phys. Rev. D, 73, 123526

\bibitem{PanSTARRS} Kaiser N. and the PanSTARRS Team, in {\it Bullettin of the Americal Astronomical Society}, page 1049, 2005
\bibitem{kenmoku} M. Kenmoku, M.,  Okamoto, Y. and Shigemoto, K., 1993, {\it Phys. Rev.} {\bf 48 D}, 578
\bibitem{Kim} Kim, A.G., Linder, E.V., Miquel, R., Mostek, N., 2004, MNRAS, 347, 909
\bibitem{Kirk} Kirkman, D., Tyler, D., Suzuki, N., O'Meara, J.M., Lubin, D., 2003, ApJS, 149, 1
\bibitem{Klein02} Kleinert H., Schmidt, H.-J., 2002, Gen. Rel. Grav.  34, 1295
\bibitem{hj} Kluske, S., Schmidt, H.J. 1996, Astron. Nachr., 37, 337
\bibitem{Knop03} Knop R.A. et al., 2003, ApJ, 598, 102
\bibitem{Koivisto} Koivisto T., 2007, Phys.Rev.D76, 043527

\bibitem{lecian} Lecian O.M. and   Montani G., arXiv:0807.4428 [gr-qc](2008)
\bibitem{Linder03} Linder, E.V., 2003, Phys. Rev. Lett., 90, 091301
\bibitem{lobo}Lobo F.S.N., arXiv: 0807.1640[gr-qc] (2008)
\bibitem{Lue} Lue, A., Scoccimarro, R., Starkman, G., 2004, Phys. Rev. D, 69, 044005
\bibitem{Lue2} Lue, A., Scoccimarro, R., Starkman, G., 2004, Phys. Rev. D, 69, 124015

\bibitem{mcd04} McDonald, P. et al. 2005, ApJ, 635, 761
\bibitem{mcgaugh} S.S. McGaugh, 2000, Ap. J. Lett. {\bf 541}, L33
\bibitem{Mendoza} Mendoza S. and Rosas-Guevara Y.M., 2007, Astron.\ Astrophys.\ 472, 367

\bibitem{navarro} Navarro, I., van Acoleyen, K. 2005, Phys. Lett. B, 622, 1
\bibitem{Netter02} Netterfield C.B. et al., 2002, ApJ, 571, 604
\bibitem{NeuBoh95} Neumann., D.~M., B\"{o}hringer, H., 1995, Astron. Astrophys., 301, 865
\bibitem{no03a} Nojiri, S., Odintsov, S.D. 2003, Phys. Lett. B, 576, 5
\bibitem{odi03} Nojiri S., Odintsov, S.D., 2003, Mod. Phys. Lett. A, 19, 627
\bibitem{nodi03} Nojiri S., Odintsov, S.D., 2003, Phys. Rev. D, 68, 12352
\bibitem{odi} Nojiri S. and Odintsov S.D. 2007, Int.\ J.\ Geom.\ Meth.\ Mod.\ Phys.\ 4, 115
\bibitem{Odintsov} Nojiri, S., Odintsov, S.D., 2007, Phys. Lett. B, 652, 343
\bibitem{Odi} Nojiri,  S., Odintsov, S.D., 2007, Phys. Lett. B, 657, 238
\bibitem{Nodi07} Nojiri S., Odintsov, S.D., Phys. Rev. D accepted (arXiv:0710.1738 [hep-th]) 2007
\bibitem{double} Nojiri, S., Odintsov, S. D., arXiv:0801.4843 [astro-ph] 2008
\bibitem{Nordin} Nordin, J., Goobar, A., J\"{o}nsson, J., 2008, JCAP, 02, 008

\bibitem{olmo} Olmo, G.J. 2005, Phys. Rev. D, 72, 083505

\bibitem{Pad03} Padmanabhan, T. 2003, Phys. Rept., 380, 235
\bibitem{PB03} Peebles, P.J.E., Rathra, B. 2003,  Rev. Mod. Phys., 75, 559
\bibitem{Perci02} Percival W.J. et al., 2002, MNRAS, 337, 1068
\bibitem{Bergliaffa} Perez Bergliaffa, S.E., 2006, Phys. Lett. B, 642, 311
\bibitem{Perlm97} Perlmutter S. et al., 1997, ApJ, 483, 565
\bibitem{Perlm99} Perlmutter S. et al., 1999, ApJ, 517, 565
\bibitem{Petto05} Pettorino, V., Baccigalupi, C., Mangano, G., 2005, JCAP, 0501, 014
\bibitem{pope04} Pope, A.C. et al. 2005, ApJ, 607, 655
\bibitem{poplawski06} Poplawski, N., 2006, Phys. Lett. B, 640, 135
\bibitem{poplawski07} Poplawski, N., 2007, Class. Quant. Grav., 24, 3013

\bibitem{Qua91} Quandt, I., Schmidt H.~J., 1991, Astron. Nachr., 312, 97

\bibitem{Rebo04} Rebolo R. et al., 2004, MNRAS, 353, 747
\bibitem{refr03} Refregier, A. 2003, ARA\&A, 41, 645
\bibitem{Riess98} Riess A.G. et al., 1998, AJ, 116, 1009
\bibitem{Riess04} Riess, A.G. et al. 2004, ApJ, 607, 665
\bibitem{R06} Riess, A.G. et al., 2007, ApJ, 659, 98

\bibitem{Sahni} Sahni, V., Starobinski, A. 2000, Int. J. Mod. Phys. D, 9, 373
\bibitem{SF} Sahni, V., Saini, T.D., Starobinsky, A.A., Alam, U., 2003, JETP Lett., 77, 201
\bibitem{sanch05} Sanchez, A.G. et al. 2006, MNRAS, 366, 189
\bibitem{sanders}  Sanders, R.H., 1990,  Ann. Rev. Astr. Ap. {\bf 2}, 1
\bibitem{Sark} Sarkar, D., Amblard, A., Holz, D.E., Cooray, A., 2008, ApJ, 678, 1
\bibitem{Schindler02} Schindler, S., 2004, Astrophys.Space Sci., 28, 419
\bibitem{Schmidt98} Schmidt B.P. et al., 1998, ApJ, 507, 46
\bibitem{hjrev} Schmidt, H.J. 2004, {\it Lectures on mathematical cosmology}, gr\,-\,qc/0407095
\bibitem{SA06} Schmidt, R.~W., Allen S.~W., 2007 MNRAS 379, 209
\bibitem{SKY} Schmidt, B.P., Keller, S.C., Francis P.J. et al., {\it Bullettin of the Americal Astronomical Society}, 37, 457, 2005
\bibitem{Sel04} Seljak, U. et al. 2005, Phys. Rev. D, 71, 103515
\bibitem{sobouti} Sobouti, Y. 2007, A \& A, 464, 921
\bibitem{sotiriou} Sotiriou, T.P. 2006, Gen. Rel. Grav., 38, 1407
\bibitem{speake} Speake, C.C. and Quinn, T.J., 1988, Phys. Rev. Lett. {\bf 61}, 1340
\bibitem{WMAP} Spergel, D.N. et al. 2003, ApJS, 148, 175
\bibitem{WMAP3} Spergel, D.N. et al., 2007, ApJS, 170, 377
\bibitem{Starobinsky} Starobinsky, A.A., 2007, JETP Lett., 86, 157
\bibitem{b10} Stelle, K., 1978, Gen. Relat. Grav., 9, 353
\bibitem{Stomp01} Stompor R. et al., 2001, ApJ, 561, L7
\bibitem{Szal03} Szalay A.S. et al., 2003, ApJ, 591, 1

\bibitem{DES} The Dark Energy Survey Collaboration, astro\,-\,ph/0510346, 2005
\bibitem{Teg03} Tegmark, M. et al. 2004, Phys. Rev. D, 69, 103501
\bibitem{Teg06} Tegmark M. et al., 2006, Phys. Rev. D, 74, 123507
\bibitem{Teg97} Tegmark, M., Taylor, A.N., Heavens, A.F., 2007, ApJ, 4802, 22
\bibitem{Tonry03} Tonry J.L. et al., 2003, ApJ, 594, 1
\bibitem{Tsuji} Tsujikawa, S., 2008,  Phys. Rev. D 77, 023507
\bibitem{LSST} Tyson, J.A. in {\it Survey and Other Telescope Technologies and Discoveries}, ed. J.A. Tyson and S.Wolff, page 10, Sidney, 2002

\bibitem{vW01} van Waerbecke, L. et al. 2001, A\&A, 374, 757
\bibitem{vnl02} Viana, P.T.P., Nichol, R.C., Liddle, A.R. 2002, ApJ, 569, 75
\bibitem{Vik05} Vikhlinin, A., Markevitch, M., Murray, S.~S., Jones, C., Forman, W., Van Speybroeck, L., 2005, ApJ, 628, 655
\bibitem{Vik06} Vikhlinin, A., Kravtsov, A., Forman, W., Jones, C., Markevitch, M., Murray, S.~S., Van Speybroeck, L., 2006, ApJ, 640, 691
\bibitem{Visser} Visser, M., 2004, Class. Quant. Grav., 21, 2603

\bibitem{Wamb} Wambsganss, J., Cen, R., Gu, X., Ostriker, J.P., 1997, ApJ, 475, L81
\bibitem{WangFlux} Wang, Y., ApJ, 536, 531
\bibitem{WM04} Wang, Y., Mukherejee, P., 2004, ApJ, 606, 654
\bibitem{W72} Weinberg, S., {\it Gravitation and cosmology}, Wiley,  New York, 1972
\bibitem{will}  Will C.M., 1993, {\it Theory and Experiments in Gravitational Physics}, Cambridge Univ. Press, Cambridge.
\bibitem{ESSENCE} Wood\,-\,Vasey, W.M., et al., 2007, ApJ, 666, 694

\bibitem{Zwi33} Zwicky, F., 1933, Helv. Phys. Acta, 6, 110

\end{thebibliography}
\end{document}